\documentclass{jpp}
\usepackage{graphicx}

\usepackage[utf8]{inputenc}
\usepackage[T1]{fontenc}
\usepackage{amsmath}
\usepackage{soul}

\newcommand\bNabla{\boldsymbol{\nabla}}

\usepackage{xcolor}
\usepackage{hyperref}

\usepackage{orcidlink}

\shorttitle{Lattice Boltzmann scheme for Hall-MHD}
\shortauthor{R. Foldes, E. Lévêque, R. Marino, E. Pietropaolo et al.}

\title{Efficient kinetic Lattice Boltzmann simulation of three-dimensional Hall-MHD Turbulence}

\author{Raffaello Foldes\corresp{\email{raffaello.foldes@ec-lyon.fr}}\aff{1,2}\orcidlink{0000-0001-7977-4823}
Emmanuel Lévêque\aff{1}\orcidlink{0000-0003-3745-4690},
 Raffaele Marino\aff{1}\orcidlink{0000-0002-6433-7767},
 Ermanno Pietropaolo\aff{2}\orcidlink{0000-0002-6633-9846},
 Alessandro De Rosis\aff{3}\orcidlink{0000-0002-4493-365X},
 {Daniele Telloni}\aff{4}\orcidlink{0000-0002-6710-8142}
 \and
 Fabio Feraco\aff{1,5}\orcidlink{0000-0002-5550-2070}
 }

\affiliation{
\aff{1}Univ Lyon, CNRS, \'Ecole Centrale de Lyon, INSA Lyon, Univ Claude Bernard Lyon I, LMFA UMR 5509, F-69134 Ecully cedex, France
\aff{2}Dipartimento di Scienze Fisiche e Chimiche, Universit\`a dell'Aquila, 67100 Coppito (AQ), Italy
\aff{3}Department of Mechanical, Aerospace and Civil Engineering, The University of Manchester, Manchester M13 9PL, United Kingdom
\aff{4}National Institute for Astrophysics - Astrophysical Observatory of Torino, Via Osservatorio 20, I-10025 Pino Torinese, Italy
\aff{5}Leibniz-Institute of Atmospheric Physics at the University of Rostock,
Schlossstrasse 6, K\"uhlungsborn, 18225, Germany}
\begin{document}

\maketitle

\begin{abstract}
Simulating plasmas in the Hall-MagnetoHydroDynamics (Hall-MHD) regime represents a valuable {approach for the investigation of} complex non-linear dynamics developing in astrophysical {frameworks} and {fusion machines}. Taking into account the Hall electric field is {computationally very challenging as} it involves {the integration of} an additional term, proportional to $\bNabla \times ((\bNabla\times\mathbf{B})\times \mathbf{B})$ in the Faraday's induction {law}. {The latter feeds back on} the magnetic field B at small scales (between the ion and electron inertial scales), {requiring} very high resolution{s} in both space and time {in order to properly describe its dynamics.} The computational {advantage provided by the} kinetic Lattice Boltzmann (LB) approach is {exploited here to develop a new} code, the \textbf{\textsc{F}}ast \textbf{\textsc{L}}attice-Boltzmann \textbf{\textsc{A}}lgorithm for \textbf{\textsc{M}}HD \textbf{\textsc{E}}xperiments (\textsc{flame}). The \textsc{flame} code integrates the plasma dynamics in lattice units coupling two kinetic schemes, one for the fluid protons (including the Lorentz force), the other to solve the induction equation describing the evolution of the magnetic field. Here, the newly developed algorithm is tested against an analytical wave-solution of the dissipative Hall-MHD equations, pointing out its stability and second-order convergence, over a wide range of the control parameters. Spectral properties of the simulated plasma are finally compared with those obtained from numerical solutions from the well-established pseudo-spectral code \textsc{ghost}. Furthermore, the LB simulations we present, varying the Hall parameter, highlight the transition from the MHD to the Hall-MHD regime, in excellent agreement with the magnetic field spectra measured in the solar wind.
\end{abstract}

\section{Introduction}
In the frame of the MHD model{,} plasma is treated as a single species quasi-neutral fluid with conductive properties sensitive to the action of the magnetic field~\citep{galtier2016}. 
In the ideal MHD description, ions and electrons are tied to the magnetic field, moving with the same velocity. The Hall-MHD model relaxes the MHD prescriptions assuming ions disunite from the magnetic field due to their inertia, while electrons remain bound to it \citep{Pandey2008}. In this framework, the resistive Ohm's law is generalized through the introduction of the Hall electric field, proportional to $\mathbf{J} \times \mathbf{B}$, where $\mathbf{J}$ and $\mathbf{B}$ denote the current density and the magnetic field, respectively. 
The Hall electric field has an effect on the magnetic field at length scales shorter than the ion inertial length $d_i=c/\omega_{pi}$ ($\omega_{pi}$ being the ion plasma frequency, $c$ the speed of light) as well as at time scales shorter than the ion cyclotron period $1/\omega_{ci}$~\citep{Huba2003}.  The scale $d_i$ corresponds to the scale at which ions and electrons decouple, and the magnetic field becomes frozen into the electron fluid rather than in the bulk plasma.
Hall-MHD has been already adopted in literature to describe a variety of astrophysical, space and laboratory environments, and to provide a detailed description of plasma dynamics. Its applications span from the star formation \citep{Norman1985,CRAL2018} to the solar atmosphere and the solar wind \citep{Galtier2007,Gonzalez2019}, and it has been used also to investigate magnetic reconnection processes \citep{Wang2001,Morales2005,Ma2018} and the dynamo action \citep{Mininni2002,Mininni2005,Gomez2010}.
A major difficulty in simulating Hall-MHD is related to the need to resolve whistler waves, evolving on fast dynamics with a phase speed $c_w(k) \propto k$ increasing linearly with the wavenumber $k$. 
In order to properly account for the propagation of the perturbations caused by the Hall effect, it is, therefore, necessary to capture those plasma waves with $\max(c_w) \propto 1/\Delta x$, at the smallest resolved wavelength $\Delta x$. The Courant-Friedrichs-Lewy (CFL) condition then yields ${\Delta t} \propto \Delta x^2$.
This scaling implies a rapid decrease of the time-step as the spatial resolution increases, which poses severe limitations in terms of computational cost. 
Nevertheless, Hall-MHD simulations have been proposed over the years in numerous studies, through the integration of the equations with pseudo-spectral \citep{Mininni2003}, finite-volume \citep{TOTH2008,CRAL2018} or hybrid particle-in-cell codes~\citep{Ma2018,Papini2019}. When dealing with turbulent flows, pseudo-spectral methods are usually recognized as the best option that allows for an equally-accurate representation of the fields at the resolved dynamical scales \citep{Patterson1971}. On the other hand, their computational cost can be prohibitive (as mentioned before) when it comes to the integration of simulations in three dimensions and for many turnover times~\citep{Huba2003}.  
{The main} purpose of the novel code that we developed here, \textsc{flame} (\textbf{\textsc{F}}ast \textbf{\textsc{L}}attice-Boltzmann \textbf{\textsc{A}}lgorithm for \textbf{\textsc{M}}HD \textbf{\textsc{E}}xperiments), is to overcome this issue. Indeed, the Lattice Boltzmann (LB) implementation provides an alternative to achieve a convenient trade-off between accuracy and computational efficiency. Unlike more traditional methods that solve the dynamics of flows at the macroscopic level, LB methods operate at an underlying mesoscopic kinetic level. The flow complexity emerges from re-iterating simple rules of collision and streaming of populations of particles moving along the links of a regular cubic lattice \citep{Kruger2016}. The connection between such an idealized representation and the macroscopic dynamics is by now well-established and accepted, placing the method on a solid theoretical and mathematical ground \citep{Shan1998}. Furthermore, due to its intrinsically discrete nature and its focus on the local dynamics, it is also computationally extremely efficient \citep{Korner2006}. A decisive contribution to make possible the simulation of ideal MHD plasmas by means of LB methods was made by \citet{DELLAR2002}, who showed that the native LB framework based on the Bhatnagar-Gross-Krook (BGK) collision \citep{BGK54} could be consistently extended to encompass both the fluid dynamics driven by the Lorentz force and the induction equation for the magnetic field.
The scheme introduced by Dellar overcomes the major limitations of previous efforts \citep{Montgomery1987,ChenMatthaeus1991,SucciVergassolaBenzi1991,Martinez1994} and fully complies with the macroscopic MHD equations in a weakly-compressible formulation (see \S\ref{sec:LB_global}). 
The first three-dimensional MHD simulations based on the scheme proposed by Dellar have been performed by~\citet{Breyiannis2004,Breyiannis2006}.
Nevertheless, it is prone to develop numerical instabilities when strong  gradients emerge in the flow, thus delaying in the community its implementation for the simulation of turbulent fluid frameworks. 
This deficiency is not exclusive to MHD simulations but rather an inherent aspect of the BGK collision operator itself.
%
%
By utilizing a so-called Multi-Relaxation-Time (MRT) operator defined in the space of moments, it becomes possible to explicitly dampen the non-hydrodynamic modes
and improve the stability \citep{Higuera_1989, BENZI1992, Humiere1994}.
Therefore, \cite{PATTISON2008} and \cite{Riley} opted to use MRT collision operators for the hydrodynamic parts of their lattice Boltzmann MHD algorithms, whereas \cite{Dellar2009} enhanced stability by considering MRT operators for both the hydrodynamic and magnetic aspects. 
An entropic stabilization has also been proposed by \cite{Vahala2018}, though leading to a more complicated scheme. 
These advances encouraged us to pursue the LB modeling to simulate Hall-MHD turbulence, an effort that has never been undertaken previously. 
In the present study, a MRT scheme based on central-moments is considered for the evolution of the velocity field, while dynamics of the magnetic field evolve under the action of a BGK collision operator, following the scheme described in \cite{DeRosis18}.
It is worth noting that \cite{Mendoza2008} had previously introduced a lattice Boltzmann algorithm for simulating two charged species along with Maxwell's equations in the Hall MHD regime, as detailed in the next section. Our approach, however, is more straightforward, neglecting the electron inertia term.
The development of \textsc{flame} was also strongly motivated by the need of the community for innovative numerical tools for the study of space plasma turbulent dynamics at scales that are by now within the reach of  high-resolution instruments on board spacecrafts, such as the ESA mission Solar Orbiter~\citep{2020A&A...642A...1M}.\\ 
The paper is organized as follows. In \S\ref{sec:HallMHD}, the Hall-MHD equations are presented in a form that is relevant for LB developments. The LB scheme implemented in \textsc{flame} is introduced and discussed in \S\ref{sec:LB_global}. The coupling between the fluid and the magnetic lattices is explained, as well as the inclusion of the Hall effect in the collision operator. The conversion from physical to lattice units is discussed in great detail. \S\ref{sec:Results} is devoted to the validation of the code against an analytical solution of the dissipative Hall-MHD equations~\citep{Xia15}. This section provides an assessment of the numerical stability and a quantitative estimation of the dispersion and dissipation errors. 
The computational efficiency is discussed in \S\ref{sec:comp_efficiency}, where GPU-accelerated simulations of the three-dimensional Orszag-Tang vortex problem are considered \citep{Orszag1979}.
In a regime of high Reynolds numbers, we show that LB simulations are able to reproduce the break in the magnetic energy spectrum at sub-ion scales, in perfect agreement with solar-wind measurements. Finally, we outline potential applications for the investigation of 
space plasmas in \S\ref{sec:sim_for_space_plasma}, and draw conclusions in \S\ref{sec:conclusion}.

\section{The Hall-MHD equations}\label{sec:HallMHD}
In this section, the Hall-MHD equations are introduced in the standard incompressible approximation and in a weakly-compressible formulation, suitable for LB developments. 

\subsection{Incompressible formulation}\label{subsec:inc_HMHD}
In this context, when we refer to the macroscopic description of the plasma what we mean is the description of the prognostic fields appearing in the model equations. Thus, at the macroscopic level, the incompressible resistive MHD equations for an electrically conductive quasi-neutral fluid consist of the incompressible Navier-Stokes equations with the addition of the Lorentz force, coupled with the resistive induction equation for the magnetic field: 
\begin{equation}\label{eq:incompressibility}
    \bNabla \cdot \mathbf{U} = 0 
\end{equation}
\begin{equation}\label{eq:NSE}
    \partial_t \mathbf{U}+(\mathbf{U} \cdot \bNabla)\mathbf{U}=\frac{1}{\rho_0}\mathbf{J}\times \mathbf{B}-\frac{1}{\rho_0}\bNabla \mathrm{p} + \nu \bNabla^2 \mathbf{U}
\end{equation}
\begin{equation}\label{eq:MFE}
    \partial_t \mathbf{B}=\bNabla \times (\mathbf{U}\times \mathbf{B} - \eta \bNabla \times \mathbf{B})
\end{equation}
\begin{equation}\label{eq:divfreeB}
    \bNabla \cdot \mathbf{B}=0
\end{equation}
where {$t$ is the time,} $\rho_0$ is the mass density of the fluid,  $\nu$ is the kinematic viscosity and $\eta$ it the magnetic resistivity. 
The electric current density is expressed as $\mathbf{J}={1}/{\mu_0}\bNabla \times \mathbf{B}$, where $\mu_0$ is the magnetic permeability in the vacuum. 
To account for the Hall effect, it is necessary to take a step back in the mathematical developments and resort to a two-fluid description that includes the fluid equations for both ions and electrons  separately. For a fully ionized plasma in which the masses of {ions} {(mainly protons)} and electrons (hereafter $i$ and $e$) are $m_e\ll m_i \approx m$, the momentum equations read as
\begin{equation}\label{eq:fluid_ions}
    \rho[\partial_t \mathbf{U} + (\mathbf{U} \cdot \bNabla)\mathbf{U}] = en(\mathbf{E}+\mathbf{U}\times \mathbf{B})-\bNabla \mathrm{p}_i + \bNabla \cdot \mathbf \sigma + \mathbf{R}
\end{equation}
\begin{equation}\label{eq:fluid_electrons}
    0 = -en(\mathbf{E}+\mathbf{U_e}\times \mathbf{B})-\bNabla \mathrm{p}_e - \mathbf{R}
\end{equation}
where {$e$ is the unit electric charge,} $\mathbf \sigma$ is the viscous stress tensor, $n$ is the particle density with $\rho=mn$, and $\mathbf{R}$ is the rate (per unit volume) of momentum exchange due to collisions between protons and electrons. 
The latter is given by $\mathbf{R}=-mn f_{ie}(\mathbf{U}-\mathbf{U_{e}})$ where $f_{ie}$ denotes the collision frequency and can be reformulated as $\mathbf{R}=-(m f_{ie}/e)\ \mathbf{J}$, with the density current $\mathbf{J}=en(\mathbf{U}-\mathbf{U_e})$.
By summing~\eqref{eq:fluid_ions} and~\eqref{eq:fluid_electrons} and assuming  {$\sigma_{\alpha \beta}=\rho \nu(\partial_\alpha U_\beta+\partial_\beta U_\alpha)$}, one obtains
\begin{equation}\label{eq:NSE_2fluid}
    \partial_t \mathbf{U}+(\mathbf{U} \cdot \bNabla)\mathbf{U}=\frac{1}{\rho}\mathbf{J}\times \mathbf{B}- \frac{1}{\rho}\bNabla \mathrm{p} + \nu \bNabla^2 \mathbf{U}.
\end{equation}
On the other hand, by replacing $\mathbf{U_e}$ by $\mathbf{U}-\mathbf{J}/ne$ and the expression for the rate of momentum exchange into \eqref{eq:fluid_electrons}, the  Ohm's law becomes
\begin{equation}\label{eq:gen_Ohm}
    \mathbf{E}=-(\mathbf{U}-\frac{1}{en}\mathbf{J})\times \mathbf{B}+\frac{1}{en}\bNabla \mathrm{p}_e + \frac{m f_{ie}}{e^2 n}\mathbf{J}.
\end{equation}
Taking the curl of this equation gives in the end an induction equation with Hall's current correction in standard physical units as
\begin{equation}\label{eq:Hall_ind}
    \partial_t \mathbf{B}=\bNabla \times [(\mathbf{U}-\alpha_H\mathbf{J})\times \mathbf{B}] + \eta \bNabla^2 \mathbf{B}
\end{equation}
where $\alpha_H=1/en$ is usually referred to as the {Hall parameter} and the magnetic resistivity $\eta = m f_{ie}/(e^2 n \mu_0)$.
Let us note that, in general, $\bNabla \times (({1}/{en)}\bNabla \mathrm{p}_e) = -({1}/{en^2})\bNabla n \times \bNabla p_e$ \citep{kulsrud}. However, in the current context, we make the assumption that the electrons are isothermal, resulting in a dynamic pressure $p_e = n T_e$, where $T_e$ is a constant plasma temperature. Therefore, $\bNabla n \times \bNabla p_e = 0$.
The Hall-MHD equations mentioned earlier include a finite ion-electron collision frequency, responsible for the $\mathbf{R}$-coupling term in \eqref{eq:fluid_ions} and \eqref{eq:fluid_electrons}. Additionally, they assume that the ion-ion collision frequency is large enough (much greater than the ion gyrofrequency) to permit the adoption of a standard Newtonian viscous stress in \eqref{eq:NSE_2fluid}. The more comprehensive \emph{Braginskii MHD} model, on the other hand, allow for the ion-ion collision frequency to be comparable to the ion gyrofrequency.
Consequently, the Hall term emerges as just one component of the anisotropic relationship between electric current and electric field, and between stress and strain rate, with a preferred direction determined by the magnetic field.
\cite{DELLAR2011} provided a first LB approach to simulating the Braginskii MHD equations by modifying the hydrodynamics collision operator to depend on the magnetic field.
Here, the main target of our simulations is represented by space plasmas providing a clear context for the use of Hall-MHD equations.

\subsection{Weakly-compressible formulation}\label{subsec:weakly_comp_HMHD}

Incompressibility is an assumption made at the macroscopic level and  cannot be implemented in the mesoscopic representation as this would imply that fluid particles move at infinite speed, in order to adapt instantaneously the pressure. 
Incompressibility can nevertheless be approached in the so-called \emph{weakly-compressible} limit, in which the speed of sound waves $c_s$ becomes much larger than the typical fluid velocity $U_0$, or equivalently, the pressure field adapts in a time shorter than the time over which the flow evolves. This regime is attained for vanishing  Mach number, $\mathrm{Ma} \equiv U_0 /c_s$. 
Consequently, the incompressible equations should be replaced with the compressible formulation
\begin{equation}\label{eq:compressibility}
    \partial_t \rho + \bNabla \cdot (\mathbf{\rho \mathbf{U}}) = 0
\end{equation}
\begin{equation}\label{eq:NSE_Mtensor}
    \partial_t (\mathbf{\rho U}) + \bNabla \cdot ( \rho \mathbf{U} \otimes \mathbf{U} + \mathrm{p} \mathbb{I} +\frac{1}{2}|\mathbf{B}|^2\mathbb{I}-\mathbf{B}\otimes \mathbf{B}) =\bNabla \cdot \mathbf{\sigma} 
\end{equation}
in which $\mathbf{\sigma}$ represents the viscous stress, the Lorentz force has been rewritten in a conservative form as the divergence of the Maxwell stress tensor $M_{\alpha \beta}=\frac{1}{2}|\mathbf{B}|^2\delta_{\alpha \beta}-B_{\alpha}B_{\beta}$\footnote{The notation 
$\left (\mathbf{a}\otimes \mathbf{b}\right )_{\alpha \beta} \equiv a_\alpha b_\beta$} is adopted, and $\mu_0$ has been absorbed by replacing $\mathbf{B}$ with $\mu_0^{1/2}\mathbf{B}$. This (standard) normalization will be assumed hereafter, which allows for simplifying the Lorentz force 
as $(\bNabla \times \mathbf{B})\times \mathbf{B}$.
The general form of the viscous stress is
\begin{equation} 
\mathbf{\sigma} = \mu ( \bNabla \mathbf U +  (\bNabla \mathbf U)^T ) +(\zeta - \frac{2}{3} \mu) (\bNabla \cdot \mathbf U) \mathbb{I} 
\end{equation}
where $\mu$ is the dynamic viscosity ($\nu=\mu/\rho$) and $\zeta$ is the bulk viscosity.
Compressibility requires resorting to an \textit{equation of state} linking pressure, mass density and temperature. Here, the low-Mach limit justifies the use of a simple isothermal relation
\begin{equation}\label{eq:EoS}
    \mathrm{p}=\rho c_s^2
\end{equation}
which is consistent with $O(\mathrm{Ma}^2)$ mass-density fluctuations. 
The induction equation describing the evolution of the magnetic field can be rewritten in the same fashion as
\begin{equation}\label{eq:Hall_MHD}
\partial_t \mathbf{B}+\bNabla \cdot \left((\mathbf{U}-\alpha_H \mathbf{J}) \otimes \mathbf{B}-\mathbf{B} \otimes (\mathbf{U}-\alpha_H \mathbf{J}) \right) = \eta \bNabla^2 \mathbf{B}.
\end{equation}
Let us remark that following the normalization of $\mathbf{B}$ by $\mu_0^{1/2}$, the Hall current $\alpha_H \mathbf{J}$ reads as ${\alpha_H}/{{\mu_0}^{1/2}}~\bNabla \times \mathbf{B}$.
In the next sections, the developed LB scheme will conform to the set of equations \eqref{eq:compressibility}, \eqref{eq:NSE_Mtensor}, \eqref{eq:EoS} and~\eqref{eq:Hall_MHD}. 
The divergence-free condition on $\mathbf{B}$ is preserved by \eqref{eq:Hall_MHD}, justifying that it is sufficient to impose $\bNabla \cdot \mathbf{B}=0$ initially. In the numerical modelling, particular attention will be paid to verify that this condition is indeed preserved with accuracy.

\section{Hall-MHD Lattice Boltzmann scheme}\label{sec:LB_global}
In this section, the standard LB method for classical fluid dynamics is briefly introduced, focusing on key steps, then it is extended to encompass Hall-MHD. Further details are provided in the appendix~\ref{sec:appendix_1}. 
A central-moment collision operator \citep{DeRosis18} and a high-connectivity D3Q27 lattice are used to integrate the dynamics of the fluid protons, while the evolution of the magnetic field is accounted by a Bhatnagar-Gross-Krook (BGK) collision operator \citep{BGK54} and a low-connectivity D3Q7 lattice. 
Our original contribution to these developments is the self-consistent integration of the Hall term in the LB scheme by suitably redefining the equilibrium state for the magnetic field.

\subsection{Lattice Boltzmann scheme for the fluid dynamics}
\label{subsec:LB_HD}

\subsubsection{Standard BGK Lattice Boltzmann scheme}
The LB method~\citep{Kruger2016} is based on the idea that fluid motions can be represented by the collective behavior of fictitious (introduced in the frame of the LB integration strategy) particle populations evolving along the links of a cubic lattice. When the lattice connectivity, which accounts for the discrete directions of propagation of the particles, is high enough to satisfy sufficient isotropy, weakly-compressible Navier-Stokes dynamics can be reproduced with an $O(\mathrm{Ma}^3)$ error.
The macroscopic variables such as the fluid density {$\rho$}, momentum {$\rho \mathbf{U}$}, or stress tensor {$\boldsymbol{\Sigma}$} are obtained as statistical moments of the particle distributions, \emph{i.e.}
\begin{eqnarray}\label{eq:LB_moments}
    \rho &=& \sum_{i=0}^{N-1} f_i \\
    \rho \mathbf{U} &=& \sum_{i=0}^{N-1} f_i \mathbf{c}_i  \\
    \boldsymbol{\Sigma} &= &\sum_{i=0}^{N-1} f_i \mathbf{c}_i \otimes \mathbf{c}_i
\end{eqnarray}
by summing over the local mass densities $f_0,\cdots,f_{N-1}$ of particles moving with velocities $\boldsymbol{c}_0,\cdots,\boldsymbol{c}_{N-1}$, respectively.
The sums replace here the integrals over $\mathbf{c}$ of the classical kinetic theory as the result of a drastic decimation in velocity of the phase space. 
From a theoretical viewpoint, the LB method is derived by expanding the solution of the continuum Boltzmann equation onto a finite basis of Hermite polynomials in velocity, and by resorting to a Gaussian quadrature formula to express the statistical moments 
\citep{He-Luo1997}.
As a consequence, the particle densities ${f_i}(\mathbf{x}, t)$ evolve according to a discrete-velocity analogue of the Boltzmann equation, which reads as
\begin{equation}\label{eq:HD_LBeq}
    \partial_t f_i + \left( \mathbf c_i \cdot \bNabla\right)  f_i= - \frac{1}{\tau}\bigg{(}f_i -f_i^{(0)}(\rho, \mathbf{U})\bigg{)}
\end{equation}
under the BGK approximation \citep{BGK54}. The latter assumes that collisions are responsible for the relaxation of the particle densities towards their equilibrium state $f_i^{(0)}(\rho, \mathbf{U})$, with a unique relaxation time $\tau=\nu/c_s^2$.

\begin{figure}
\includegraphics[width=\textwidth]{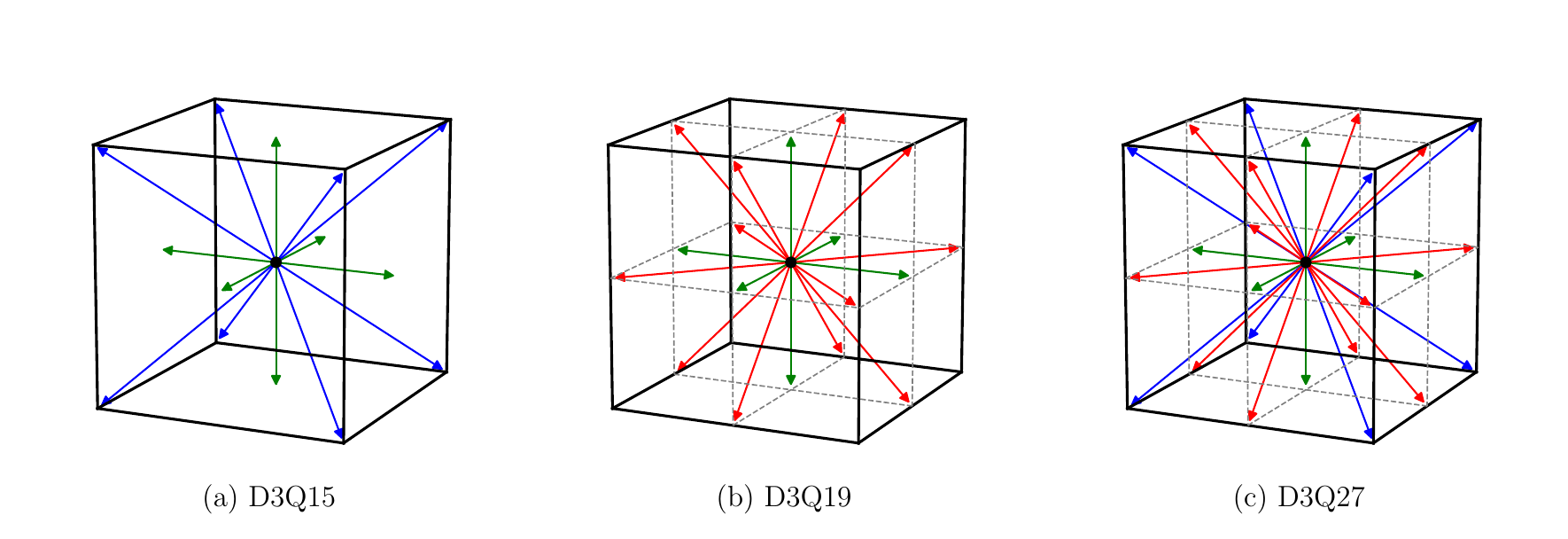}
\caption{Typical cubic lattices with 15, 19 and 27 velocities, respectively. At each lattice node, the  microscopic velocities point towards the centre (black), the 6 centres of faces (green), the 12 centres of the edges (red) or the 8 corners (blue) of a cube.
The arrows represent the local displacements $\boldsymbol{c}_i \Delta t$ of  particles from a lattice node to a neighbouring node during exactly one time-step. In the present study, a D3Q27 lattice that is more appropriate to simulate strongly non-linear fluid dynamics is considered \citep{SILVA2014}.
}
\label{fig:lattice_types}
\end{figure}

The \emph{Lattice} keyword refers to the discretization in space and time of (\ref{eq:HD_LBeq}) with a set of microscopic velocity $\boldsymbol{c}_0,~\cdots,~\boldsymbol{c}_{N-1}$ chosen in a way such that particles travel from a lattice node to a neighbour lattice node in exactly one time-step (see Fig.~\ref{fig:lattice_types}). 

The LB scheme 
can be expressed simply by using the change of variables \[\bar{f}_i = f_i + {\Delta t}/{2\tau} (f_i-f_i^{(0)}) \] originally introduced in  \citep{He1998}, as
\begin{equation}\label{eq:HD_LBscheme}
    \bar{f}_i(\mathbf{x}+\mathbf{c}_i \Delta t,t+\Delta t) = \bar{f}_i(\mathbf{x},t) - \omega\bigg{(}\bar{f}_i(\mathbf{x},t)-{f}^{(0)}_i(\rho,\mathbf{U})(\mathbf{x},t)\bigg{)}
\end{equation}
where the discrete distribution functions $\bar{f}_i(\mathbf{x},t)$ depend on the three spatial coordinates $\mathbf{x}$ and on time $t$.
This change of variable comes from the trapezoidal rule used to approximate the integral of the collision operator (right-hand side of (\ref{eq:HD_LBeq})) between $t$ and $t+\Delta t$ \citep{Kruger2016}.
It also calls for a redefinition of the relaxation time as $\tau + \Delta t/2$~\citep{Henon1987} so that
\begin{equation}
    \frac{1}{\omega}=\bigg{(}\frac{\nu}{c_s^2 \Delta t}+\frac{1}{2}\bigg{)}
\end{equation}
where the speed of sound $c_s$ is linked to the lattice spacing by $\Delta x/\Delta t = \sqrt{3}c_s$ for the D3Q27 lattice. 
The expressions of the mass density and fluid momentum as statistical moments remain unchanged with

\begin{eqnarray}\label{eq:LB_moments2}
    \rho &=& \sum_{i=0}^{N-1} \bar f_i \\
    \mathrm{and}\quad \rho \mathbf{U} &=& \sum_{i=0}^{N-1}  \bar f_i \mathbf{c}_i.
\end{eqnarray}

In practice, \eqref{eq:HD_LBscheme} is  divided into a two-step algorithm with a streaming step consecutive to a local collision operation, \emph{i.e.}

\begin{eqnarray}\label{eq:stream}
        \bar{f}_i(\mathbf{x}+\mathbf{c}_i\Delta t,t+\Delta t) &=& \bar{f}_i^*(\mathbf{x},t) \\ 
        \bar{f}^*_i(\mathbf{x},t) &=& \bar{f}_i(\mathbf{x},t) - \omega \left( \bar{f}_i(\mathbf{x},t)-{f}^{(0)}_i(\rho,\mathbf{U})(\mathbf{x},t) \right ) \label{eq:collision}.
\end{eqnarray}

To complete the algorithm, the particle densities at the equilibrium ${f}_i^{(0)}$ need to be specified. By construction, ${f}_i^{(0)}$ is defined as a truncated Hermite expansion of the continuous Maxwell-Boltzmann distribution evaluated in $\mathbf{c_i}$, which reads as 
\begin{equation}
    f_i^{(0)}(\rho, \mathbf{U}) = w_i \rho \left( 1  
    +\frac{\boldsymbol{c}_i\cdot \mathbf{U}}{c_s^2}
    +\frac{(\boldsymbol{c}_i\cdot \mathbf{U})^2}{2c_s^4}
    - \frac{\mathbf{U}\cdot \mathbf{U}}{2c_s^2}
    + \cdots
    \right)
\end{equation}
with the weights $w_\mathrm{center}=8/27$, $w_\mathrm{face}=2/27$, $w_\mathrm{edge}=1/54$ and $w_\mathrm{corner}=1/216$ for the D3Q27 lattice~\citep{He-Luo1997}. 
An expansion truncated at the second order is {enough} to recover the Navier-Stokes equations with an $O(\mathrm{Ma}^3)$ error. However, several groups \citep{malaspinas2015increasing, coreixas2017recursive, COREIXAS_PRE_100_2019, derosisHermite} have recently shown that accounting for high-order terms results in a gain in accuracy and stability. In our code, ${f}_i^{(0)}$ has been developed up to the sixth order. 
The extension of the standard LB algorithm to encompass the Lorentz force is straightforward and relies on the fundamental property that the second-order statistical moment at equilibrium gives the conservative part of the stress tensor. Therefore, incorporating the Lorentz force in the equation describing the fluid dynamics, or equivalently, the Maxwell tensor in the stress tensor amounts to upgrading the equilibrium state as
\begin{equation}
    {f}_i^{\mathrm{mhd}(0)}(\rho, \mathbf{U}, \mathbf{B}) =
    {f}_i^{(0)}(\rho, \mathbf{U}) +
    \frac{w_i}{2c_s^4} \left((\mathbf{B}\cdot \mathbf B) (\boldsymbol{c}_i\cdot\boldsymbol{c}_i)-(\mathbf{c}_i\cdot\mathbf{B})^2\right)
\end{equation}
so that the second-order moment becomes,
\begin{equation}
     \boldsymbol{\Sigma}^{\mathrm{mhd}(0)} = \sum_{i=0}^{N-1} f_i^{\mathrm{mhd}(0)}\mathbf{c}_i \otimes \mathbf{c}_i = \rho \mathbf{U} \otimes \mathbf{U} + \mathrm{p} \mathbb{I} +\frac{1}{2}|\mathbf{B}|^2\mathbb{I}-\mathbf{B}\otimes \mathbf{B}.
\end{equation}
This concludes the introduction of the standard BGK-LB algorithm for MHD.

\subsubsection{Central-moment Lattice Boltzmann scheme}

Despite its simplicity, effectiveness and large popularity, the BGK-LB scheme is known to suffer from numerical instability when large velocity gradients develop in the flow. 
This issue made it necessary to adapt either the numerical discretization of~\eqref{eq:HD_LBeq} or the collision operator \citep{Kruger2016}. If the former leads to more stable schemes, accuracy is also considerably degraded. This drawback motivated the remarkable efforts made towards developing collision operators with improved stability, as recently reviewed by \cite{COREIXAS_PRE_100_2019}. 
Moment-based collision operators rely on relaxing statistical moments rather than distributions. In addition, different relaxation times can be chosen to individually over-damp non-hydrodynamic moments (mainly responsible for instabilities) while ensuring the correct relaxation of hydrodynamic moments, \emph{e.g.} density, velocity or stress tensor. By doing so stability can be considerably enhanced while preserving physical consistency.  
Nevertheless, 
{due to the strongly nonlinear character of turbulent dynamics, spurious dissipative effects can occur as a result of the numerical integration of fluid-like equations over a very large number of grid points and of time-steps, as is the case for Hall-MHD turbulence}.

A significant reduction of {dissipation} artifacts developing in turbulence simulations
can be obtained by considering statistical moments expressed in the reference frame of the moving fluid rather than in the laboratory inertial frame, referring to a class of so-called central-moment (CM) collision operators \citep{Geier2006,geier2007properties,GEIERcumulant,DeRosis18}.    
This is the very framework adopted {in lay-outing our code} (details are given in the appendix \ref{sec:appendix_1}).
A key ingredient of CM-LB schemes is the shift of the particle velocities by the local fluid velocity {that defines a} new set of local microscopic velocities $\bar{\mathbf{c}}_i=\mathbf{c}_i-\mathbf{U}$ {used for the CMs evaluation}. 
Here, we consider the set of CMs {as} formally {defined} by
\begin{equation}
    |\mathrm{k}\rangle \equiv [\mathrm{k}_0 \cdots\mathrm{k}_{26}]^\top=\mathrm{T}^\top|{\bar f}\rangle
\end{equation}
where the transformation matrix $\mathrm{T}$ applies to the set of distributions $| \bar f \rangle \equiv [\bar f_0 \cdots \bar f_{26}]^\top$ 
and is explicitly defined by the column vectors
\begin{equation*}\label{eq:transf_matrix}
    \begin{aligned}
    &|\mathrm{T}_0\rangle=|1\rangle \\
    &|\mathrm{T}_1\rangle;|\mathrm{T}_2\rangle;|\mathrm{T}_3\rangle=[\bar{c}_{ix}]^\top;[\bar{c}_{iy}]^\top;[\bar{c}_{iz}]^\top\\
    &|\mathrm{T}_4\rangle;|\mathrm{T}_5\rangle;|\mathrm{T}_6\rangle=[\bar{c}_{ix}\bar{c}_{iy}]^\top;[\bar{c}_{ix}\bar{c}_{iz}]^\top;[\bar{c}_{iy}\bar{c}_{iz}]^\top\\
    &|\mathrm{T}_7\rangle;|\mathrm{T}_8\rangle;|\mathrm{T}_9\rangle=[\bar{c}_{ix}^2-\bar{c}_{iy}^2]^\top;[\bar{c}_{ix}^2-\bar{c}_{iz}^2]^\top;[\bar{c}_{ix}^2+\bar{c}_{iy}^2+\bar{c}_{iz}^2]^\top\\
    &|\mathrm{T}_{10}\rangle;|\mathrm{T}_{11}\rangle;|\mathrm{T}_{12}\rangle=[\bar{c}_{ix}\bar{c}_{iy}^2+\bar{c}_{ix}\bar{c}_{iz}^2]^\top;[\bar{c}_{ix}\bar{c}_{iy}^2+\bar{c}_{iy}\bar{c}_{iz}^2]^\top;[\bar{c}_{ix}^2\bar{c}_{iy}+\bar{c}_{iy}^2\bar{c}_{iz}]^\top\\
    &|\mathrm{T}_{13}\rangle;|\mathrm{T}_{14}\rangle;|\mathrm{T}_{15}\rangle=[\bar{c}_{ix}\bar{c}_{iy}^2-\bar{c}_{ix}\bar{c}_{iz}^2]^\top;[\bar{c}_{ix}\bar{c}_{iy}^2-\bar{c}_{iy}\bar{c}_{iz}^2]^\top;[\bar{c}_{ix}^2\bar{c}_{iy}-\bar{c}_{iy}^2\bar{c}_{iz}]^\top\\
    &|\mathrm{T}_{16}\rangle=[\bar{c}_{ix}\bar{c}_{iy}\bar{c}_{iz}]^\top \\
    &|\mathrm{T}_{17}\rangle;|\mathrm{T}_{18}\rangle;|\mathrm{T}_{19}\rangle=[\bar{c}_{ix}^2\bar{c}_{iy}^2+\bar{c}_{ix}^2\bar{c}_{iz}^2+\bar{c}_{iy}^2\bar{c}_{iz}^2]^\top;[\bar{c}_{ix}^2\bar{c}_{iy}^2+\bar{c}_{ix}^2\bar{c}_{iz}^2-\bar{c}_{iy}^2\bar{c}_{iz}^2]^\top;[\bar{c}_{ix}^2\bar{c}_{iy}^2-\bar{c}_{ix}^2\bar{c}_{iz}^2]^\top\\
    &|\mathrm{T}_{20}\rangle;|\mathrm{T}_{21}\rangle;|\mathrm{T}_{22}\rangle=[\bar{c}_{ix}^2\bar{c}_{iy}\bar{c}_{iz}]^\top;[\bar{c}_{ix}\bar{c}_{iy}^2\bar{c}_{iz}]^\top;[\bar{c}_{ix}\bar{c}_{iy}\bar{c}_{iz}^2]^\top\\
    &|\mathrm{T}_{23}\rangle;|\mathrm{T}_{24}\rangle;|\mathrm{T}_{25}\rangle=[\bar{c}_{ix}\bar{c}_{iy}^2\bar{c}_{iz}^2]^\top;[\bar{c}_{ix}^2\bar{c}_{iy}\bar{c}_{iz}^2]^\top;[\bar{c}_{ix}^2\bar{c}_{iy}^2\bar{c}_{iz}]^\top\\
    &|\mathrm{T}_{26}\rangle=[\bar{c}_{ix}^2\bar{c}_{iy}^2\bar{c}_{iz}^2]^\top.
    \end{aligned}
\end{equation*}
This set of vectors forms a simple relevant basis ($\mathrm{T}$ is reversible) allowing for a suitable separation between hydrodynamic and non-hydrodynamic moments 
\citep{CM3D-derosis-2017}. 
In the space of CMs, the collision step (\ref{eq:collision}) now generalizes as 

\begin{equation}
    \label{eq:cm_collision_step}
    |\mathrm{k}^*\rangle =|\mathrm{k}\rangle -S \bigg{(}|\mathrm{k}\rangle-|\mathrm{k}^{(0)}\rangle\bigg{)} 
    \quad \mathrm{with}~~|\mathrm{k}^{(0)}\rangle =\mathrm{T}^\top|{f^{\mathrm{mhd}(0)}}\rangle 
\end{equation}
where $S$ is a diagonal matrix applied to each moment individually. Let us point out that 
the BGK collision is recovered by taking $S=\omega \mathbb{Id}$. 
A proper choice for $S$  is given by
\begin{equation}\label{eq:collision_matrix}
    S=\mathrm{diag}[1,1,1,1,\omega,\omega,\omega,\omega,\omega,1,...,1]    
\end{equation}
which ensures that mass and momentum are conserved by the collision operator and that kinematic viscosity is suitably taken into account. 
The bulk viscosity can be set separately from the shear viscosity and, here,
it is implicitly defined by taking the trace of the second-order post-collision central-moment at equilibrium.
{Eventually,} the post-collision distributions are obtained by returning {to} the space of {the} distributions {through}
\begin{equation}
    |\bar f^*\rangle = {\mathrm{T}^{-1}}^\top|\mathrm{k}^*\rangle
\end{equation}
before moving on to the streaming step (\ref{eq:stream}).

\subsection{Vector-valued Lattice Boltzmann scheme for the magnetic field}\label{sec:LB_BGK_HMHD}

We now {present the} LB scheme for the magnetic field introduced by \cite{DELLAR2002}{, here} extended to encompass the Hall effect {in simulating MHD turbulent plasmas}.
Following {the} works {previously done} by \cite{Croisille95} and \cite{Bouchut99},  \cite{DELLAR2002} proposed {a decomposition of} the magnetic field as: 

\begin{equation}\label{eq:mag_field_LB}
    \mathbf{B}(\mathbf{x},t)=\sum_{i=0}^{M-1} \mathbf{\bar g}_i(\mathbf{x},t)
\end{equation}
where the sum spans a set of vector-valued densities 
$\mathbf{g}_0,\cdots,\mathbf{g}_{M-1}$ associated with the microscopic velocities $\boldsymbol{\xi}_0, \cdots,\boldsymbol{\xi}_{M-1}$. 

The magnetic field $\mathbf{B}$ is here {provided} by the zeroth-order moment of $\mathbf{\bar g}_i$  {hinting} that a lattice with low connectivity should {suffice to simulate its dynamics}.
In practice, a D3Q7 lattice with only seven velocities (see green arrows in Fig.~\ref{fig:lattice_types}) {shall} prove {to be} satisfactory {in reproducing the magnetic field of} Hall-MHD turbulent {plasmas}. 
Analogously to {the} fluid {case}, a LB scheme can be derived {in order} to simulate the induction equation in the form 
\begin{equation}\label{eq:MHD_LBscheme}
    \mathbf{\bar g}_i(\mathbf{x}+\boldsymbol{\xi}_i\Delta t,t+\Delta t) = \mathbf{\bar g}_i(\mathbf{x},t) - \omega_B\bigg{(}\mathbf{\bar g}_i(\mathbf{x},t)-\mathbf{g}^{(0)}_i(\mathbf{U},\mathbf{B})(\mathbf{x},t)\bigg{)}
\end{equation}
where the relaxation parameter $\omega_\mathrm{m}$ is now related to the magnetic resistivity $\eta$ by
\begin{equation}\label{eq:mag_relaxation_time}
    \frac{1}{\omega_B}=\bigg{(}\frac{\eta}{C^2\Delta t}+\frac{1}{2}\bigg{)}
\end{equation}
with $\Delta x/\Delta t=2C$ for the D3Q7 lattice.  
In practice, it is desirable that the nodes of the D3Q7 and D3Q27 lattices coincide so that the macroscopic quantities such as $\mathbf{u}$, $\mathbf{B}$ or $\mathbf{J}$ may be exchanged between the two lattices without interpolation. 
This constraint imposes that
\begin{equation}
    2 C = \sqrt{3} c_s.
    \label{eq:lattice_coincidence}
\end{equation}

%

In the context of ideal MHD, the densities at equilibrium are given by  
\begin{equation}
    g_{i\alpha}^{(0)} (\mathbf{U}, \mathbf{B})=W_i\left(B_{\alpha}+\frac{1}{C^2}\xi_{i\beta}(U_{\beta}B_{\alpha}-B_{\beta}U_{\alpha})\right)
\end{equation}
with $W_\mathrm{center}=1/4$ and $W_\mathrm{face}=1/8$ for a D3Q7 lattice. {By doing so}, the first-order moment 
\begin{equation}\label{eq:MHD_tensors}
    \sum_{i=0}^{M-1}\xi_{i\alpha} g^{(0)}_{i\beta}=\mathbf{U} \otimes \mathbf{B} - \mathbf{B}\otimes \mathbf{U}
\end{equation}
{would} suitably reconstruct the transport term of the induction equation. 
Including the Hall correction in {this} equation {is thus equivalent} to upgrading the equilibrium densities, so that 
\begin{equation}\label{eq:Hall_eq_tensor}
    \boldsymbol{\mathrm{\Lambda}}^{ (0)}_{\alpha\beta}=\sum_{i=0}^{M-1}\xi_{i\alpha}g^{\mathrm{Hall} (0)}_{i\beta}=(\mathbf{U} - \alpha_H \mathbf{J}) \otimes \mathbf{B} -  \mathbf{B} \otimes (\mathbf{U} - \alpha_H \mathbf{J})
\end{equation}
which is obviously possible by now considering
\begin{equation}
    g_{i\alpha}^{\mathrm{Hall} (0)} (\mathbf{U}, \mathbf{B}, \mathbf{J}) = W_i\left(B_{\alpha}+\frac{1}{C^2}\xi_{i\beta}((U_{\beta} - \alpha_H J_\beta)B_{\alpha}-B_{\beta}(U_{\alpha}-\alpha_H J_\alpha))\right).
    \label{eq:gHall_eq}
\end{equation}
Nevertheless, $\mathbf{J}$ needs to be computed, possibly from the densities, in this expression. 
Note that although equilibrium distributions have been expanded up to the sixth order in $\mathbf{U}$, they have only been extended up to the second order in $\mathbf{B}$, which may sound contradictory. However, it is important to recognize that there is no continuous distribution for the magnetic field that is analogous to the Maxwell-Boltzmann distribution for the velocity. Thus, only the first two orders of the expansion can be reconstructed by matching the moments to the terms of the induction equation. Attempting to consider higher-order expansions would open a large variety of possibilities for defining non-physical moments, which is beyond the scope of the present work.
An essential benefit of the LB framework is that the spatial derivatives of the magnetic field, thus $\mathbf{J}$, are self-consistently obtained (within an $O(\mathrm{Ma}^3)$ error) from the first-order moment of the densities as
\begin{equation}\label{eq:LB_density_current}
\mathrm{J}_{\gamma} = \varepsilon_{\alpha\beta\gamma} \frac{\partial B_{\alpha}}{\partial x_{\beta}}=-\varepsilon_{\alpha\beta\gamma} \frac{\omega_B}{C^2}\bigg{(}\mathbf{\boldsymbol{\Lambda}}_{\alpha\beta}-\boldsymbol{\mathrm{\Lambda}}_{\alpha\beta}^{(0)}\bigg{)} 
\end{equation}
where $\varepsilon_{\alpha\beta\gamma}$ is the Levi-Civita tensor and  
$ \boldsymbol{\mathrm{\Lambda}}_{\alpha\beta}=\sum_{i=0}^{M-1}\xi_{i\alpha}\bar g_{i\beta}$ \citep{DELLAR2002}.

By replacing~\eqref{eq:Hall_eq_tensor} in~\eqref{eq:LB_density_current} we obtain 
a linear system readily solvable to obtain 
the current density $\mathbf{J}$, namely
\begin{equation}\label{eq:soe_current_density}
    \left(\mathbb{I}+\frac{2\alpha_H\omega_B}{C^2}\mathbf{M}\right)\mathbf{J}= \mathbf{J_0}
\end{equation}
where
\begin{equation}\label{eq:soe_objects}
     \mathbf{M}=\begin{bmatrix}
       0 & B_z & -B_y \\[0.3em]
       -B_z & 0 & B_x \\[0.3em]
       B_y & -B_x & 0           
     \end{bmatrix}\; \mathrm{and}\;\;
     \mathbf{J_0}=\begin{bmatrix}
       \boldsymbol{\mathrm{\Lambda}}_{yz}-\boldsymbol{\mathrm{\Lambda}}_{zy} - 2\left(U_yB_z-U_zB_y\right) \\[0.3em]
       \boldsymbol{\mathrm{\Lambda}}_{zx}-\boldsymbol{\mathrm{\Lambda}}_{xz} - 2\left(U_zB_x-U_xB_z\right) \\[0.3em]
       \boldsymbol{\mathrm{\Lambda}}_{xy}-\boldsymbol{\mathrm{\Lambda}}_{yx} - 2\left(U_xB_y-U_yB_x\right)          
     \end{bmatrix}.
\end{equation}
Obviously, a solution 
exists {only} if it is possible to 
invert
$\widetilde{\mathbf{M}}=\mathbb{I}+\left(2\alpha_H\omega_B/C^2\right)\mathbf{M}$. It can be easily verified that $\mathrm{det}(\widetilde{\mathbf{M}})\neq0$, which proves this solution exists and is unique. The current density obtained by solving  \eqref{eq:soe_current_density} can then be used to compute the equilibrium densities \eqref{eq:gHall_eq} and proceed to the collision operation. 
%
It is fair to mention that in a similar vein, \cite{DELLAR2013115} introduced a modification of the collision operator to incorporate MHD current-dependent resistivity, with the current being derived from the non-equilibrium components of the magnetic distribution functions.
The expression of~\eqref{eq:LB_density_current} also provides a consistent approximation of the divergence of the magnetic field. Indeed by taking the trace of the magnetic tensor, one obtains
\begin{equation}\label{eq:LB_B_div}
    \bNabla \cdot \mathbf{B} \simeq -\frac{\omega_B}{C^2}\mathrm{Tr}\left(\boldsymbol{\Lambda}_{\alpha\beta}\right)
\end{equation}
by noticing that $\mathrm{Tr}\left(\boldsymbol{\Lambda}_{\alpha\beta}^{(0)}\right)=0$. 
Furthermore, the ${O}(\mathrm{Ma}^3)$ correction cancels out by taking the trace. Therefore, this correction is pushed to a higher order, so that the divergence-free $\bNabla \cdot \mathbf{B}=0$ corresponds with high accuracy to the condition $\mathrm{Tr}\left(\boldsymbol{\Lambda}_{\alpha\beta}\right)=0$ in the LB framework \citep{DELLAR2002}. In practice, we have checked in our LB simulations that this condition was maintained {throughout the runs,} to machine round-off error. 

\subsection{Dimensionless formulation}
\label{sec:unit_conversion}

In the following, the {Hall-MHD} equations are {re-arranged} in a dimensionless form {in terms of the}  control parameter $\epsilon_H$,  associated with the Hall parameter $\alpha_H = 1/ne$. This {control} parameter is then recast in lattice units for practical LB {purposes}.
Physical quantities in lattice units are hereafter indicated with the superscript $^\mathrm{lbm}$.
In lattice units, the lattice spacing $\Delta x$ and the time-step $\Delta t$ of the scheme define the units of length and time, respectively. 
In order to obtain a dimensionless induction equation, let us normalise the magnetic field with a reference value, $B_0$, the fluid velocity with $U_0$, the current density with $B_0/L_0$, the length with $L_0$ and the time with $L_0/U_0$. {Leveraging these characteristic quantities, \eqref{eq:Hall_ind} can be written in {a} dimensionless form as
\begin{equation}\label{eq:adim_Hall_ind}
    \bigg{(}\frac{U_0 B_0}{L_0}\bigg{)}\partial_t \mathbf{b}=\frac{1}{L_0}\bNabla \times \bigg{[}\bigg{(}U_0\mathbf{u}-\frac{\alpha_H B_0}{\sqrt{\mu_0} L_0}\bNabla \times \mathbf{b}\bigg{)}\times (B_0\mathbf{b})\bigg{]} + \frac{\eta B_0}{L_0^2} \bNabla^2 \mathbf{b}
\end{equation}
Dimensionless fields are here indicated by lowercase letters. 
This equation {can be reduced to}
\begin{equation}\label{eq:adim_Hall_ind2}
    \partial_t \mathbf{b}=\bNabla \times \bigg{[}\bigg{(}\mathbf{u}-\epsilon_H\bNabla \times \mathbf{b}\bigg{)}\times \mathbf{b}\bigg{]} + \frac{1}{\mathrm{Re}_m} \bNabla^2 \mathbf{b}
\end{equation}
by defining the magnetic Reynolds number $\mathrm{Re}_m=U_0L_0/\eta$  and the dimensionless Hall parameter
\begin{equation}\label{eq:hall_param_pu}
    \epsilon_H = \frac{\alpha_H B_0}{\sqrt{\mu_0} L_0 U_0}.
\end{equation}
We can {treat in the same fashion} the fluid momentum equation, where the reference scales are the same as those {used to adimensionalize} the induction equation. Therefore,
\begin{equation}\label{eq:adim_NSE2}
    \rho \frac{U_0^2}{L_0}\left[\partial_t \mathbf{u}+\left(\mathbf{u}\cdot \bNabla\right)\mathbf{u}\right] = -\frac{1}{L_0}\bNabla \rho c_s^2 + \rho\nu \frac{U_0}{L_0^2}\bNabla^2 \mathbf{u}+\frac{B_0^2}{L_0}\left(\bNabla\times \mathbf{b}\right)\times\mathbf{b}
\end{equation}
which gives
\begin{equation}
    \label{eq:normalised_momentum}
    \partial_t \mathbf{u}+\left(\mathbf{u}\cdot \bNabla\right)\mathbf{u} = -\frac{1}{\mathrm{Ma}^2} \frac{1}{\rho} {\bNabla \rho} + \frac{1}{\mathrm{Re}}\bNabla^2 \mathbf{u}+ \left( \frac{V_A}{U_0} \right)^2 \left (\bNabla\times \mathbf{b} \right )\times\mathbf{b}
\end{equation}
where the control parameters are the Mach number $\mathrm{Ma}=U_0/c_s$, the (fluid) Reynolds number $\mathrm{Re}=U_0L_0/\nu$ and $V_A/U_0$, the Alfv\'en velocity {being} $V_A=B_0/\sqrt{\rho}$. 
The Hall number $\epsilon_H$ is given in lattice units by
\begin{equation}\label{eq:hall_param_lu}
    \epsilon_H^\mathrm{lbm} = \frac{\alpha_H}{\sqrt{\mu_0}} \frac{\left [ B_0/B^\mathrm{lbm}_0 \right ]}{\left [ L_0/L^\mathrm{lbm}_0 \right ] \left [U_0/U^\mathrm{lbm}_0 \right ]}=\epsilon_H \frac{U^\mathrm{lbm}_0 L_0^\mathrm{lbm}}{B_0^\mathrm{lbm}}.
\end{equation}

If one considers that the reference velocity $U_0$ corresponds to the Alfv\'en velocity ($U_0 = V_A$)  
and $\rho\simeq 1$ for simplicity,
one obtains that $U_0^\mathrm{lbm}= B_0^\mathrm{lbm}$ and 
\begin{equation}\label{eq:hall_param_lu2}
    \epsilon_H^\mathrm{lbm} = \epsilon_H L_0^\mathrm{lbm} = \epsilon_H N
\end{equation}
with $N=L_0/\Delta x$ being the number of lattice points per reference length $L_0$.
The Hall {parameter} $\epsilon_H$ can also be {obtained} as the ratio of two reference scales as  
\begin{equation}\label{eq:Hall_param}
    \epsilon_H = \frac{V_A}{U_0L_0}\sqrt{\frac{m}{\mu_0ne^2}}=\frac{L_H}{L_0}
\end{equation}
with
\begin{equation}\label{eq:Hall_param2}
    L_H = \frac{V_A}{U_0}\sqrt{\frac{m}{\mu_0ne^2}}.
\end{equation}
In lattice units, 
\begin{equation}
\epsilon_H^\mathrm{lbm} = \frac{L_H}{L_0} ~ \frac{U^\mathrm{lbm}_0 L_0^\mathrm{lbm}}{B_0^\mathrm{lbm}} = \frac{L_H}{\Delta x} ~ \frac{U^\mathrm{lbm}_0}{B_0^\mathrm{lbm}} = \frac{U^\mathrm{lbm}_0 L_H^\mathrm{lbm}}{B_0^\mathrm{lbm}}.
\end{equation}
If $U_0=V_A$, $L_H$ is equal to the ion inertial length $d_i$
(or ion skin depth). In that case, $\epsilon_H^\mathrm{lbm} = L_H^\mathrm{lbm}$ and {this} corresponds to the number of lattice points per ion inertial length. {It is assumed that} the dynamics {of a MHD plasma develops under the influence of} the Hall effect at scales $\ell$ smaller that $L_H$.

\subsection{CFL condition for Hall-MHD turbulence}
The Courant–Friedrichs–Lewy (CFL) condition \citep{lewy1928} determines, for an explicit time-marching scheme, the maximum time-step for convergence, as 
\begin{equation}
    \Delta t \le \Delta x/c_\mathrm{max}
\end{equation}  
where $c_\mathrm{max}$ refers to the largest speed at which a signal propagates in the solution. In the context of Hall-MHD, $c_\mathrm{max}$ should be identified with the largest phase speed of {the} whistler waves. 
When the plasma dynamics {in the direction of} the magnetic field $B$ is dominant, the phase speed of {the} whistler waves varies as $c_w(k)=k V_A^2/\omega_{ci}$  with the wavenumber $k$; $V_A$ is the Alfv\'en velocity, {while}  $\omega_{ci}$ is the ion cyclotron frequency. 
In physical units, $\omega_{ci}=eB/m_i$ and $V_A=B/\sqrt{\mu_0 nm_i}$, $m_i$ being the mass of ions and the Hall parameter {being} $\alpha_H=1/ne$. Therefore, one obtains that the time-step decreases quadratically with the grid spacing, as
\begin{equation}
    \Delta t \le \frac{\mu_0 (\Delta x)^2}{\pi ~\alpha_H B}
\end{equation}
{assuming} the {largest} attainable wavenumber {to be} $k_\mathrm{max}= \pi/\Delta x$ in {the} context of {the} Hall-MHD turbulence.
This condition {can be rewritten} accounting for the rescaling of the magnetic field by ${\mu_0}^{1/2}$ 
\begin{equation}
    \Delta t \le \frac{ (\Delta x)^2}{\pi~(\alpha_H/\sqrt{\mu_0}) ~B_0}
\end{equation}
which finally yields in lattice units {to}
\begin{equation}
    1 \le \frac{ 1}{\pi~ \epsilon_H^\mathrm{lbm} L_0^\mathrm{lbm} U_0^\mathrm{lbm}} = \frac{1}{\pi} ~\frac{B_0^\mathrm{lbm}}{\epsilon_H \left(N {U_0^\mathrm{lbm}}\right)^2}
\end{equation}
where \eqref{eq:hall_param_pu} is used to {retrieve} $\alpha_H/\sqrt{\mu_0}$, and $N=L_0^\mathrm{lbm}$. 
If $U_0=V_A$, the CFL condition for whistler waves can {be reformulated} as a condition on the Mach number ${\mathrm{Ma}=\sqrt{3} U_0^\mathrm{lbm}}$, which is {in turn} written {as}
\begin{equation}
    \mathrm{Ma}\le \left(\frac{\sqrt{3}}{\pi} \right)\frac{1}{\epsilon_H N^2}.
\end{equation}\label{eq:CFL_LBM}
This condition 
recalls the quadratic dependence of the time step on the resolution obtained with conventional CFD methods \citep{Gomez2010}. It also confirms that Hall-MHD turbulence is computationally very demanding due to the presence of whistler waves. 

\begin{figure}
\includegraphics[width=\textwidth]{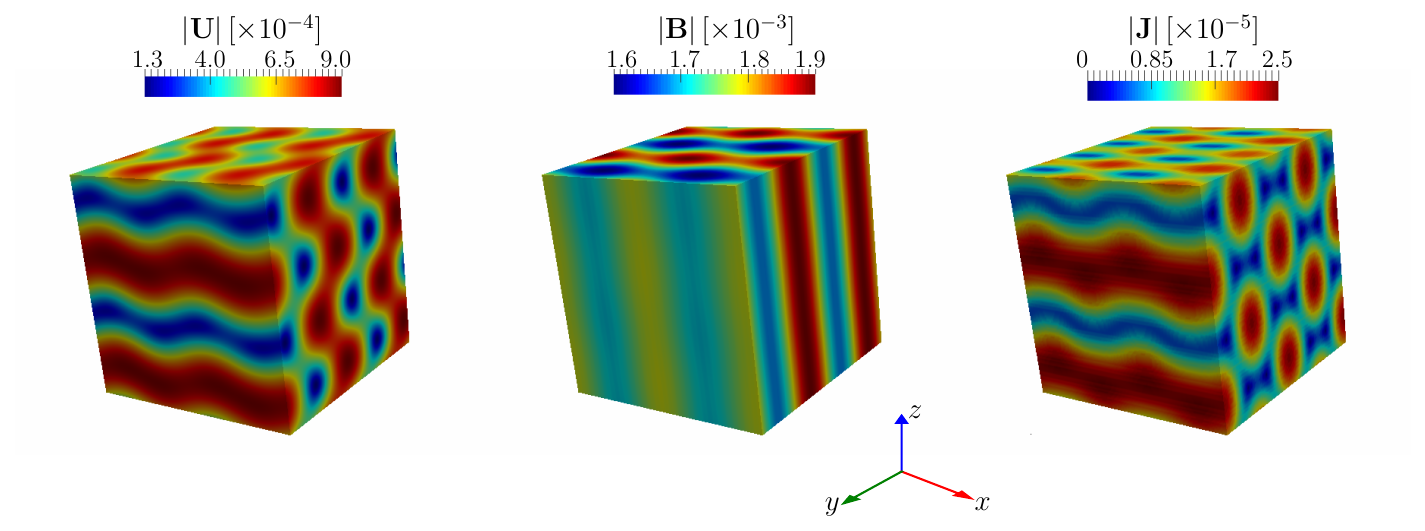}
\caption{Three-dimensional rendering of the initial condition as indicated in~\eqref{eq:Xia_IC} and~\eqref{eq:Xia_IC2}. The magnitudes of the fluid velocity (left), magnetic field (center) and density of electric current (right) are here displayed for $N=128$.
}
\label{fig:3D_rendering}
\end{figure}

\section{Results}\label{sec:Results}
{Our} LB scheme and code \textsc{flame} {is now} validated against {the} analytical solution of the incompressible and dissipative Hall-MHD equations {proposed in}~\citep{Xia15}. {The latter is used as a} benchmark {to evaluate} accuracy and convergence {of the numerical solutions} for different values of the control parameters {(in a regime of low Reynolds numbers in which the aforementioned analytical solution holds)}. 
A {further} validation was {done focusing on}  the MHD {range of scales, this time in a regime of} high Reynolds numbers.
{The solutions of the MHD dynamics produced by \textsc{flame} were compared in this case with those obtained} with a well-established pseudo-spectral solver, widely used for turbulent plasma simulations, namely the Geophysical High-Order Suite for Turbulence (\textsc{ghost}, \cite{Mininni2011,rosenberg20}). 
Finally, the {physical consistency of the output and the computational performance} were evaluated when accounting explicitly for the Hall effect in the turbulent regime. 
This allowed us to {assess} the reliability of our code in simulating {the multi-scale dynamics generated by} turbulent flows at high Reynolds numbers, and in reproducing the transition from {the} MHD to {the} Hall-MHD spectral range (at sub-ion scales).

The \textsc{flame} code relies on a multi-GPU (Graphic Processing Unit) implementation of the LB scheme in order to reach high resolution {that optimizes the} computational times. Massive multi-threading is handled {within} the OpenCL (Open Computing Language) framework, allowing {a high portability of the code}. The spatial domain is split along a single direction and each GPU is assigned a sub-domain. A one-to-one mapping operates between the host CPU processes and the GPUs. Therefore, the exchange of boundary nodes between the GPUs is handled through memory transfers with the CPU processes and a message-passing interface (MPI) between {the} latter. 
Turbulence simulations were run on a {cluster} equipped with NVIDIA A100-40Gb GPU {cards, hosted} at the CINECA {super}computing center (Italy). 
}

\begin{table}
  \begin{center}
\def~{\hphantom{0}}
  \begin{tabular}{lc}
       Parameters                     & Values                                            \\ [3pt]
Resolution: $N$                            & $32,\ 64,\ 96,\ 128$                                 \\
Mach number: $\mathrm{Ma}~ [\times 10^{-2}]$ & $1.0,\ 0.7,\ 0.5,\ 0.3$                                \\
Kinematic viscosity: $\nu ~[\times 10^{-3}]$~~~        & $1.0,\ 0.5,\ 0.33,\ 0.25,\ 0.2,\ 0.17,\ 0.14,\ 0.13,\ 0.11$ \\ 

  \end{tabular}
  \caption{Parameters of LB simulations.
  The magnetic Prandtl number is kept unitary. 
  The kinematic viscosity is given in dimensionless units, \emph{i.e.} normalised by $U_0 L_0$, which means that the Reynolds number $\mathrm{Re}=1/\nu$.
  }
\label{tab:params}
  \end{center}
\end{table}

\subsection{Exact solution of the dissipative Hall-MHD}

{Due to their high computational cost, the availability in the literature of plasma simulations reproducing the Hall-MHD range of scales (in three dimensions) is much less than for the MHD case.} 
{Moreover, Hall-MHD simulations are in general performed using} pseudo-spectral codes \citep{Ferrand2022,Meyrand2012,Gomez2010,Yadav2022}, which integrate {of course} the dynamical equations in the Fourier space. 
Interestingly, \cite{Mahajan05} derived {an analytical} solution for the non-dissipative Hall-MHD equations, then extended by \cite{Xia15} {with the inclusion of} dissipative effects. 
This solution {is} used in the following to test the stability and convergence of \textsc{flame}. 
{Encompassing} dissipative effects \cite{Xia15}, {this analytical solution} {allowed us to quantify as well the numerical dissipation spuriously introduced by our scheme.}  

The solution provided by \cite{Xia15} is rewritten in a dimensionless form (see \S\ref{sec:unit_conversion}) as \begin{equation}
    \mathbf{u}(\mathbf{x},t) = \mathbf{u}^\prime(\mathbf{x},t) \quad \mathrm{and} \quad \mathbf{b}(\mathbf{x},t) = 
    \hat{\mathbf{e}}_z 
    + \mathbf{b}^\prime(\mathbf{x},t)
\end{equation} where the fluctuating velocity and magnetic fields are damped circular polarized waves 
given respectively by

\begin{equation}
\begin{aligned}
    \mathbf{u}^\prime(\mathbf{x},t) &=  \big[ B (\hat{\mathbf{e}}_x + i\hat{\mathbf{e}}_y) \exp(i \mathrm{k}z - i\omega_\pm t)\\
    &+ C (\hat{\mathbf{e}}_y + i\hat{\mathbf{e}}_z) \exp(i \mathrm{k}x)\\
    &+ A (\hat{\mathbf{e}}_z + i\hat{\mathbf{e}}_x) \exp(i \mathrm{k}y) \big] e^{-\nu \mathrm{k}^2 t}
\end{aligned}\label{eq:Xia_IC_orig}
\end{equation}
and 
\begin{equation}
    \mathbf{b}^\prime(\mathbf{x},t) = \alpha_\pm \mathbf{u}^\prime(\mathbf{x},t)
\end{equation} 
in complex notations.  
The amplitudes $A$, $B$, and $C$ are arbitrary real values.
The ambient magnetic field here {is} assumed to be oriented along the unit vector $\hat{\mathbf{e}}_z$.
Since the dynamical equations only consist of real variables, either the imaginary part or {the} real part is a solution. 
The pulsation 
$\omega_\pm = -\alpha_\pm \mathrm{k}$,
where $\alpha_\pm$ depends itself on the wavenumber $\mathrm{k}$ as
\begin{equation}
    \alpha_\pm= -\frac{1}{2} \epsilon_H \mathrm{k} \pm \sqrt{\frac{\epsilon_H^2\mathrm{k}^2}{4}+1}.
\end{equation}
The magnetic Prandtl number is {assumed} 
equal to unity {in obtaining} this solution and  the reference velocity is assumed equal to the Alfv\'en velocity, \emph{i.e.} $U_0=V_A$ in \eqref{eq:normalised_momentum}. 
Finally, it {is worth mentioning that this analytical} solution {holds} in a strictly incompressible framework, which{, given the intrinsically compressible nature of the LB scheme, prescribes that} our simulations {must be run at} a (very) low Mach number so that relative density fluctuations {generated by the code} remain negligible. In our investigations, the Hall-MHD equations {have been} integrated in a cubic box of size $L_0=2\pi$. 
The evolution of the velocity field is deterministic from the initial condition
\begin{equation}
\begin{aligned}
    &\mathrm{u}^\mathrm{lbm}_x(\mathbf{x},0) = {U_0^\mathrm{lbm}} \left(B\sin \left(\frac{4\pi z_k}{N}\right)+A\cos \left(\frac{4\pi y_j}{N}\right) \right)\\
    &\mathrm{u}^\mathrm{lbm}_y(\mathbf{x},0) = {U_0^\mathrm{lbm}} \left(B\cos\left(\frac{4\pi z_k}{N}\right) + C\sin \left(\frac{4\pi x_i}{N}\right) \right)\\
    &\mathrm{u}^\mathrm{lbm}_z(\mathbf{x},0) = {U_0^\mathrm{lbm}} \left(C\cos \left(\frac{4\pi x_i}{N}\right) + A\sin \left(\frac{4\pi y_j}{N}\right) \right)
\end{aligned}\label{eq:Xia_IC}
\end{equation}
expressed in lattice units with $A=0.3$, $B=0.2$, $C=0.1$ and $N$ being the number of lattice nodes per reference length $L_0$. 
The reference velocity $U_0^\mathrm{lbm}$ is related by construction to the Mach number through $U_0^\mathrm{lbm} = \mathrm{Ma}/\sqrt{3}$. 
The magnetic field is initially proportional to the fluid velocity with
\begin{equation}
\begin{aligned}
    &\mathrm{b}^\mathrm{lbm}_x(\mathbf{x},0) =  \alpha_+ \mathrm{u}^\mathrm{lbm}_x(\mathbf{x},0)\\
    &\mathrm{b}^\mathrm{lbm}_y(\mathbf{x},0) = \alpha_+ \mathrm{u}^\mathrm{lbm}_y(\mathbf{x},0)\\
    &\mathrm{b}^\mathrm{lbm}_z(\mathbf{x},0) = \alpha_+ \mathrm{u}^\mathrm{lbm}_z(\mathbf{x},0) + U_0^\mathrm{lbm}
\end{aligned}\label{eq:Xia_IC2}
\end{equation}
since $U_0=V_A$. For {sake of} simplicity, the initial density is {set to one everywhere in the space}.
The (normalized) Hall parameter is fixed at $\epsilon_H=1$, that is $\epsilon_H^\mathrm{lbm}=N$ according to \eqref{eq:hall_param_lu2}. 
This value ensures that the solution is affected by the Hall effect with $L_H = L_0$ from \eqref{eq:Hall_param}. 
A three-dimensional rendering of the initial conditions expressed in \eqref{eq:Xia_IC} and \eqref{eq:Xia_IC2} is {displayed} in Fig.~\ref{fig:3D_rendering}. With this {initialization}, the current density $\mathbf{J}=\bNabla\times \mathbf{B}$ is non-zero at $t=0$. 
The parameters {used in} the different simulations are reported in  Tab.~\ref{tab:params}. The Mach number is {always} small enough {for the plasma} to approach the incompressible limit and {in order to} reduce the {intrinsic} discretization error of the LB method. The CFL condition imposed by this solution is also satisfied. 
Finally, let us mention that analogous simulations were performed with the phase speed $\alpha_-$ yielding very similar results on accuracy and stability. However, the phase speed is much larger in the latter case,  requiring a significant reduction of the Mach number (with $\epsilon_H=1$). Results {obtained} for $\alpha_+$ and the velocity field only ($\mathbf{b} = \alpha_+ \mathbf{u} + \hat{\mathbf{e}}_z$) are presented in the following.

\begin{figure}
\includegraphics[width=0.8\textwidth]{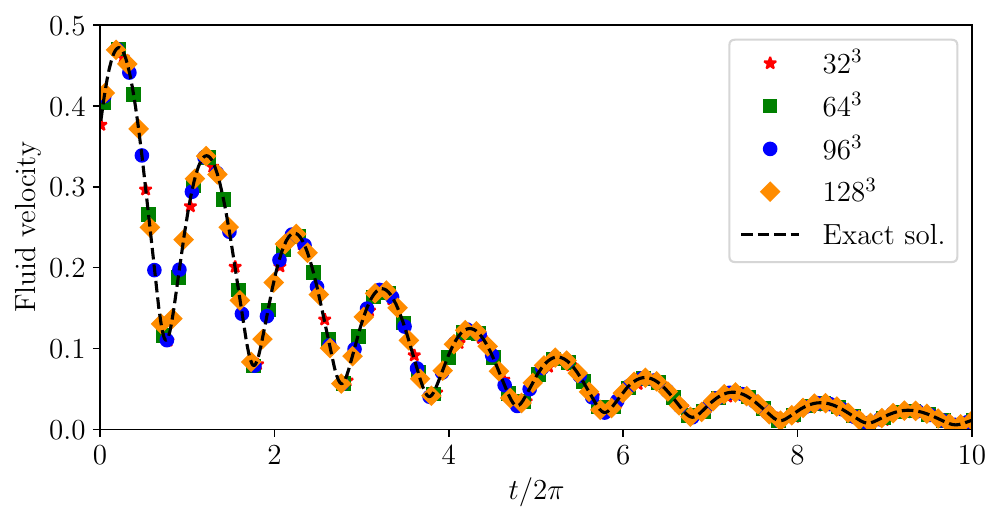}
\centering
\caption{Temporal evolution of the velocity magnitude $|\mathbf{u}|(\mathbf{0},t)$. Comparison between the analytical (dashed line) and the numerical solutions (symbols) at different lattice resolutions. The Mach number $\mathrm{Ma}=0.003$ and the kinematic viscosity $\nu=3.3\cdot 10^{-4}$. 
}
\label{fig:sol_comparison_ex1}
\end{figure}

\subsection{Stability and incompressibility}

\begin{figure}
\includegraphics[width=\textwidth]{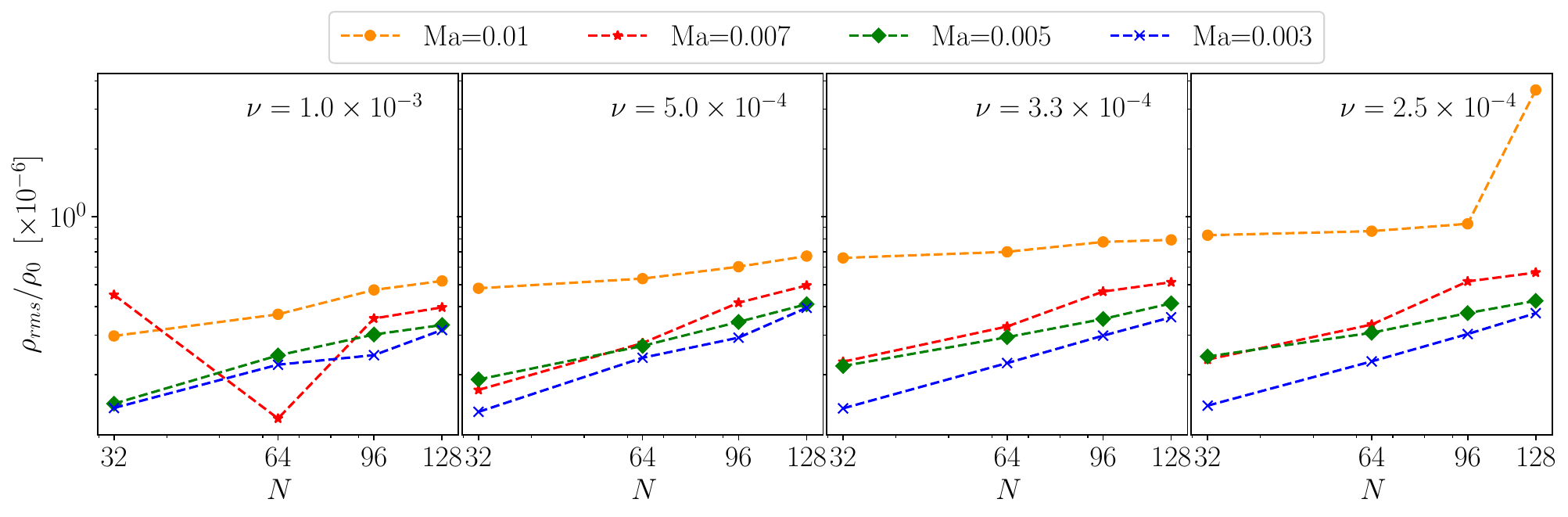}
\caption{Relative density fluctuations at different values of the Mach number ($\mathrm{Ma}$) and kinematic viscosity ($\nu$) as a function of the resolution ($N$). In our simulations, the reference density $\rho_0$ is fixed at unity.}
\label{fig:stab_example2}
\end{figure}

The stability {of the scheme} was tested exploring 
the parameter space defined {through} the Mach number, the lattice resolution and the kinematic viscosity (see Tab.~\ref{tab:params}). 
The analytical solution introduced by \cite{Xia15} {is such that the} nonlinear terms {in} the incompressible dissipative Hall MHD equations {are strictly zero}. In practice, physical instabilities triggered by numerical errors {do} naturally develop and grow in time in simulations {whenever} the viscosity is too small, eventually {inducing the} transition to a turbulent state. 
Therefore, the numerical stability and accuracy of \textsc{flame} were {assessed} in {runs in which} the viscosity was sufficiently high to prevent {such} transition to turbulence. 
\begin{figure}
\includegraphics[width=\textwidth]{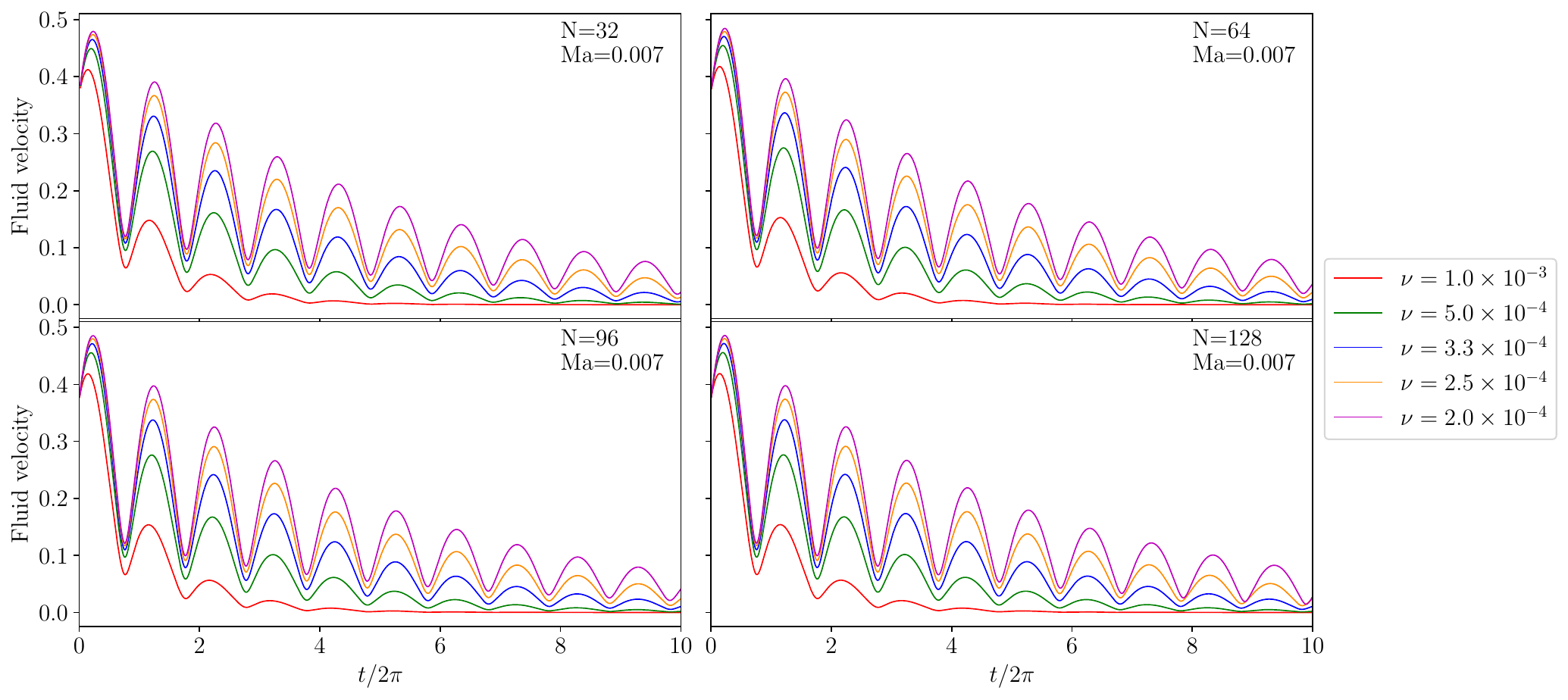}
\caption{Temporal evolution of the velocity magnitude $|\mathbf{u}|(\mathbf{0},t)$ for different values of the resolution ($N$) and viscosity ($\nu$) at fixed Mach number $\mathrm{Ma}=0.007$.}
\label{fig:stab_example1}
\end{figure}
{The} typical temporal evolution of the velocity at a fixed location {in the simulation domain} is shown in Fig.~\ref{fig:sol_comparison_ex1}. The solution appears as a damped wave propagating in the direction of the ambient magnetic field. The amplitude and the phase of the solution are well captured in the LB simulation. The results obtained for different resolutions and viscosity values at Mach number $\mathrm{Ma}=0.007$ are shown in Fig.~\ref{fig:stab_example1} {for} 10 wave periods. 
All simulations remained numerically stable in the explored range of parameters.
The temporal averages of relative density fluctuations at different values of Mach number and kinematic viscosity are displayed in Fig.~\ref{fig:stab_example2}. 
The level of these relative fluctuations is typically of order $10^{-7}$ -- $10^{-6}$' indicating {a very good convergence towards the} incompressible limit {in all the} simulations {presented}. Furthermore, the results confirm that the amplitude of density fluctuations decreases with the Mach number.

\subsection{Dispersion and dissipation errors}
\begin{figure}
\includegraphics[width=\textwidth]{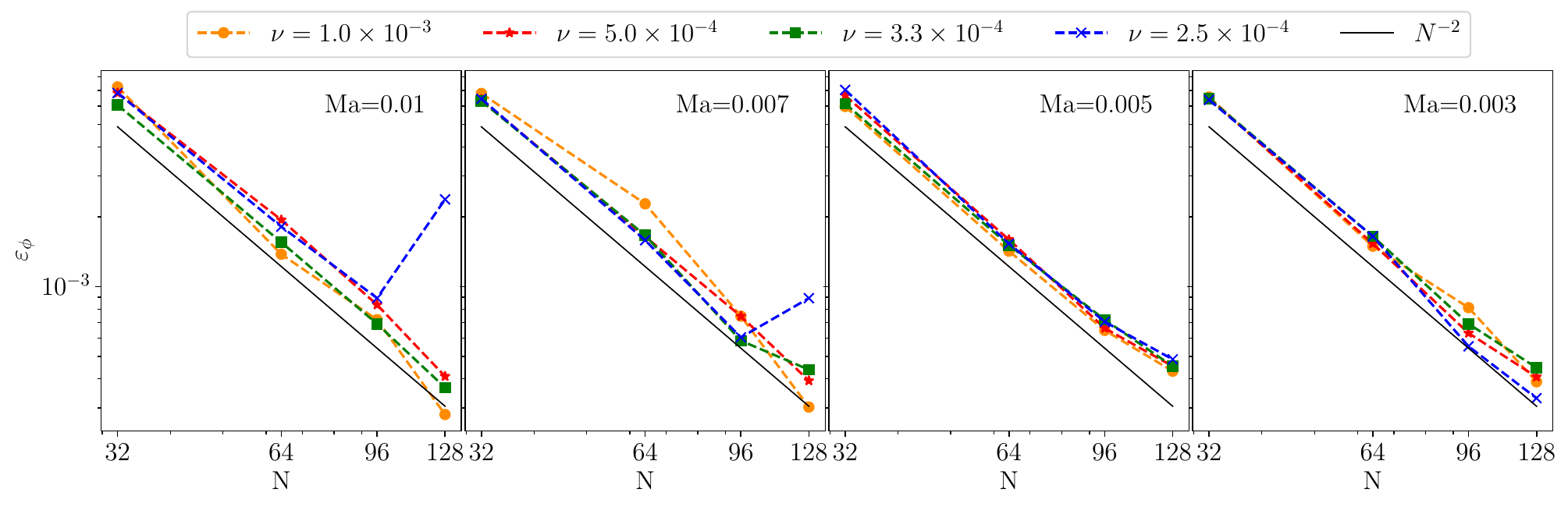}
\caption{Relative dispersion error as defined by \eqref{eq:dispersion_error} for different values of the kinematic viscosity $\nu$ and the Mach number $\mathrm{Ma}$. The error decreases as $N^{-2}$ as expected for a LB scheme.}
\label{fig:dispersion_error}
\end{figure}

\begin{figure}
\includegraphics[width=\textwidth]{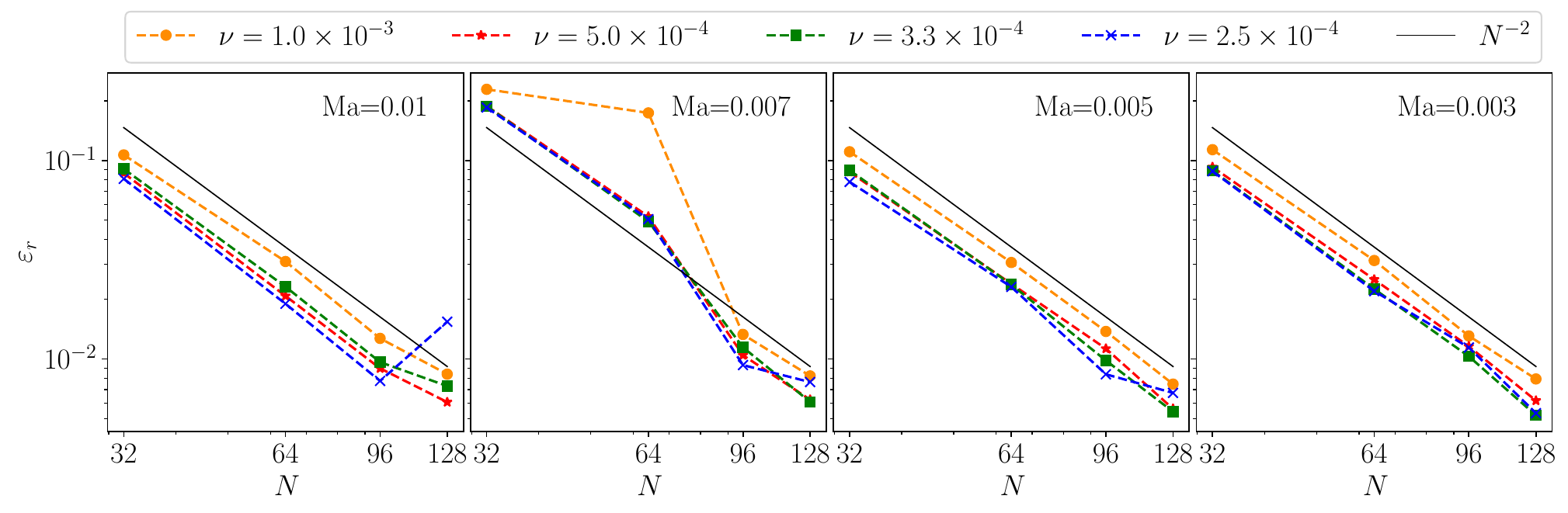}
\caption{Acoustic scaling ($\mathrm{Ma}$ constant) of the relative dissipation error for different values of the kinematic viscosity ($\nu$) at fixed Mach number ($\mathrm{Ma}$). 
The error decreases as $N^{-2}$.
}
\label{fig:acoustic_scaling}
\end{figure}

\begin{figure}
\includegraphics[width=\textwidth]{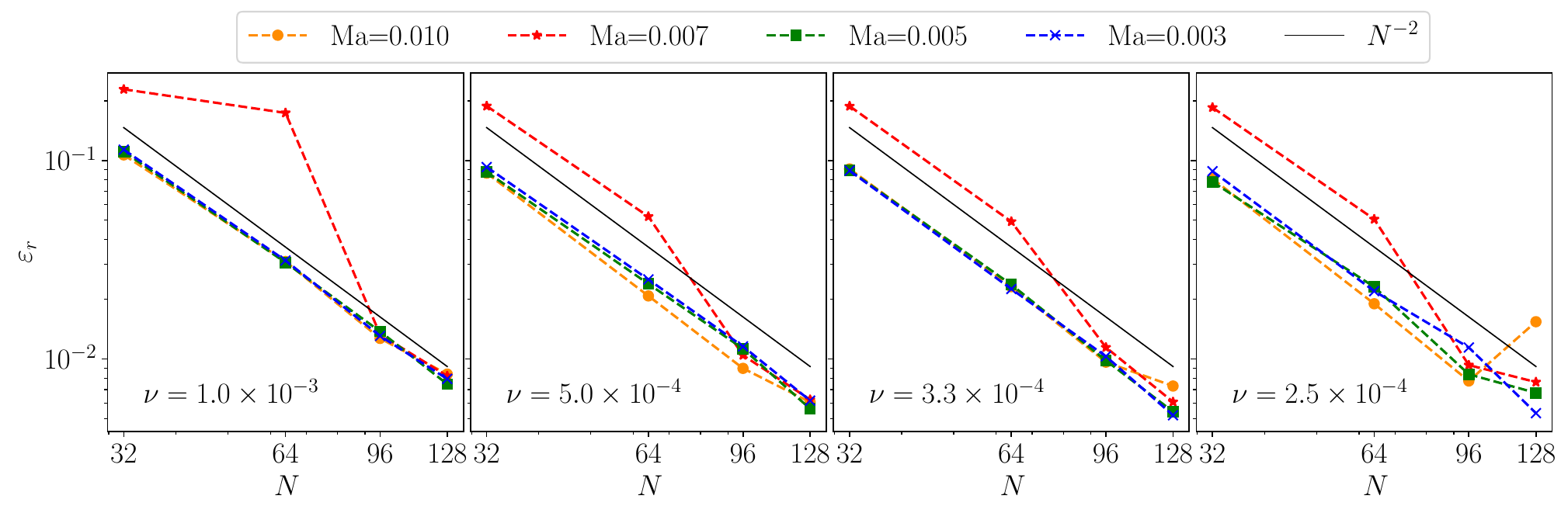}
\caption{Diffusive scaling ($\nu$ constant) of the (relative) dissipation error for different values of the Mach number ($\mathrm{Ma}$) at fixed kinematic viscosity ($\nu$).}
\label{fig:diffusive_scaling}
\end{figure}

The dispersion and dissipation errors of the LB scheme implemented in \textsc{flame} are now {assessed}. 
In this analysis, the dispersion error (or phase error) is computed by evaluating the shift in time between the {local} maxima of the numerical solution and the analytical wave solution (see Fig.~\ref{fig:sol_comparison_ex1}). 
Therefore, {tagging as} $t^\mathrm{max}_{i}$ and $\bar t^\mathrm{max}_i$ the positions in time of the maxima of the numerical  
and analytical solution (at a fixed location) {respectively}, the average value of the relative dispersion error {can be} defined as
\begin{equation}\label{eq:dispersion_error}
    \varepsilon_{\phi}=
    1-\frac{1}{M}\sum_{i=0}^{M-1}\frac{t^\mathrm{max}_{i+1}-t^\mathrm{max}_i}{\bar t^\mathrm{max}_{i+1}-\bar t^\mathrm{max}_i}
\end{equation}
over $M$ oscillating periods. {For practical purposes} we have used $M=10$. 
As expected, it {can be} observed in Fig.~\ref{fig:dispersion_error} {how} the dispersion error is very small and decreases {as} the resolution $N$ of the grid {increases}, {showing} a power-scaling law close to $1/N^2$. This confirms a second-order accuracy of the LB scheme. 
{We also} found that the dispersion error {exhibits a rather} constant behavior when changing the Mach number, and does not seem to be affected by the value of the kinematic viscosity either. 
Let us remark that some results differ from the global trend, certainly due to the premise of (physical) instabilities at the lowest viscosity. 
After {synchronizing the} phases {of} numerical and analytical solutions, the (relative) dissipation error is evaluated by comparing the velocity magnitude of the {two} solutions, \emph{i.e.}
\begin{equation}\label{eq:dissipation_error_new}
    \varepsilon_r=\left[\frac{1}{M}\sum_{i=0}^{M-1}\left(\frac{  \mathrm{u}(t_i)- \mathrm{\bar u}(t_i) }{ \mathrm{\bar u}(t_i)}\right)^2 \right]^{1/2}.
\end{equation}
The dissipation error {provides} a first measure of the numerical dissipation.
Two different scaling behaviors are considered, namely the so-called acoustic and diffusive scaling \citep{Kruger2016}. The acoustic scaling consists {in} keeping the Mach number fixed {while monitoring} the convergence rate of the error $\varepsilon_r$ for different Reynolds numbers, as a function of the resolution (see Fig.~\ref{fig:acoustic_scaling}). 
On the other hand, the diffusive scaling is obtained by keeping the lattice viscosity fixed (see Fig.~\ref{fig:diffusive_scaling}).
The behavior of the numerical solution is consistent between the two regimes, showing a convergence of the dissipation error with respect to the grid resolution $\propto 1/N^2$, as expected for a second-order scheme. 

\begin{figure}
\includegraphics[width=\textwidth]{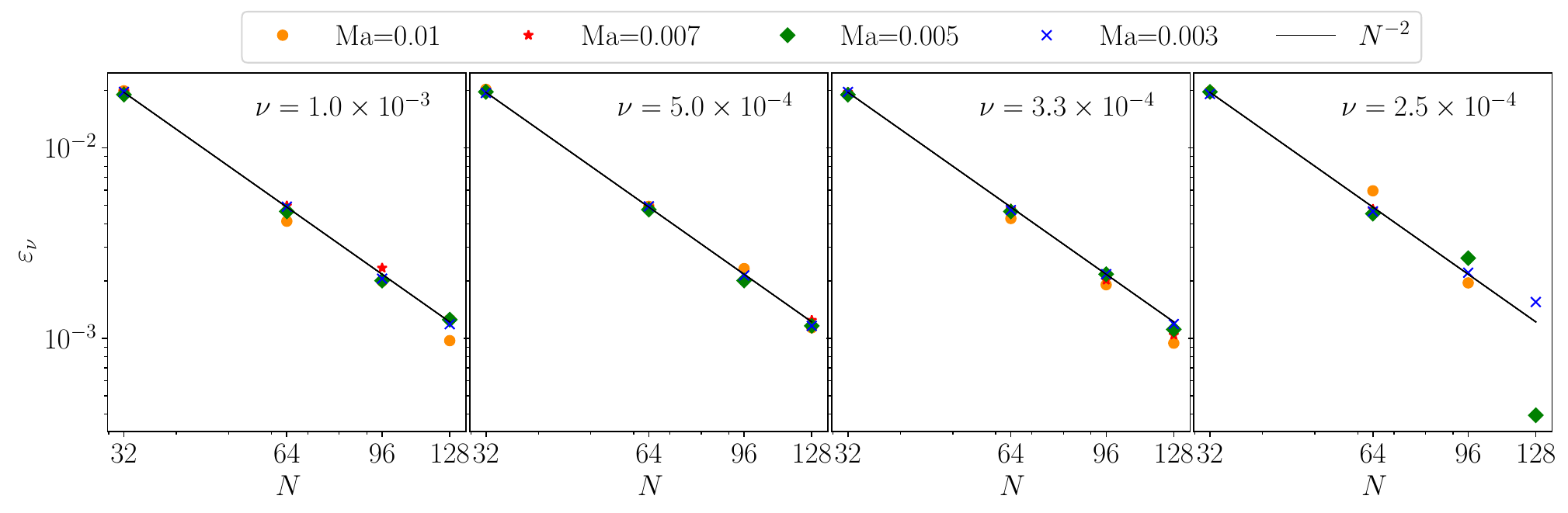}
\caption{Scaling of the ratio between the numerical and kinematic viscosities $\varepsilon_{\nu}=\nu^\mathrm{num}/\nu$ with the grid resolution ($N$) for different Mach numbers ($\mathrm{Ma}$) and kinematic viscosity. The second-order accuracy of the LB algorithm is highlighted by the black lines, \emph{i.e.} $\varepsilon_{\nu} \sim O(\Delta x^2)$.}
\label{fig:numerical_dissipation}
\end{figure}

One of the advantages of dealing with a \emph{dissipative} solution of the Hall-MHD equations is the possibility to identify an effective viscosity $\tilde \nu$ related to the damping  $\propto \exp(-\tilde \nu k^2 t)$ of the numerical solution.
By decomposing $\tilde \nu$ into the sum of a physical and a (spurious) numerical viscosity, $\tilde \nu = \nu + \nu_\mathrm{num}$, the ratio between these two contributions {reads as}
\begin{equation}\label{eq:viscosity_error}
    \varepsilon_{\nu} = \frac{\nu_\mathrm{num}}{\nu} = \frac{\tilde \nu-\nu}{\nu}.
\end{equation}
The results obtained for the viscosity error $\varepsilon_{\nu}$ are shown in Fig.~\ref{fig:numerical_dissipation}.
{Here, we} found that the numerical viscosity represents only a small percentage of the 
{estimated total} viscosity, and {it} decreases as $1/N^2$ with the resolution, which is once again consistent with a second-order accuracy of the LB scheme. Interestingly, it is observed that the (relative) viscosity error {is} independent {from} the physical viscosity and the Mach number, {whereas it} only depends on the lattice resolution. 

Finally, despite~\cite{DELLAR2002} showed that a D3Q7 lattice was sufficient to reliably account for the dynamics of each component of the magnetic field, {in order to check the validity of this statement,} LB simulations with enhanced connectivity have been performed {here} to investigate {whether} a more isotropic representation of the magnetic densities would {significantly} improve the level of accuracy of the algorithm \citep{SILVA2014}. Interestingly, our results showed no significant improvement when upgrading the magnetic lattice to D3Q15 or D3Q27 lattices (see Fig.~\ref{fig:lattice_types}), {thus confirming what was reported in \cite{DELLAR2002}}. A plausible explanation {of this} lies {on} the fact that the magnetic field is represented as a zeroth order moment of densities for each component (see~\eqref{eq:mag_field_LB}). Therefore, a few degrees of freedom are certainly sufficient to accurately reconstruct the moments and describe the magnetic field dynamics. 

\subsection{Comparison with pseudo-spectral simulations of MHD turbulence}
In this section, comparisons are made {between the dynamics of MHD plasmas simulated with \textsc{flame} and the outputs obtained} with the \textsc{ghost} pseudo-spectral solver for high-resolution simulations, {when both codes perform the same decaying test run initialized with the classical} Orszag-Tang (OT) vortex problem \citep{Orszag1979}. 
{Indeed,} the OT solution is often considered as a prototypical flow to study freely evolving MHD turbulence.
The \textsc{ghost} solver {has been} widely used {to tackle a variety of problems related to both} geophysical {fluids} and {space plasmas}~\citep{Marino2013,Pouquet2013,Marino2014,Marino2015,Mininni2002,Mininni2003,Mininni2006,Gomez2010,pouquet_helicity}.
It is a well-established community code available on {\url{https://github.com/pmininni/GHOST}}. \textsc{ghost} is a hybrid MPI/OpenMP/CUDA-parallelized {framework that hosts a variety of solvers having also GPU capability}, {delivering} high performance, robust results and an optimal scaling up to hundreds of thousand {computing cores}. 
It relies on a second-order Runge-Kutta scheme for time integration {and is de-aliased based on the classical two-third rule}. As a pseudo-spectral de-aliased code, it provides {very high} accuracy in {resolving the} spatial scales \citep{Patterson1971}. 
The OT vortex problem {prescribes the following initialization for}  the velocity and magnetic fields:
\begin{eqnarray*}
    \mathbf{U(\mathbf{x},0})&=&U_0\left[-2\sin{y}~;~2\sin{x}~;~0\right]\\
    \mathbf{B(\mathbf{x},0})&=&B_0\left[-2 \sin{2y}+\sin{z}~;~ 2\sin{x}+\sin{z} ~;~ \sin{x}+\sin{y}\right]
\end{eqnarray*}
with $U_0=1$ and $B_0=0.8$ in a cubic box of size $2\pi$. 

In the simulation {performed here}, the Reynolds number attains {values up to} $\mathrm{Re}=UL/\nu\simeq1600$ when the flow reaches its peak of dissipation.
The {small-scale} energy {dissipation} is defined as  
$\epsilon=-\nu\langle|\nabla \times \mathbf{U}|^2\rangle-\eta\langle |\mathbf{J}|^2\rangle$ and encompasses both the kinetic and magnetic dissipation with $\nu/\eta= 1$.   
In the definition of the Reynolds number, $U$ refers to the r.m.s velocity and $L=2\pi\int k^{-1}E_v(k)dk/\int E_v(k)dk$ is the integral length scale, where $E_v(k)$ is the energy spectrum of the velocity field. The Mach number is fixed at $\mathrm{Ma}=0.025$. The number of grid points in each direction is $N=512$.

\begin{figure}
\includegraphics[width=\textwidth]{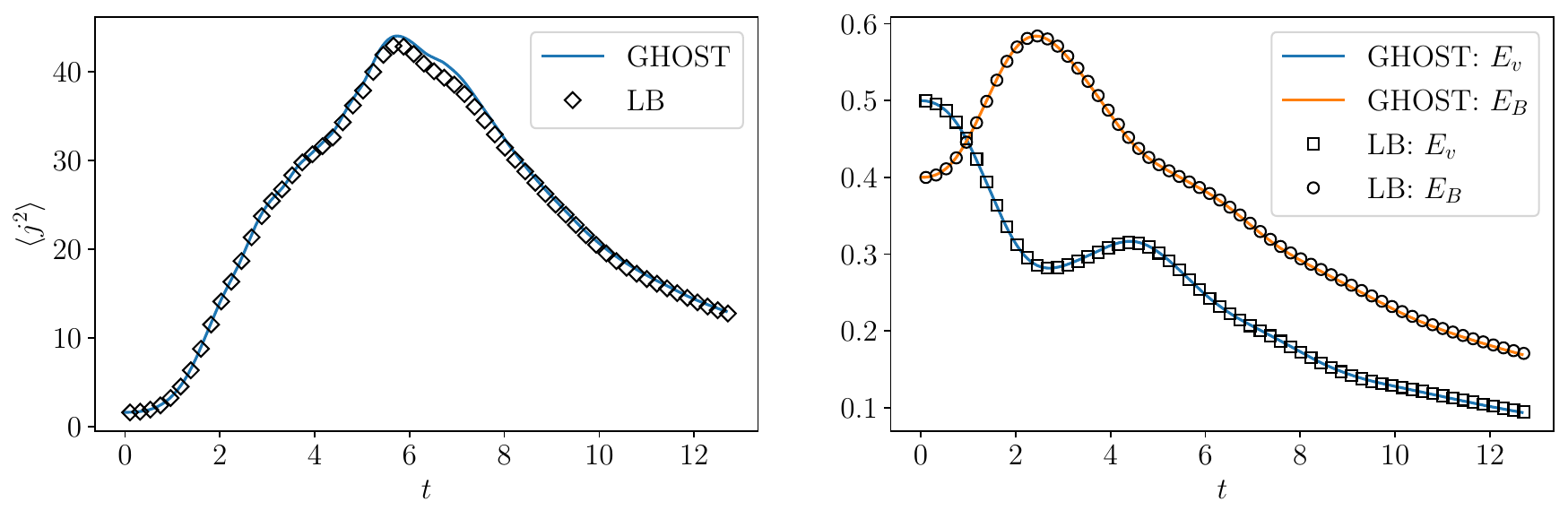}
\caption{(Left) Time evolution of the mean magnetic dissipation $\propto \langle |\mathbf{J}|^2 \rangle$ in freely-evolving MHD turbulence for a LB simulation ($N=512$) and a pseudo-spectral simulation (N=$512$) performed with the \textsc{ghost} solver. (Right) Time evolution of the mean kinetic ($E_v$) and magnetic $(E_B$) energies.}
\label{fig:LB_GHOST_time_comp}
\end{figure}

The time evolution of the mean magnetic dissipation, as well as the kinetic and magnetic energies, are shown in 
Fig.~\ref{fig:LB_GHOST_time_comp} {for two realizations of} LB and pseudo-spectral simulations of the same {OT} problem. 
{The simple visual inspection of the runs shows that} the agreement between \textsc{flame} and \textsc{ghost} is very satisfactory {for the cases under study}. Only a slight underestimation of the magnetic dissipation {in the \textsc{flame} run can be}  observed {for a few time steps} after the peak {of the current density $\mathbf{J}=\nabla \times \mathbf{B}$}. Let us recall at this stage that $\mathbf{J}$ is directly {obtained} from the magnetic densities in the LB simulation, and is not {inferred} by differentiating the magnetic field. 

\begin{figure}
\includegraphics[width=\textwidth]{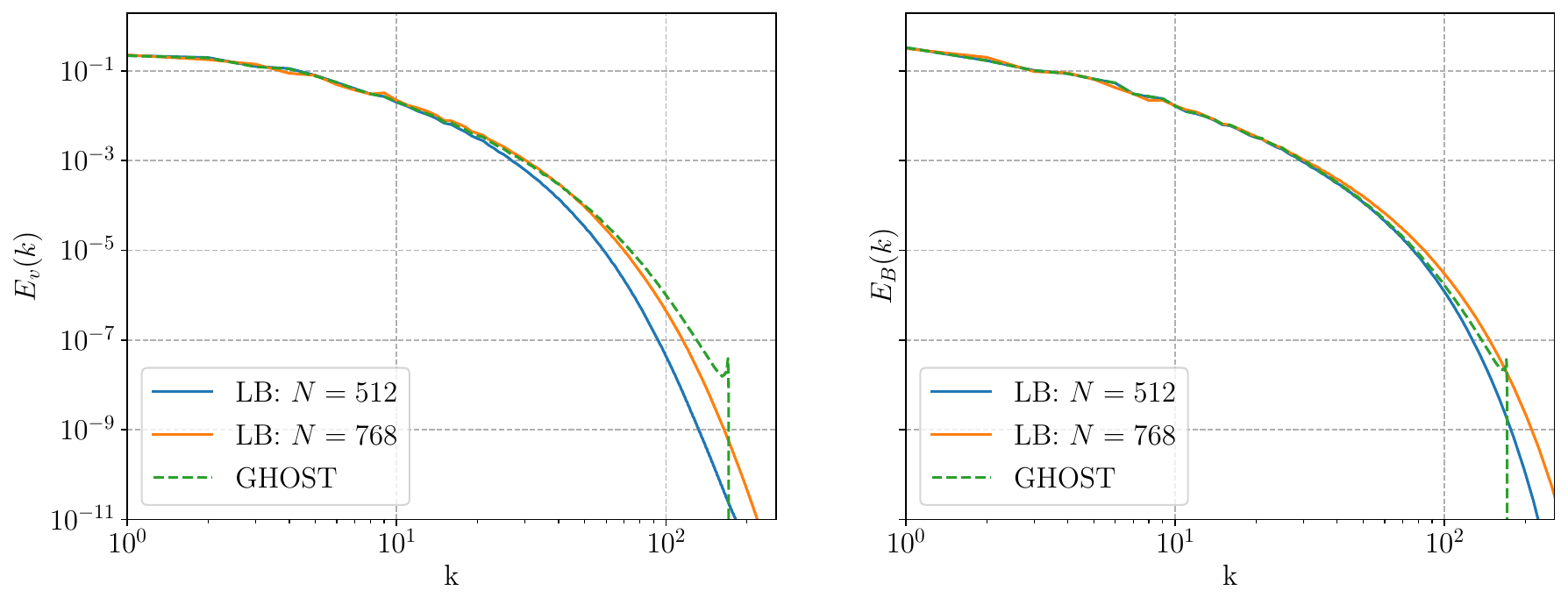}
\caption{Kinetic (left) and magnetic (right) energy spectra of MHD turbulence at the peak of magnetic dissipation. The spectra are normalised by the total kinetic and magnetic energies respectively. Comparison between LB simulations at two different resolutions ($N=512$, $N=768$) and a  de-aliased pseudo-spectral simulation ($N=512$) performed with the \textsc{ghost} solver.}
\label{fig:LB_GHOST_spectra_comp}
\end{figure}
A more detailed comparison is {provided by looking at the Fourier decomposition of the fields obtained with the two codes}. 
The kinetic and magnetic energy spectra are displayed in Fig.\ref{fig:LB_GHOST_spectra_comp} 
at the peak of {the} magnetic dissipation. 
The kinetic energy spectrum of the LB simulation {seems} over-damped at high wavenumbers. This is related to a known drawback of the moment-based collision operator, which ensures higher stability (compared to the standard BGK collision operator) but at the cost of an enhanced numerical dissipation \citep{COREIXAS_PRE_100_2019}. 
However, when increasing the spatial resolution to $N=768$, the numerical dissipation is reduced and the spectrum {of the \textsc{flame} run gets} very close to {that of the} pseudo-spectral solution. 
This observation is qualitatively consistent with  the statement made in \citet{Shen2018} that LB needs about twice the spatial resolution of a pseudo-spectral simulation to achieve similar accuracy in turbulent flows.

\begin{figure}
\includegraphics[width=\textwidth]{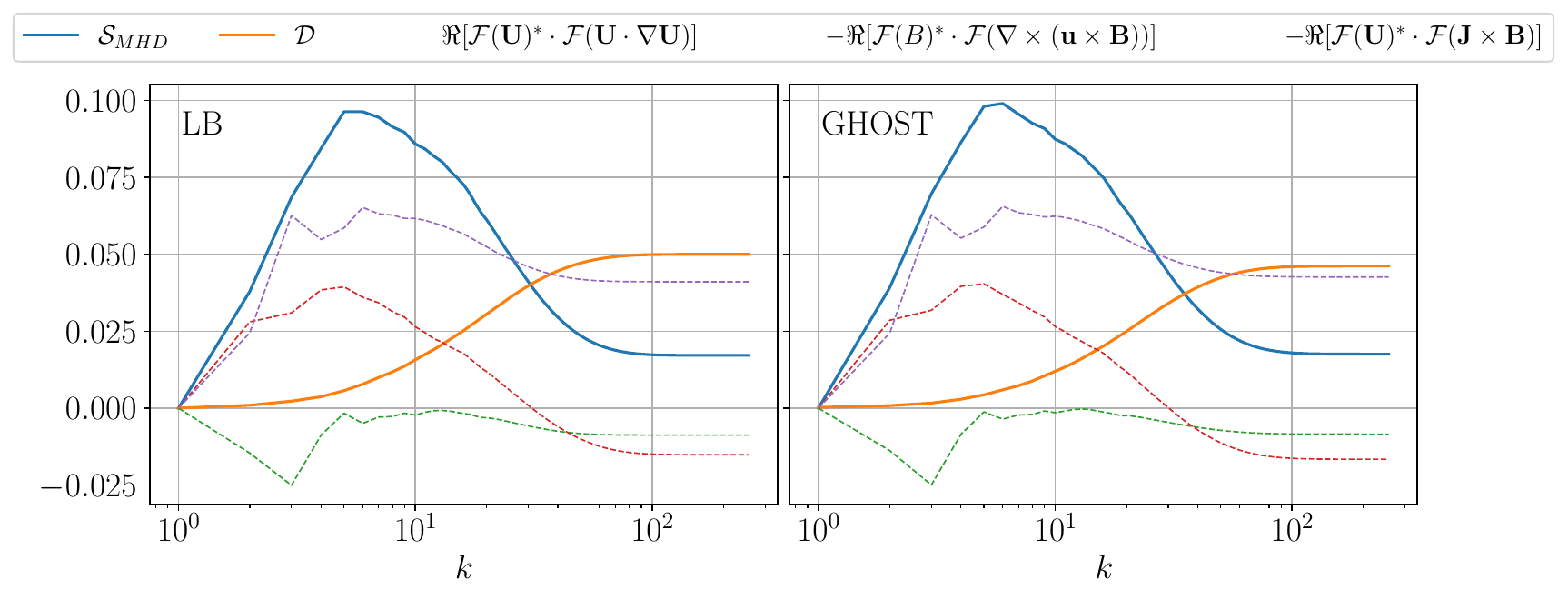}
\caption{Various third-order energy fluxes across wavenumber $k$ as reported in~\eqref{eq:mhd_en_trans}.  Comparison is made between  LB (left) and  de-aliased pseudo-spectral (right) \textsc{ghost} simulations. 
Let us notice that  $  \partial_t \sum_{|\mathbf{k}^\prime|<k} E(\mathbf{k}^\prime) = -\mathcal{S}_\mathrm{MHD}(k) -\mathcal{D}(k) <0$ as expected for freely-decaying MHD turbulence.}
\label{fig:LB_GHOST_energy_comp}
\end{figure}

Concerning the magnetic energy spectrum, the results from both simulations perfectly match, reflecting the fact that {the} BGK collision operator adopted for the magnetic scheme does not add numerical dissipation (as compared to the pseudo-spectral simulation). 
It should also be noted that, while the maximum wave-number is $k_\mathrm{max}=N/3$ (due to the 2/3 rule for de-aliasing) in pseudo-spectral simulations, the range of resolved scales reaches the Nyquist cut-off $k_\mathrm{max}=N/2$ in LB simulations.
Particular attention is now paid to the wavenumber-by-wavenumber energy budget of the MHD equations. Starting from \eqref{eq:NSE_Mtensor} and~\eqref{eq:Hall_MHD}, the (total) energy flux across wavenumber $k$ can be  {defined} as 
\begin{equation}
    \mathcal{S}_\mathrm{MHD}(k) =  \sum_{|\mathbf{k'}|<k} \Re{\left[\mathcal{F}({\mathbf{U}})^*\cdot \left( \mathcal{F}({\mathbf{U}\cdot \nabla \mathbf{U}}) - \mathcal{F}({\mathbf{J}\times \mathbf{B}}) \right) - \mathcal{F}({\mathbf{B}})^*\cdot \mathcal{F}(\nabla \times ({\mathbf{U}\times \mathbf{B}}))\right]} 
\end{equation}
whereas the (total) dissipation in the range $[0,k[$ is given by
\begin{equation}
    \mathcal{D}(k) = \sum_{|\mathbf{k'}|<k}  \nu k'^{2}|\mathcal{F}({\mathbf{U}})|^2+\eta k'^{2}|\mathcal{F}({\mathbf{B}})|^2
\label{eq:mhd_en_trans}
\end{equation}
where $\mathcal{F}(\cdot)$
means the Fourier transform and $^*$ is the complex conjugate.
The wavenumber-by-wavenumber energy budget then writes 
\begin{equation}
    \partial_t \sum_{|\mathbf{k}^\prime|<k} E(\mathbf{k}^\prime) = -\mathcal{S}_\mathrm{MHD}(k) -\mathcal{D}(k).
\end{equation}
{We would like to} mention that the contribution of the pressure term (not shown here) is negligible in {the context of these simulations}.
The {fluxes} obtained for the LB and pseudo-spectral {OT implementations}  (with $N=512$) are {displayed in detail} in Fig.~\ref{fig:LB_GHOST_energy_comp}. 
A  satisfactory agreement is {observed} in particular for the non-linear energy transfer terms, {over} the entire range of resolved wavenumbers. 
The slight over-dissipative {nature} of the LB scheme is again evidenced {in the output of} the dissipation term $\mathcal{D}(\mathrm{k})$ at very high wavenumbers.

\section{High-Resolution simulations of 3D Hall-MHD {plasmas}}\label{sec:comp_efficiency}

\begin{table}
  \begin{center}
\def~{\hphantom{0}}
  \begin{tabular}{cccccccc}
      Run & $N$                      & $\mathrm{Ma}\,[\times 10^{-4}]$ & $\mathrm{Re}$ & $\mathrm{Pr}_m$ & $\epsilon_H$ & $t_\mathrm{tot}/\tau_{0}$ & $t_\mathrm{peak}/\tau_{0}$ \\ [3pt]
I   & 512                      & 7.0         & 4400                & 1      & 0.0025   & 48.2 & 31.8                \\
II  & 512                      & 1.0         & 5240                & 1      & 0.01      & 35.8 & 33.6           \\
III & 512                      & 0.625         & 7150                & 1      & 0.025     & 29.6 & 26.5           \\ 
IV & 768                      & 0.6         &    6000            & 1      & 0.015     &  36.6 &       32.0     \\ 
  \end{tabular}
  \caption{Parameters of Hall-MHD turbulence runs. $\mathrm{Re}$, $\mathrm{Ma}$ and $\mathrm{Pr}_m$ denote respectively the Reynolds number (at the peak of magnetic dissipation), the initial Mach number and the magnetic Prandtl number. 
  The (dimensionless) Hall parameter is $\epsilon_H$. The number of grid points per dimension is $N$. The total duration of the run is $t_\mathrm{tot}/\tau_{0}$ and the time at which the peak of current density occurs is $t_\mathrm{peak}/\tau_{0}$ in units of the reference time scale $L_0/U_0$. The  Mach number satisfies the CFL condition \eqref{eq:CFL_LBM} imposed by whistler waves.}
\label{tab:runs}
  \end{center}
\end{table}

\begin{figure}
\centering
\includegraphics[width=0.8\textwidth]{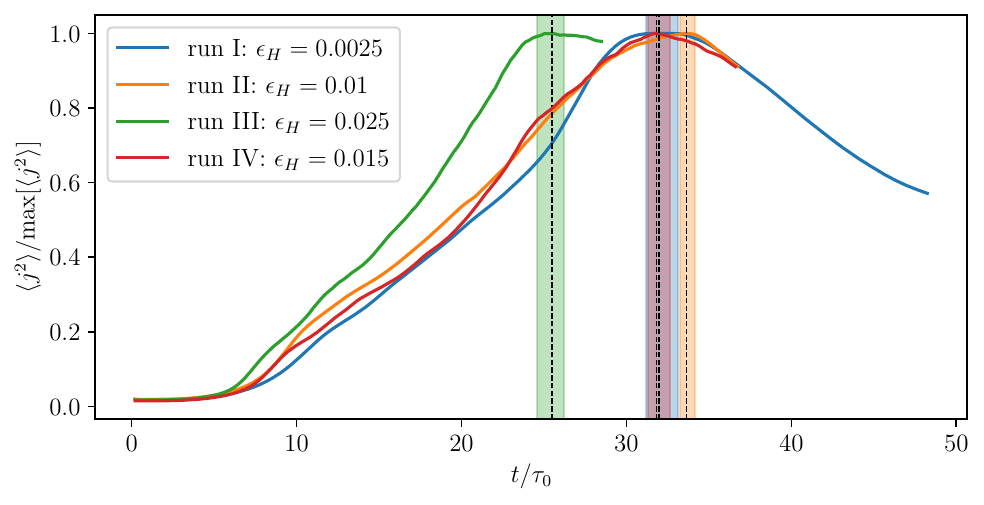}
\caption{Evolution of the magnetic dissipation over time for the three simulations performed with the OT initial condition (see Tab.\ref{tab:runs}). 
The shaded areas around the peak of the current density (black dashed line) correspond to the range over which the energy spectra in Fig.~\ref{fig:spectra_OT} have been averaged.}
\label{fig:current_density_OT}
\end{figure}

\begin{figure}
\centering
\includegraphics[width=\textwidth]{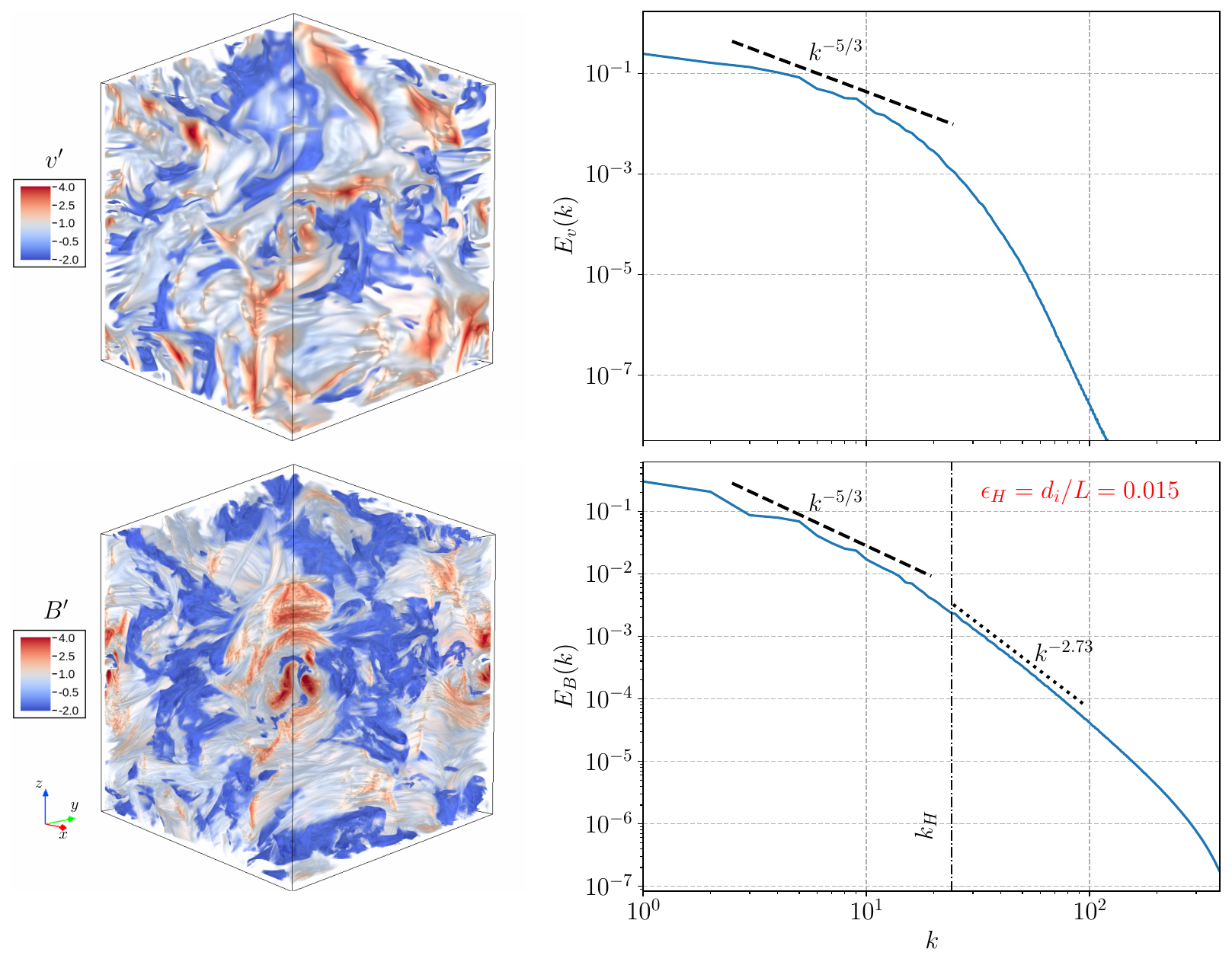}
\caption{Results of a 3D Hall-MHD simulation at $N=768^3$ taken at the peak of dissipation. The upper panels show a three-dimensional visualization of the normalized velocity field magnitude $v^\prime=(v-\bar{v})/\sigma_v$ with $v=|\mathbf{v}|$ (upper left) and the relative power density spectrum (upper right). The same in the two bottom panels for the normalized magnetic field $B^\prime$. The slopes $k^{-5/3}$ and $k^{-2.73}$ are given as a reference.}
\label{fig:OT_3D_rendering}
\end{figure}

\begin{figure}
\centering
\includegraphics[width=\textwidth]{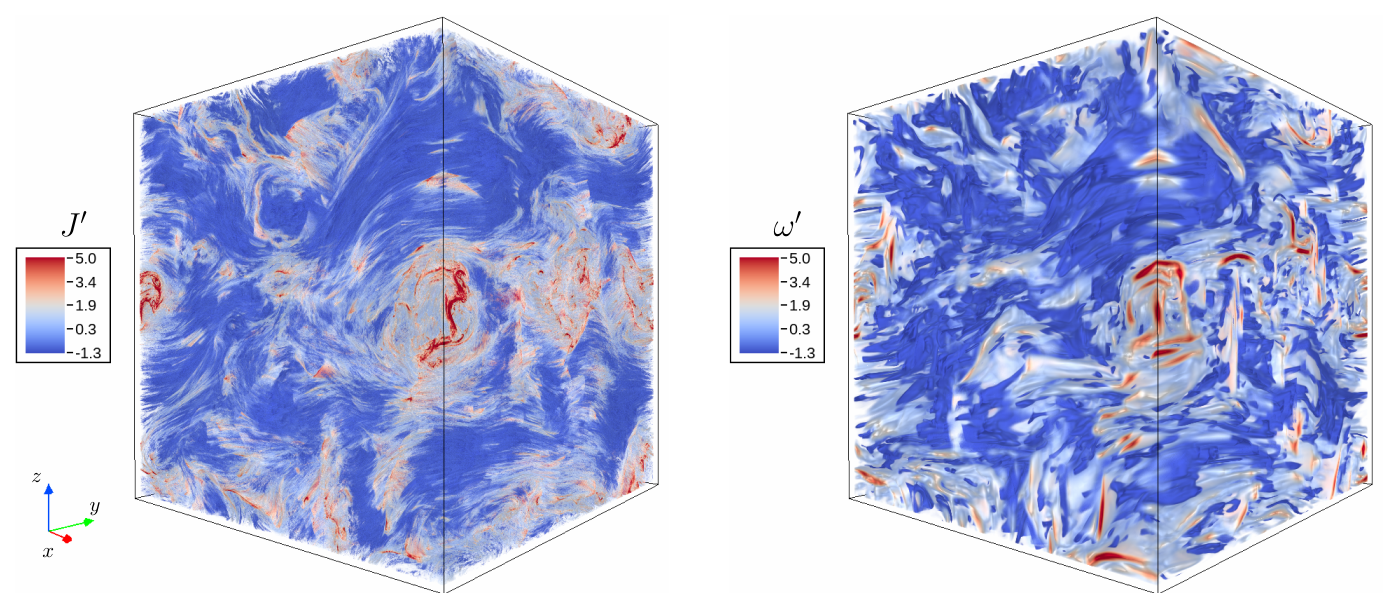}
\caption{Three-dimensional rendering of the normalized current density $J^\prime=(J-\bar{J})/\sigma_J$, with $J=|\mathbf{J}|$ (left), and vorticity $\omega^\prime$ (right) taken at the same time as in Fig.~\ref{fig:OT_3D_rendering}.}
\label{fig:OT_3D_rendering2}
\end{figure}

\begin{figure}
\includegraphics[width=\textwidth]{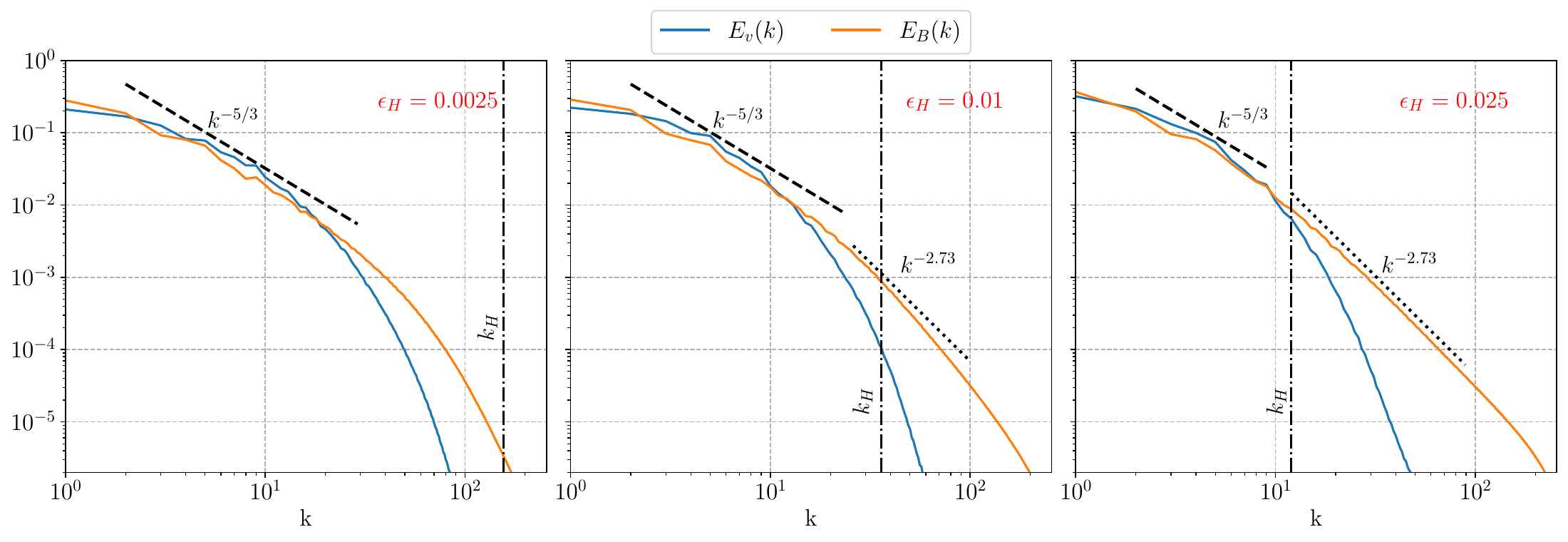}
\caption{Kinetic (blue) and magnetic (orange) spectra corresponding to three different Hall-MHD turbulence regimes (see Tab.~\ref{tab:runs}).
The black dashed and dotted lines indicate respectively the $-5/3$ and $-2.73$ slopes in reference to solar wind measurements \citep{Kiyani2015}.
All the spectra are taken at the peak of magnetic dissipation. The (black) vertical dash-dotted lines indicate the cross-over wavenumbers between MHD and Hall-MHD regimes given by $k_H=2\pi/(\epsilon_H L_0)$ with $L_0=2\pi\int E_v(k)k^{-1}dk/\int E_v(k)dk$.}
\label{fig:spectra_OT}
\end{figure}

{\textsc{flame} was used to simulate plasma dynamics in a regime in which the Hall-MHD term is non-negligible. In particular, the governing equations have been integrated in} a triply periodic cubic lattice of size $L=2\pi$ with resolution $512^3$ and $768^3$,  {initialized with the} OT vortex as described in the previous section, {for different values of the Hall parameter (see Tab.~\ref{tab:runs})}. 
%
 {The Mach number was} adjusted to the Hall parameter {in order to accommodate} the CFL condition {based on the time-scale of the} whistler waves (see~\ref{eq:CFL_LBM}).
The Reynolds number is {estimated here} at the peak of magnetic dissipation (indicated by the vertical dashed lines in Fig~\ref{fig:current_density_OT}), at which {the plasma is assumed to have reached a fully} developed turbulent state. 
For a  $512^3$ lattice dimension, {only three} GPUs were used in parallel, resulting in a computational speed of about 20 iterations per second, or equivalently, in $2.7$ BLUPS (Billions of Lattice-node Updates Per Second). This led to a wall-clock computational time of 10, 55 and 69 hours respectively for the three runs {indicated in Tab.~\ref{tab:runs}} to pass the peak of magnetic-energy dissipation. The computational times reported above are comprehensive of the time required to transfer the three-dimensional vector fields ($\mathbf{u}$, $\mathbf{B}$ and $\mathbf{J}$) between the CPUs and GPUs, and perform post-processing operations such as the tracking of the mean kinetic and magnetic energies, and mean energy dissipation rates. All computations were performed in double precision. 
A rendering of the large-scale fields $\mathbf{u}$ and $\mathbf{B}$ is {shown} in Fig.~\ref{fig:OT_3D_rendering} for the simulation at {the highest} resolution
(run IV in Tab.~\ref{tab:runs}), {again} taken at the peak of {the} magnetic-energy dissipation. The three-dimensional visualization is {displayed} together with the kinetic and magnetic {energy} spectra, the latter showing two regimes above and below the ion inertial length $d_i$. At the same time, the small-scale {activity visible in Fig.~\ref{fig:OT_3D_rendering2}} for the electric current density $\mathbf{J}$ and the vorticity $\boldsymbol{\omega}=\bNabla\times \mathbf{u}$, {emphasizes the presence of}
 current sheets, Kelvin-Helmholtz instabilities and vortices, {emerging as the disordered} structures characteristic of the Hall effect~\citep{Miura2014}. 
Furthermore, {we have found that} increasing the intensity of the Hall effect produces a faster development of turbulence in {the} plasmas {under study} due to the presence of {both} whistler and Hall-drift waves, propagating {quicker} than the Alfv\'en waves in {the} ideal MHD~\citep{Huba2003}. 
This is consistent with the behavior captured in Fig.~\ref{fig:current_density_OT} for the three runs,  increasing {the} Hall parameter.
The kinetic and magnetic energy spectra {averaged over a time interval (around the peak of dissipation, as indicated by the shaded areas in Fig.~\ref{fig:current_density_OT})} are plotted in Fig.~\ref{fig:spectra_OT} for each run at resolution $512^3$. 
As expected, {increasing the value of} the Hall parameter $\epsilon_H$ (indicated by the vertical dash-dotted line in Fig.~\ref{fig:spectra_OT}) {produces a shift} of the Hall length-scale $L_H$ towards {larger} scales,  hence {a shrink of the} Kolmorogov's $k^{-5/3}$ {power law range in} both kinetic and magnetic energy spectra. 
{A very} surprising and promising {feature} of these simulations is the behavior of the magnetic spectra in the Hall-MHD regime. {In fact,} at wavenumbers $k > k_H$, the spectrum develops (as $\epsilon_H$ increases) a power-law scaling that is in perfect agreement with the $k^{-2.73}$ scaling obtained from {the spectral analysis of} solar wind measurements at sub-ion scales, {as reported in} \citep{Kiyani2015}. 

{In the MHD regime, the time step of the} (compressible) Lattice Boltzmann {runs} is constrained {by the need for resolving} sound waves. Therefore, the time-step of an LB simulation is {typically much smaller} compared to the time-step of {equivalent} (incompressible) pseudo-spectral simulation, the ratio between the two time-steps being typically the Mach number \citep{Horstmann2022}. 
{Therefore, in the case of the MHD, the advantage for our} LB {scheme} in terms of turn-around times is not {that} big compared to {standard} pseudo-spectral simulations. {The situation is different when it comes to the simulation of plasmas in the Hall-MHD regime}, where the time-steps of the two methods are identically constrained by the speed of whistler waves. In {this case}, the efficiency of the LB scheme (exploiting the computational power of GPU accelerators) becomes a major advantage with respect to pseudo-spectral simulations,  {leading} to wall-clock turn-around times that are significantly smaller {for LB schemes, and for \textsc{flame} in particular}.
Finally, {we would like to mention} that {an} extension {of \textsc{flame} allowing the simulation of} {the} electron MHD {dynamics} {would simply consist in} modifying the equilibrium distributions for the magnetic field in the LB scheme, by neglecting the bulk velocity $\mathbf{U}$ with respect to the Hall current $\alpha_H \mathbf{J}$ in \eqref{eq:gHall_eq}.

\section{Hall-MHD simulations for space plasma turbulence investigations}
\label{sec:sim_for_space_plasma}
\begin{figure}
	\begin{center}
		\includegraphics[width=\textwidth]{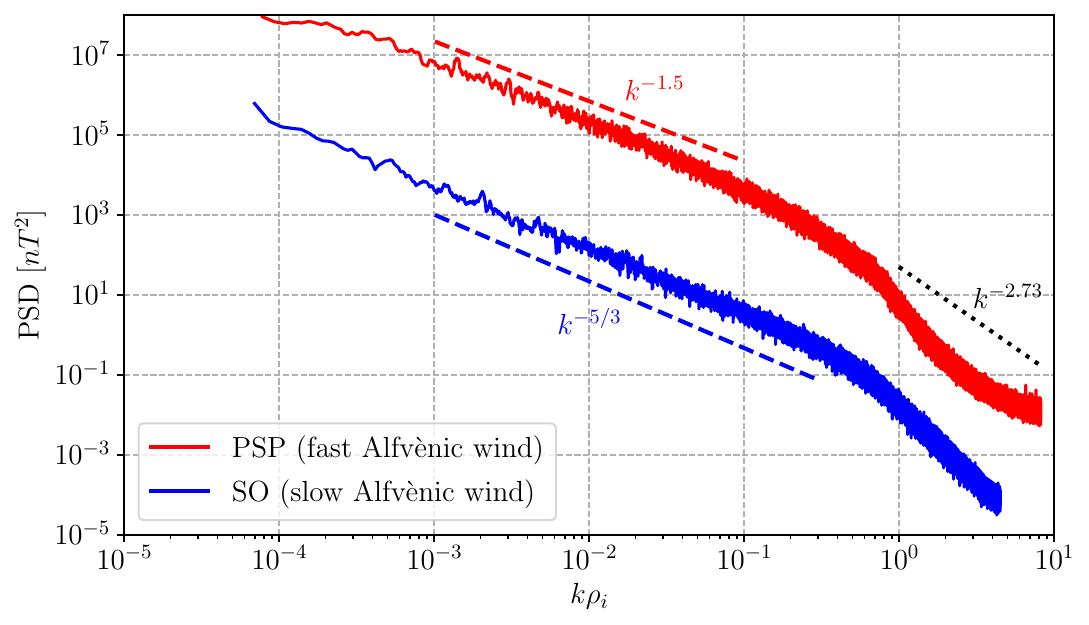}
	\end{center}
	\caption{Power Spectra Density (PSD) of the magnetic field fluctuations observed by PSP (red) and SO (blue) on November $20$, $2021$ and July $14-15$, $2020$, respectively. The $k^{-1.67}$, $k^{-1.5}$ and $k^{-2.73}$ scalings are shown for reference as colored lines}
	\label{fig:solo_psp_spectra}
\end{figure}

Space plasmas, whose dynamics involve the turbulent transport of the energy from very large \citep{Adhikari_2015,adhikari_proceed} down to the very small scales \citep{Cerri_2016} due to their large Reynolds numbers \cite{Matthaeus, Tulasi}, do actually develop well-defined MHD and Hall-MHD power-law spectral ranges, with a distinct transition between them. This clearly emerges from the observations performed with plasma and magnetic field instruments on board of two of the most recent space missions: Solar Orbiter \citep[SO;][]{2020A&A...642A...1M} and Parker Solar Probe \citep[PSP;][]{2016SSRv..204....7F}. Fig.~\ref{fig:solo_psp_spectra} shows the trace spectra of the magnetic field fluctuations measured by the PSP/FIELDS \citep{2016SSRv..204...49B} and SO/MAG \citep{2020A&A...642A...9H} magnetometers on board these state of the art spacecrafts. In particular, the PSP (red) magnetic field sample, measured on November $20$, $2021$, is relative to the fast solar wind plasma stream coming from an equatorial coronal hole, while the SO (blue) sample is relative to a low-speed solar wind stream measured on July $14-15$, $2020$, whose origin was identified in a coronal streamer and pseudo-streamer configuration \citep{2021A&A...656A..21D}. In Tab.~\ref{tab:solo_psp_spectra} we report the characteristic parameters of  these solar wind samples. It is worth recalling that the ion gyroradius $\rho_i=v_{T,i}/\omega_{ci}$ (with $v_{T,i}$ ion thermal speed) and inertial length $d_i=c/\omega_{pi}$ are defined in terms of ion cyclotron $\omega_{ci}$ and plasma frequency $\omega_{pi}$, respectively, the latter being in general significantly larger than the former. For values of density $n$, temperature $T$ and $\beta$ typical for space plasmas, the relation $\rho_i\lesssim d_i$ is valid. However, it has been remarked by several authors how, for $\beta \sim \mathcal{O}(1)$, these characteristic length scales are comparable $\rho_i\simeq d_i$~\citep{Alex2008,Alex2009,Sar2009,Kiyani2015}. Thus, in the solar wind the breaking point identifying the transition between the end of the MHD range and the beginning of the range where plasma kinetic effects become relevant, in the magnetic field spectrum, at the sub-ion scales, is often referred as occurring either at the ion gyroradius or at the inertial length scale, when $\beta \sim 1$.
In spite of the different speeds, both the SO and the PSP solar wind samples we considered here are Alfv\'enic, i.e., they are characterized by a high correlation between velocity and magnetic field fluctuations \citep[see][and references therein, for a comprehensive review on the solar wind turbulence]{2013LRSP...10....2B}. A clear frequency break is observed at $k\rho_{i}\sim1$ separating fluid and kinetic scales, as shown in Fig.\ref{fig:solo_psp_spectra}, marking the transition from the MHD turbulent inertial range (where energy is adiabatically transferred to smaller and smaller scales), that is characterized by a Kolmogorov-like spectrum \cite{Marino2011,Marino2012,Marino2023}, to a range where the kinetic effects begin to dominate and in which the energy gets dissipated (at the bottom of such range), ultimately heating the solar wind plasma \cite{Marino2008}. As is known from spacecraft observations, fluid and kinetic scales in the solar wind are characterized by different power-law spectral exponents. Features of these spectral ranges mostly depend on the distance from the Sun at which observations are made, i.e., on the observed stage of evolution of the solar wind turbulence \citep[see, e.g.,][]{2021ApJ...912L..21T,2022ApJ...938L...8T}. The physical phenomena as well as the governing parameters controlling the evolution of turbulence in the interplanetary space are still matter of investigation. Nonlinear interactions \citep{2013LRSP...10....2B}, expansion-driven magnetic \citep{2021A&A...650A..21S} and velocity shears \cite{Marino2012}, as well as the parametric decay of Alfv\'en waves \citep{1996PhPl....3.4427M}, all certainly play some role. However, to date, there is not a clear consensus on how turbulence evolves from a spectrum resembling the one predicted by the Iroshnikov-Kraichnan phenomenology \citep{1963AZh....40..742I,1965PhFl....8.1385K} to a Kolmogorov-like spectrum \citep{1941DoSSR..30..301K} as the solar wind expands from regions within the solar corona, or very close to it, to the outer heliosphere. Moreover, the slope of the magnetic-field spectrum beyond the ion skin depth (or ion inertial length) is highly variable, with power-law exponents ranging from $\sim-4$ to $\sim-2$ \citep{2006ApJ...645L..85S,2014ApJ...793L..15B}, being also affected by the redistribution of the magnetic field energy at the (larger) fluid scales: in general, the larger is overall the power spectral density (PSD) within the MHD inertial range, the steeper is the spectrum at the kinetic scales. A number of dissipative wave-particle mechanisms are supposedly involved in the energy transfer and dissipation at the very small scales. Among these, cyclotron-resonant dissipation certainly plays an important role \citep[see, e.g.,][]{2014ApJ...787L..24B,2019ApJ...885L...5T}, though the way energy is first brought to the small scales then dissipated in the collision-less solar wind plasma is still a matter of debate. Both the evolution of turbulence in the heliosphere and how energy is dissipated in the solar wind, are major open questions in the space plasma community that could be effectively targeted by means of numerical investigations produced with \textsc{flame}, which allows capturing the transition between MHD and Hall-MHD regimes (Fig.\ref{fig:spectra_OT}), like the more standard pseudo-spectral codes. Another puzzle of solar and space plasma dynamics that can be tackled with our LB code is how magnetic switchbacks observed in the solar corona as well as in the solar wind do contribute to the local heating of the plasma. The switchbacks are intermittent magnetic-field polarity reversals widely observed in the heliosphere \citep{2019Natur.576..237B} and in the solar corona \citep{2022ApJ...936L..25T}, that are thought to play a major role in the acceleration and heating of the solar wind. However, characterizing their contribution to the plasma energetics among other plasma processes is a challenging task, for which it is important to run highly accurate $3$D Hall-MHD numerical simulations, able to resolve an extended dynamical range with the largest possible scale separation.
A first implementation of \textsc{flame} aiming at demonstrating the decaying nature of solar wind turbulence has been presented in~\citep{Sorriso2023}, where a direct comparison between the simulated fields and the observations performed by the Helios 2 spacecraft is proposed, showing very good agreement.

\begin{table}
  \begin{center}
\def~{\hphantom{0}}
  \begin{tabular}{cccccccc}
      Probe & $\bar{\rho}\,[cm^3]$ & $\bar{V}\,[km/s]$ & $\bar{T}\,[MK]$ & $\bar{B}\,[n\mathrm{T}]$ & $\beta$ & $d_i\,[km]$ & $D_{sun}\,[\mathrm{AU}]$\\ [3pt]
PSP   & 419                      & 622         & 1.89                & 332      & 0.50   & 57.5 & 0.09\\
SO  & 16                      & 429        & 1.50                & 6.76     & 3.59     & 11.1 & 0.64\\
  \end{tabular}
  \caption{Main solar wind parameters computed at time where the solar wind samples used to produce the power spectra in Fig.~\ref{fig:solo_psp_spectra} have been collected. Here we report solar wind density $\bar{\rho}$, solar wind bulk velocity $\bar{V}$, proton temperature $\bar{T}$, average magnetic field $\bar{B}$, ion inertial length $d_i$ and the distance of the spacecraft (SO and PSP) from the Sun $D_{sun}$.}
\label{tab:solo_psp_spectra}
  \end{center}
\end{table}
%
%
%
%

%
%

\section{Conclusions}\label{sec:conclusion}
The LB approach {extends the} horizon for the numerical investigation of plasma dynamics. Stability issues, which have long been a handicap for the {implementation} of the LB method to {investigate} turbulent flows, are now {mostly} solved thanks to the use of improved collision operators {that do not compromise} the accuracy {of the numerical solutions}. 
Furthermore, the computational efficiency of the LB schemes on many-core devices such as GPUs allows for advantageous turn-around times. 
{A major} advantage of dealing with a kinetic representation {at the level of the numerical scheme}, is that the derivatives of the magnetic field are directly embedded in the solution,  allowing for an intrinsically accurate description of the current density {since it does not require further implementations of a} differentiation scheme. 
{The study presented here} shows that the LB approach provides a valuable and efficient numerical tool to simulate Hall-MHD {plasma} turbulence. 
Furthermore, the  LB approach a priori allows us to add {complexity to the plasma,} such as thermal effects, multi-species, complex geometries, etc. at the cost of new coupled lattice dynamics and boundary conditions, therefore preserving the computational performance. 
Extended MHD codes currently utilized for tokamak applications such as NIMROD \citep{NIMROD}, BOUT++ \citep{bout++} or JOREK \citep{JOREK} rely on implicit or semi-implicit timestepping to ensure stability with longer time-steps than what is imposed by the CFL condition for explicit timestepping.
In our (explicit) scheme, the time-step was established by default to meet this later condition according to  \eqref{eq:CFL_LBM}. It would be valuable to investigate the extent to which this constraint on $\Delta t$ could be relaxed while maintaining stability, due to the magnetic diffusivity (and to a lesser extent the numerical diffusivity) taming  whistler waves at high frequencies. Preliminary tests indicate that there may indeed be room to increase the time-step in the context of Hall-MHD turbulence.
The preliminary results provided by a simple benchmark based on the OT vortex problem {anticipate} that our code {will be able to deliver excellent performances with the simulation of} astrophysical and space plasmas {in which} the Hall term {is expected to} play a significant role {in} the dynamics of the system. 
Indeed, in plasmas as well as in anisotropic fluids, turbulence has to compete with waves in transferring energy across the scales \citep{Marino2015a,herbert16}. The interplay of waves and turbulence is responsible for the emergence of new characteristic length scales and the existence of different regimes \citep{Marino2013,Feraco2018} in which various forms of energy can cascade to small or to large scales \citep{Marino2014}, or undergoing a dual energy cascade \citep{Pouquet2013,Marino2015}. The computational efficiency of our LB model will allow us to run simulations of fluids and plasmas separating regimes (in terms of spatial and temporal scales) where different physical phenomena dominate. 
We proved, {though in a simplified} configuration, that our {LB} model 
is able to capture {the} physical effects {produced} by the Hall term, such as faster dynamics due to the interplay of whistler waves and turbulence, the breakdown of the Kolmogorov spectrum at sub-ion scales and a behavior of the magnetic energy spectrum at that scales which has been already observed in solar wind measurements. All that provides {\textsc{flame}} with the potential to become a {powerful} tool {for the} investigation {of} magnetohydrodynamic plasmas in a variety of configurations of interest for heliospheric and magnetospheric studies. 
While an incompressible (or weakly-compressible) formulation is usually justified in the context of a space plasma~\citep{Andres2022,Brodiano2023}, it would be worth accounting explicitly for compressible effects to reach a more comprehensive representation of its dynamics. Adding compressibility effects would require resorting to an additional equation of state and a coupled lattice scheme to deal with the temperature or density space-and-time evolutions.
The present {analysis and tests were performed} using a benchmark configuration (OT vortex problem) which is isotropic,  hence does not {embed} the  anisotropy {introduced by} the ambient magnetic field {in which} the solar wind {develops its dynamics}. However,  LB simulations of Hall-MHD flows {performed with \textsc{flame}} will be suitable to investigate {plasmas immersed in a background magnetic field,} at scales that are nowadays within the reach of the high-resolution instruments on board of the latest solar and magnetospheric missions, such as Solar Orbiter or Parker Solar Probe. 

\section*{Acknowledgements}

We gratefully thank Dr. Emmanuel Quemener for the technical support provided during the development of our multi-GPU code at the Centre Blaise Pascal computer testing platform of the École Normale de Lyon (France).
Most of the simulations were ran on HPC facilities at the École Centrale de Lyon (PMCS2I), in Ecully (France), that is supported by the Auvergne-Rhône-Alpes region through the GRANT CPRT07-13 CIRA and the national Equip@Meso grant (ANR-10-EQPX-29-01).
We acknowledge as well CINECA (Italy) that provided cpu time under the ISCRA initiative (to perform high-resolution simulations of Hall-MHD turbulence) as well as support in the frame of the project LaB-HMHD - HP10C4HXCB. 
R.M., R.F. and F.F. acknowledge support from the project ``EVENTFUL'' (ANR-20-CE30-0011), funded by the French ``Agence Nationale de la Recherche'' - ANR through the program AAPG-2020. The collaboration of R.F. and R.M. was facilitated by support from the International Space Science Institute in ISSI Team 556.
We kindly acknowledge the two anonymous referees for the relevant and interesting remarks which helped to significantly improve the presentation of our results.

\section*{Data availability}
The data that support the findings of this study are available from the corresponding author, R.F., upon reasonable request.

\section*{Competing interests}
The authors report no conflict of interests.

\appendix

\section{Central-moment-based LB scheme for fluid dynamics}\label{sec:appendix_1}
For the fluid, the discretization (in velocity) of the phase space refers to the D3Q27 lattice. The set of adopted microscopic velocities $\{\mathbf{c}_i\}_{i=0,..,26}$ is defined in Cartesian components by
\begin{equation*}\label{eq:LB_velocities_app}
\begin{split}
    |c_x\rangle = [0,-1,0,0,-1,-1,-1,-1,0,0,-1,-1,-1,-1,1,0,0,1,1,1,1,0,0,1,1,1,1]^\top \\
    |c_y\rangle = [0,0,-1,0,-1,1,0,0,-1,-1,-1,-1,1,1,0,1,0,1,-1,0,0,1,1,1,1,-1,-1]^\top \\
    |c_z\rangle = [0,0,0,-1,0,0,-1,1,-1,1,-1,1,-1,1,0,0,1,0,0,1,-1,1,-1,1,-1,1,-1]^\top
\end{split}
\end{equation*}

The equilibrium densities (without accounting for the Lorentz force) are developed up to the sixth-order as
\begin{equation*}\label{eq:Hermite_exp}
  \begin{split}
    {f}_i^{(0)}(\rho, \mathbf{U})={}& w_i\rho \bigg{\{} 1+\frac{\mathbf{c}_i\cdot \mathbf{U}}{c_s^2} +
    \frac{1}{2c_s^4} \bigg{[} H^{(2)}_{ixx}U^2_x+H^{(2)}_{iyy}U^2_y+H^{(2)}_{izz}U^2_z+2\bigg{(}H^{(2)}_{ixy}U_xU_y+\\
    & + H^{(2)}_{ixz}U_x U_z+H^{(2)}_{iyz}U_y U_z\bigg{)}\bigg{]}+\frac{1}{2c_s^6}\bigg{[}H^{(3)}_{ixxy}U^2_x U_y+H^{(3)}_{ixxz}U^2_x U_z+H^{(3)}_{ixyy}U_x U^2_y+ \\ 
    & +H^{(3)}_{ixzz}U_x U^2_z+H^{(3)}_{iyzz}U_y U^2_z+H^{(3)}_{iyyz}U^2_y U_z+2H^{(3)}_{ixyz}U_x U_y U_z \bigg{]}+\frac{1}{4c_s^8}\bigg{[}H^{(4)}_{ixxyy}U^2_x U^2_y+\\ 
    & +H^{(4)}_{ixxzz}U^2_x U^2_z+H^{(4)}_{iyyzz}U^2_y U^2_z+2\bigg{(}H^{(4)}_{ixyzz}U_x U_y U^2_z+H^{(4)}_{ixyyx}U_x U^2_y U_z+\\
    & +H^{(4)}_{ixxyz}U^2_x U_y U_y\bigg{)}\bigg{]}+\frac{1}{4c_s^{10}}\bigg{[}H^{(5)}_{ixxyzz}U^2_x U_y U^2_z+H^{(5)}_{ixxyyz}U^2_x U^2_y U_z+\\
    & +H^{(5)}_{ixyyzz}U_x U^2_y U^2_z\bigg{]}+\frac{1}{8c_s^{12}}H^{(6)}_{ixxyyzz}U^2_x U^2_y U^2_z\bigg{\}}
\end{split}
\end{equation*}
where the weights $w_i$ are related to the lattice connectivity with $w_\mathrm{center}=8/27$, $w_\mathrm{face}=2/27$, $w_\mathrm{edge}=8/27$ and $w_\mathrm{corner}=1/216$ for the D3Q27 lattice (see Fig.~\ref{fig:lattice_types}), and $H^{(n)}_i$ refers to the n$^\mathrm{th}$-order Hermite polynomial tensor in velocity $\mathbf{c}_i$. 
The Lorentz force is eventually taken into account by upgrading the densities as
\begin{equation*}
    {f}_i^{\mathrm{mhd}(0)}(\rho, \mathbf{U}, \mathbf{B}) =
    {f}_i^{(0)}(\rho, \mathbf{U}) +
    \frac{w_i}{2c_s^4} \left((\mathbf{B}\cdot \mathbf B) (\boldsymbol{c}_i\cdot\boldsymbol{c}_i)-(\mathbf{c}_i\cdot\mathbf{B})^2\right).
\end{equation*}

The set of central moments $\mathrm{k_i}$ is computed by applying the (invertible) transformation matrix $\mathrm{T}$ with the column vectors
\begin{equation*}\label{eq:transf_matrix_app}
    \begin{aligned}
    &|\mathrm{T}_0\rangle=|1\rangle \\
    &|\mathrm{T}_1\rangle;|\mathrm{T}_2\rangle;|\mathrm{T}_3\rangle=[\bar{c}_{ix}]^\top;[\bar{c}_{iy}]^\top;[\bar{c}_{iz}]^\top\\
    &|\mathrm{T}_4\rangle;|\mathrm{T}_5\rangle;|\mathrm{T}_6\rangle=[\bar{c}_{ix}\bar{c}_{iy}]^\top;[\bar{c}_{ix}\bar{c}_{iz}]^\top;[\bar{c}_{iy}\bar{c}_{iz}]^\top\\
    &|\mathrm{T}_7\rangle;|\mathrm{T}_8\rangle;|\mathrm{T}_9\rangle=[\bar{c}_{ix}^2-\bar{c}_{iy}^2]^\top;[\bar{c}_{ix}^2-\bar{c}_{iz}^2]^\top;[\bar{c}_{ix}^2+\bar{c}_{iy}^2+\bar{c}_{iz}^2]^\top\\
    &|\mathrm{T}_{10}\rangle;|\mathrm{T}_{11}\rangle;|\mathrm{T}_{12}\rangle=[\bar{c}_{ix}\bar{c}_{iy}^2+\bar{c}_{ix}\bar{c}_{iz}^2]^\top;[\bar{c}_{ix}\bar{c}_{iy}^2+\bar{c}_{iy}\bar{c}_{iz}^2]^\top;[\bar{c}_{ix}^2\bar{c}_{iy}+\bar{c}_{iy}^2\bar{c}_{iz}]^\top\\
    &|\mathrm{T}_{13}\rangle;|\mathrm{T}_{14}\rangle;|\mathrm{T}_{15}\rangle=[\bar{c}_{ix}\bar{c}_{iy}^2-\bar{c}_{ix}\bar{c}_{iz}^2]^\top;[\bar{c}_{ix}\bar{c}_{iy}^2-\bar{c}_{iy}\bar{c}_{iz}^2]^\top;[\bar{c}_{ix}^2\bar{c}_{iy}-\bar{c}_{iy}^2\bar{c}_{iz}]^\top\\
    &|\mathrm{T}_{16}\rangle=[\bar{c}_{ix}\bar{c}_{iy}\bar{c}_{iz}]^\top \\
    &|\mathrm{T}_{17}\rangle;|\mathrm{T}_{18}\rangle;|\mathrm{T}_{19}\rangle=[\bar{c}_{ix}^2\bar{c}_{iy}^2+\bar{c}_{ix}^2\bar{c}_{iz}^2+\bar{c}_{iy}^2\bar{c}_{iz}^2]^\top;[\bar{c}_{ix}^2\bar{c}_{iy}^2+\bar{c}_{ix}^2\bar{c}_{iz}^2-\bar{c}_{iy}^2\bar{c}_{iz}^2]^\top;[\bar{c}_{ix}^2\bar{c}_{iy}^2-\bar{c}_{ix}^2\bar{c}_{iz}^2]^\top\\
    &|\mathrm{T}_{20}\rangle;|\mathrm{T}_{21}\rangle;|\mathrm{T}_{22}\rangle=[\bar{c}_{ix}^2\bar{c}_{iy}\bar{c}_{iz}]^\top;[\bar{c}_{ix}\bar{c}_{iy}^2\bar{c}_{iz}]^\top;[\bar{c}_{ix}\bar{c}_{iy}\bar{c}_{iz}^2]^\top\\
    &|\mathrm{T}_{23}\rangle;|\mathrm{T}_{24}\rangle;|\mathrm{T}_{25}\rangle=[\bar{c}_{ix}\bar{c}_{iy}^2\bar{c}_{iz}^2]^\top;[\bar{c}_{ix}^2\bar{c}_{iy}\bar{c}_{iz}^2]^\top;[\bar{c}_{ix}^2\bar{c}_{iy}^2\bar{c}_{iz}]^\top\\
    &|\mathrm{T}_{26}\rangle=[\bar{c}_{ix}^2\bar{c}_{iy}^2\bar{c}_{iz}^2]^\top 
    \end{aligned}
\end{equation*}
where $\bar{\mathbf{c}}_i=\mathbf{c}_i-\mathbf{U}$ is the set of microscopic velocities obtained by the shift of particle velocities by the local fluid velocity. The $27\times 27$ collision matrix $S$ for the central moments is a diagonal matrix with the respective relaxation rates
\begin{equation*}\label{eq:collision_matrix_app}
    S=\mathrm{diag}[1,1,1,1,\omega,\omega,\omega,\omega,\omega,1,...,1],
\end{equation*}
which leads to
\begin{equation*}\label{eq:post_collision_moments}
    \begin{aligned}
    &\mathrm{k}_{0\cdots 3}^*=\langle\mathrm{T}_{0\cdots 3}|{f^{\mathrm{mhd}(0)}}\rangle \\
    &\mathrm{k}_{4\cdots 8}^*=\omega \langle\mathrm{T}_{4\cdots 8}|{f^{\mathrm{mhd}(0)}}\rangle + (1-\omega)\langle\mathrm{T}_{4\cdots 8}|{\bar f}\rangle\\
    &\mathrm{k}_{9\cdots 26}^*=\langle\mathrm{T}_{9\cdots 26}|{f^{\mathrm{mhd}(0)}}\rangle.
    \end{aligned}
\end{equation*}

\section{Calculation of the electric current density}\label{sec:appendix_2}
The electric current is obtained by solving the linear system~\eqref{eq:soe_current_density}. By using \eqref{eq:soe_objects}, this system can be re-expressed as
\begin{equation*}\label{eq:final_system}
     \widetilde{\mathbf{M}}\begin{bmatrix}
      J_x \\[0.3em]
      J_y \\[0.3em]
      J_z           
     \end{bmatrix}=
     -\frac{1}{2\alpha_H}
     \begin{bmatrix}
       {\mathrm{\Lambda}}_{yz}-{\mathrm{\Lambda}}_{zy} - 2\left(U_yB_z-U_zB_y\right) \\[0.3em]
       {\mathrm{\Lambda}}_{zx}-{\mathrm{\Lambda}}_{xz} - 2\left(U_zB_x-U_xB_z\right) \\[0.3em]
       {\mathrm{\Lambda}}_{xy}-{\mathrm{\Lambda}}_{yx} - 2\left(U_xB_y-U_yB_x\right)           
     \end{bmatrix}
\end{equation*}
where $ {\mathrm{\Lambda}}_{\alpha\beta}=\sum_{i=0}^{M-1}\xi_{i\alpha}\bar g_{i\beta}$  and $\widetilde{\mathbf{M}}$ is the invertible matrix
\begin{equation*}
    \widetilde{\mathbf{M}}=\begin{bmatrix}
       C^2/2\alpha_H\omega_B & B_z & -B_y \\[0.3em]
       -B_z & C^2/2\alpha_H\omega_B & B_x \\[0.3em]
       B_y & -B_x & C^2/2\alpha_H\omega_B           
     \end{bmatrix}
\end{equation*}

where $C$ represents the characteristic speed of \emph{magnetic} particles on the D3Q7 lattice
and $\omega_B$ is the relaxation pulsation~\eqref{eq:mag_relaxation_time} associated with the BGK collision operator for the magnetic field. The expression for the three components of the electric current density obtained by solving the previous linear system reads as 
\begin{equation*}\label{eq:current_density_explicit1}
  \begin{split}
    J_x={}& \frac{1}{D}(-2C^4\omega_BB_yU_z+2C^4\omega_BB_zU_y-C^4\omega_B\Lambda_{yz}+C^4\omega_B\Lambda_{zy}-4C^2\alpha_H\omega_B^2B_xB_yU_y-\\
    & -4C^2\alpha_H\omega_B^2B_xB_zU_z+4C^2\alpha_H\omega_B^2B_y^2U_x-2C^2\alpha_H\omega_B^2\Lambda_{xy}B_y+2C^2\alpha_H\omega_B^2\Lambda_{yx}B_y+\\
    &+4C^2\alpha_H\omega_B^2B_z^2U_x-2C^2\alpha_H\omega_B^2\Lambda_{xz}B_z+2C^2\alpha_H\omega_B^2\Lambda_{zx}B_z-4\alpha_H^2\omega_B^3\Lambda_{yz}B_x^2+\\
    &+4\alpha_H^2\omega_B^3\Lambda_{zy}B_x^2+4\alpha_H^2\omega_B^3\Lambda_{xz}B_xB_y-4\alpha_H^2\omega_B^3\Lambda_{zx}B_xB_y-4\alpha_H^2\omega_B^3\Lambda_{xy}B_xB_z+\\
    &+4\alpha_H^2\omega_B^3\Lambda_{yx}B_xB_z)
\end{split}
\end{equation*}
\begin{equation*}\label{eq:current_density_explicit2}
  \begin{split}
    J_y={}& \frac{1}{D}(2C^4\omega_BB_xU_z+2C^4\omega_BB_zU_x+C^4\omega_B\Lambda_{xz}-C^4\omega_B\Lambda_{zx}+4C^2\alpha_H\omega_B^2B_x^2U_y-\\
    & -4C^2\alpha_H\omega_B^2B_xB_yU_x+2C^2\alpha_H\omega_B^2\Lambda_{xy}B_x-2C^2\alpha_H\omega_B^2\Lambda_{yx}B_x-4C^2\alpha_H\omega_B^2B_yB_zU_z+\\
    &+4C^2\alpha_H\omega_B^2B_z^2U_y-2C^2\alpha_H\omega_B^2\Lambda_{yz}B_z+2C^2\alpha_H\omega_B^2\Lambda_{zy}B_z-4\alpha_H^2\omega_B^3\Lambda_{yz}B_xB_y+\\
    &+4\alpha_H^2\omega_B^3\Lambda_{zy}B_xB_y+4\alpha_H^2\omega_B^3\Lambda_{xz}B_y^2-4\alpha_H^2\omega_B^3\Lambda_{zx}B_y^2-4\alpha_H^2\omega_B^3\Lambda_{xy}B_yB_z+\\
    &+4\alpha_H^2\omega_B^3\Lambda_{yx}B_yB_z)
\end{split}
\end{equation*}
\begin{equation}\label{eq:current_density_explicit3}
  \begin{split}
    J_z={}& \frac{1}{D}(-2C^4\omega_BB_xU_y+2C^4\omega_BB_yU_x-C^4\omega_B\Lambda_{xy}+C^4\omega_B\Lambda_{yx}+4C^2\alpha_H\omega_B^2B_x^2U_z-\\
    & -4C^2\alpha_H\omega_B^2B_xB_zU_x+2C^2\alpha_H\omega_B^2\Lambda_{xz}B_x-2C^2\alpha_H\omega_B^2\Lambda_{zx}B_x+4C^2\alpha_H\omega_B^2B_y^2U_z-\\
    &-4C^2\alpha_H\omega_B^2B_yB_zU_y+2C^2\alpha_H\omega_B^2\Lambda_{yz}B_y-2C^2\alpha_H\omega_B^2\Lambda_{zy}B_y-4\alpha_H^2\omega_B^3\Lambda_{yz}B_xB_z+\\
    &+4\alpha_H^2\omega_B^3\Lambda_{zy}B_xB_y+4\alpha_H^2\omega_B^3\Lambda_{xz}B_yB_z-4\alpha_H^2\omega_B^3\Lambda_{zx}B_yB_z-4\alpha_H^2\omega_B^3\Lambda_{xy}B_z^2+\\
    &+4\alpha_H^2\omega_B^3\Lambda_{yx}B_z^2)
\end{split}
\end{equation}
where $D=C^6+4C^2\alpha_H^2\omega_B^2|\mathbf{B}|^2>0$. 

\bibliographystyle{jpp}
\bibliography{biblio}

\begin{thebibliography}{113}
\expandafter\ifx\csname natexlab\endcsname\relax\def\natexlab#1{#1}\fi
\def\au#1{#1} \def\ed#1{#1} \def\yr#1{#1}\def\at#1{#1}\def\jt#1{\textit{#1}}
  \def\bt#1{#1}\def\bvol#1{\textbf{#1}} \def\vol#1{#1} \def\pg#1{#1}
  \def\publ#1{#1}\def\arxiv#1{#1}\def\org#1{#1}\def\st#1{\textit{#1}}

\bibitem[Adhikari {\em et~al.\/}(2015{\natexlab{{\em a\/}}})Adhikari, Zank,
  Bruno, Telloni, Hunana, Dosch, Marino \& Hu]{Adhikari_2015}
{\sc \au{Adhikari, L.}, \au{Zank, G.~P.}, \au{Bruno, R.}, \au{Telloni, D.},
  \au{Hunana, P.}, \au{Dosch, A.}, \au{Marino, R.} \& \au{Hu, Q.}}
  \yr{2015{\natexlab{{\em a\/}}}}  \at{The transport of low-frequency
  turbulence in astrophysical flows. ii. solutions for the super-alfvÃ‰nic
  solar wind}.  \jt{The Astrophysical Journal}  \bvol{805}~(1),  \pg{63}.

\bibitem[Adhikari {\em et~al.\/}(2015{\natexlab{{\em b\/}}})Adhikari, Zank,
  Bruno, Telloni, Hunana, Dosch, Marino \& Hu]{adhikari_proceed}
{\sc \au{Adhikari, L}, \au{Zank, G~P}, \au{Bruno, R}, \au{Telloni, D},
  \au{Hunana, P}, \au{Dosch, A}, \au{Marino, R} \& \au{Hu, Q}}
  \yr{2015{\natexlab{{\em b\/}}}}  \at{The transport of low-frequency
  turbulence in the super-{A}lfvénic solar wind}.  \jt{Journal of Physics:
  Conference Series}  \bvol{642}~(1),  \pg{012001}.

\bibitem[Alexandrova {\em et~al.\/}(2008)Alexandrova, Lacombe \&
  Mangeney]{Alex2008}
{\sc \au{Alexandrova, O.}, \au{Lacombe, C.} \& \au{Mangeney, A.}} \yr{2008}
  \at{Spectra and anisotropy of magnetic fluctuations in the {E}arth's
  magnetosheath: {C}luster observations}.  \jt{Ann. Geophysicae}
  \bvol{26}~(11),  \pg{3585--3596}.

\bibitem[Alexandrova {\em et~al.\/}(2009)Alexandrova, Saur, Lacombe, Mangeney,
  Mitchell, Schwartz \& Robert]{Alex2009}
{\sc \au{Alexandrova, O.}, \au{Saur, J.}, \au{Lacombe, C.}, \au{Mangeney, A.},
  \au{Mitchell, J.}, \au{Schwartz, S.~J.} \& \au{Robert, P.}} \yr{2009}
  \at{Universality of solar-wind turbulent spectrum from {MHD} to electron
  scales}.  \jt{Phys. Rev. Lett.}  \bvol{103},  \pg{165003}.

\bibitem[{Andr{\'e}s} {\em et~al.\/}(2022){Andr{\'e}s}, {Sahraoui}, {Huang},
  {Hadid} \& {Galtier}]{Andres2022}
{\sc \au{{Andr{\'e}s}, N.}, \au{{Sahraoui}, F.}, \au{{Huang}, S.}, \au{{Hadid},
  L.~Z.} \& \au{{Galtier}, S.}} \yr{2022}  \at{{The incompressible energy
  cascade rate in anisotropic solar wind turbulence}}.  \jt{Astron. \&
  Astroph.}  \bvol{661},  \pg{A116},  \arxiv{arXiv: 2112.13748}.

\bibitem[{Bale} {\em et~al.\/}(2019){Bale}, {Badman}, {Bonnell}, {Bowen},
  {Burgess}, {Case}, {Cattell}, {Chandran}, {Chaston}, {Chen}, {Drake}, {de
  Wit}, {Eastwood}, {Ergun}, {Farrell}, {Fong}, {Goetz}, {Goldstein},
  {Goodrich}, {Harvey}, {Horbury}, {Howes}, {Kasper}, {Kellogg}, {Klimchuk},
  {Korreck}, {Krasnoselskikh}, {Krucker}, {Laker}, {Larson}, {MacDowall},
  {Maksimovic}, {Malaspina}, {Martinez-Oliveros}, {McComas}, {Meyer-Vernet},
  {Moncuquet}, {Mozer}, {Phan}, {Pulupa}, {Raouafi}, {Salem}, {Stansby},
  {Stevens}, {Szabo}, {Velli}, {Woolley} \& {Wygant}]{2019Natur.576..237B}
{\sc \au{{Bale}, S.~D.}, \au{{Badman}, S.~T.}, \au{{Bonnell}, J.~W.},
  \au{{Bowen}, T.~A.}, \au{{Burgess}, D.}, \au{{Case}, A.~W.}, \au{{Cattell},
  C.~A.}, \au{{Chandran}, B.~D.~G.}, \au{{Chaston}, C.~C.}, \au{{Chen},
  C.~H.~K.}, \au{{Drake}, J.~F.}, \au{{de Wit}, T.~Dudok}, \au{{Eastwood},
  J.~P.}, \au{{Ergun}, R.~E.}, \au{{Farrell}, W.~M.}, \au{{Fong}, C.},
  \au{{Goetz}, K.}, \au{{Goldstein}, M.}, \au{{Goodrich}, K.~A.}, \au{{Harvey},
  P.~R.}, \au{{Horbury}, T.~S.}, \au{{Howes}, G.~G.}, \au{{Kasper}, J.~C.},
  \au{{Kellogg}, P.~J.}, \au{{Klimchuk}, J.~A.}, \au{{Korreck}, K.~E.},
  \au{{Krasnoselskikh}, V.~V.}, \au{{Krucker}, S.}, \au{{Laker}, R.},
  \au{{Larson}, D.~E.}, \au{{MacDowall}, R.~J.}, \au{{Maksimovic}, M.},
  \au{{Malaspina}, D.~M.}, \au{{Martinez-Oliveros}, J.}, \au{{McComas}, D.~J.},
  \au{{Meyer-Vernet}, N.}, \au{{Moncuquet}, M.}, \au{{Mozer}, F.~S.},
  \au{{Phan}, T.~D.}, \au{{Pulupa}, M.}, \au{{Raouafi}, N.~E.}, \au{{Salem},
  C.}, \au{{Stansby}, D.}, \au{{Stevens}, M.}, \au{{Szabo}, A.}, \au{{Velli},
  M.}, \au{{Woolley}, T.} \& \au{{Wygant}, J.~R.}} \yr{2019}  \at{{Highly
  structured slow solar wind emerging from an equatorial coronal hole}}.
  \jt{Nat.}  \bvol{576}~(7786),  \pg{237--242}.

\bibitem[{Bale} {\em et~al.\/}(2016){Bale}, {Goetz}, {Harvey}, {Turin},
  {Bonnell}, {Dudok de Wit}, {Ergun}, {MacDowall}, {Pulupa}, {Andre}, {Bolton},
  {Bougeret}, {Bowen}, {Burgess}, {Cattell}, {Chandran}, {Chaston}, {Chen},
  {Choi}, {Connerney}, {Cranmer}, {Diaz-Aguado}, {Donakowski}, {Drake},
  {Farrell}, {Fergeau}, {Fermin}, {Fischer}, {Fox}, {Glaser}, {Goldstein},
  {Gordon}, {Hanson}, {Harris}, {Hayes}, {Hinze}, {Hollweg}, {Horbury},
  {Howard}, {Hoxie}, {Jannet}, {Karlsson}, {Kasper}, {Kellogg}, {Kien},
  {Klimchuk}, {Krasnoselskikh}, {Krucker}, {Lynch}, {Maksimovic}, {Malaspina},
  {Marker}, {Martin}, {Martinez-Oliveros}, {McCauley}, {McComas}, {McDonald},
  {Meyer-Vernet}, {Moncuquet}, {Monson}, {Mozer}, {Murphy}, {Odom},
  {Oliverson}, {Olson}, {Parker}, {Pankow}, {Phan}, {Quataert}, {Quinn},
  {Ruplin}, {Salem}, {Seitz}, {Sheppard}, {Siy}, {Stevens}, {Summers}, {Szabo},
  {Timofeeva}, {Vaivads}, {Velli}, {Yehle}, {Werthimer} \&
  {Wygant}]{2016SSRv..204...49B}
{\sc \au{{Bale}, S.~D.}, \au{{Goetz}, K.}, \au{{Harvey}, P.~R.}, \au{{Turin},
  P.}, \au{{Bonnell}, J.~W.}, \au{{Dudok de Wit}, T.}, \au{{Ergun}, R.~E.},
  \au{{MacDowall}, R.~J.}, \au{{Pulupa}, M.}, \au{{Andre}, M.}, \au{{Bolton},
  M.}, \au{{Bougeret}, J.~L.}, \au{{Bowen}, T.~A.}, \au{{Burgess}, D.},
  \au{{Cattell}, C.~A.}, \au{{Chandran}, B.~D.~G.}, \au{{Chaston}, C.~C.},
  \au{{Chen}, C.~H.~K.}, \au{{Choi}, M.~K.}, \au{{Connerney}, J.~E.},
  \au{{Cranmer}, S.}, \au{{Diaz-Aguado}, M.}, \au{{Donakowski}, W.},
  \au{{Drake}, J.~F.}, \au{{Farrell}, W.~M.}, \au{{Fergeau}, P.}, \au{{Fermin},
  J.}, \au{{Fischer}, J.}, \au{{Fox}, N.}, \au{{Glaser}, D.}, \au{{Goldstein},
  M.}, \au{{Gordon}, D.}, \au{{Hanson}, E.}, \au{{Harris}, S.~E.}, \au{{Hayes},
  L.~M.}, \au{{Hinze}, J.~J.}, \au{{Hollweg}, J.~V.}, \au{{Horbury}, T.~S.},
  \au{{Howard}, R.~A.}, \au{{Hoxie}, V.}, \au{{Jannet}, G.}, \au{{Karlsson},
  M.}, \au{{Kasper}, J.~C.}, \au{{Kellogg}, P.~J.}, \au{{Kien}, M.},
  \au{{Klimchuk}, J.~A.}, \au{{Krasnoselskikh}, V.~V.}, \au{{Krucker}, S.},
  \au{{Lynch}, J.~J.}, \au{{Maksimovic}, M.}, \au{{Malaspina}, D.~M.},
  \au{{Marker}, S.}, \au{{Martin}, P.}, \au{{Martinez-Oliveros}, J.},
  \au{{McCauley}, J.}, \au{{McComas}, D.~J.}, \au{{McDonald}, T.},
  \au{{Meyer-Vernet}, N.}, \au{{Moncuquet}, M.}, \au{{Monson}, S.~J.},
  \au{{Mozer}, F.~S.}, \au{{Murphy}, S.~D.}, \au{{Odom}, J.}, \au{{Oliverson},
  R.}, \au{{Olson}, J.}, \au{{Parker}, E.~N.}, \au{{Pankow}, D.}, \au{{Phan},
  T.}, \au{{Quataert}, E.}, \au{{Quinn}, T.}, \au{{Ruplin}, S.~W.},
  \au{{Salem}, C.}, \au{{Seitz}, D.}, \au{{Sheppard}, D.~A.}, \au{{Siy}, A.},
  \au{{Stevens}, K.}, \au{{Summers}, D.}, \au{{Szabo}, A.}, \au{{Timofeeva},
  M.}, \au{{Vaivads}, A.}, \au{{Velli}, M.}, \au{{Yehle}, A.}, \au{{Werthimer},
  D.} \& \au{{Wygant}, J.~R.}} \yr{2016}  \at{{The {FIELDS} instrument suite
  for {S}olar {P}robe {P}lus. Measuring the coronal plasma and magnetic field,
  plasma waves and turbulence, and radio signatures of solar transients}}.
  \jt{Space Sci. Rev.}  \bvol{204}~(1-4),  \pg{49--82}.

\bibitem[Benzi {\em et~al.\/}(1992)Benzi, Succi \& Vergassola]{BENZI1992}
{\sc \au{Benzi, R.}, \au{Succi, S.} \& \au{Vergassola, M.}} \yr{1992}  \at{The
  lattice {B}oltzmann equation: {T}heory and applications}.  \jt{Phys. Rep.}
  \bvol{222}~(3),  \pg{145--197}.

\bibitem[Bhatnagar {\em et~al.\/}(1954)Bhatnagar, Gross \& Krook]{BGK54}
{\sc \au{Bhatnagar, P.~L.}, \au{Gross, E.~P.} \& \au{Krook, M.}} \yr{1954}
  \at{A model for collision processes in gases. {I}. {S}mall amplitude
  processes in charged and neutral one-component systems}.  \jt{Phys. Rev.}
  \bvol{94},  \pg{511--525}.

\bibitem[Bouchut(1999)]{Bouchut99}
{\sc \au{Bouchut, François}} \yr{1999}  \at{Construction of {BGK} models with
  a family of kinetic entropies for a given system of conservation laws}.
  \jt{J. Stat. Phys.}  \bvol{95},  \pg{113--170}.

\bibitem[Breyiannis \& Valougeorgis(2004)]{Breyiannis2004}
{\sc \au{Breyiannis, G.} \& \au{Valougeorgis, D.}} \yr{2004}  \at{Lattice
  kinetic simulations in three-dimensional magnetohydrodynamics}.  \jt{Phys.
  Rev. E}  \bvol{69},  \pg{065702}.

\bibitem[Breyiannis \& Valougeorgis(2006)]{Breyiannis2006}
{\sc \au{Breyiannis, George} \& \au{Valougeorgis, Dimitris}} \yr{2006}
  \at{Lattice kinetic simulations of {3-D} {MHD} turbulence}.  \jt{Comp. \&
  Fl.}  \bvol{35},  \pg{920--924}.

\bibitem[Brodiano {\em et~al.\/}(2023)Brodiano, Dmitruk \&
  Andrés]{Brodiano2023}
{\sc \au{Brodiano, M.}, \au{Dmitruk, P.} \& \au{Andrés, N.}} \yr{2023}  \at{A
  statistical study of the compressible energy cascade rate in solar wind
  turbulence: Parker solar probe observations}.  \jt{Physics of Plasmas}
  \bvol{30}~(3),  \pg{032903}.

\bibitem[{Bruno} \& {Carbone}(2013)]{2013LRSP...10....2B}
{\sc \au{{Bruno}, Roberto} \& \au{{Carbone}, Vincenzo}} \yr{2013}  \at{{The
  solar wind as a turbulence laboratory}}.  \jt{Living Rev. Sol. Phys.}
  \bvol{10}~(1),  \pg{2}.

\bibitem[{Bruno} \& {Trenchi}(2014)]{2014ApJ...787L..24B}
{\sc \au{{Bruno}, R.} \& \au{{Trenchi}, L.}} \yr{2014}  \at{{Radial dependence
  of the frequency break between fluid and kinetic scales in the solar wind
  fluctuations}}.  \jt{Astroph. J. Lett.}  \bvol{787}~(2),  \pg{L24}.

\bibitem[{Bruno} {\em et~al.\/}(2014){Bruno}, {Trenchi} \&
  {Telloni}]{2014ApJ...793L..15B}
{\sc \au{{Bruno}, R.}, \au{{Trenchi}, L.} \& \au{{Telloni}, D.}} \yr{2014}
  \at{{Spectral slope variation at proton scales from fast to slow solar
  wind}}.  \jt{Astroph. J. Lett.}  \bvol{793}~(1),  \pg{L15}.

\bibitem[Cerri {\em et~al.\/}(2016)Cerri, Califano, Jenko, Told \&
  Rincon]{Cerri_2016}
{\sc \au{Cerri, S.~S.}, \au{Califano, F.}, \au{Jenko, F.}, \au{Told, D.} \&
  \au{Rincon, F.}} \yr{2016}  \at{Subproton-scale cascades in solar wind
  turbulence: Driven hybrid-kinetic simulations}.  \jt{The Astrophysical
  Journal Letters}  \bvol{822}~(1),  \pg{L12}.

\bibitem[Chen {\em et~al.\/}(1991)Chen, Chen, Martnez \&
  Matthaeus]{ChenMatthaeus1991}
{\sc \au{Chen, Shiyi}, \au{Chen, Hudong}, \au{Martnez, Daniel} \&
  \au{Matthaeus, William}} \yr{1991}  \at{Lattice {B}oltzmann model for
  simulation of magnetohydrodynamics}.  \jt{Phys. Rev. Lett.}  \bvol{67},
  \pg{3776--3779}.

\bibitem[Coreixas {\em et~al.\/}(2019)Coreixas, Chopard \&
  Latt]{COREIXAS_PRE_100_2019}
{\sc \au{Coreixas, Christophe}, \au{Chopard, Bastien} \& \au{Latt, Jonas}}
  \yr{2019}  \at{Comprehensive comparison of collision models in the lattice
  {B}oltzmann framework: {T}heoretical investigations}.  \jt{Phys. Rev. E}
  \bvol{100},  \pg{033305}.

\bibitem[Coreixas {\em et~al.\/}(2017)Coreixas, Wissocq, Puigt, Boussuge \&
  Sagaut]{coreixas2017recursive}
{\sc \au{Coreixas, Christophe}, \au{Wissocq, Gauthier}, \au{Puigt, Guillaume},
  \au{Boussuge, Jean-Fran{\c{c}}ois} \& \au{Sagaut, Pierre}} \yr{2017}
  \at{Recursive regularization step for high-order lattice {B}oltzmann
  methods}.  \jt{Phys. Rev. E}  \bvol{96}~(3),  \pg{033306}.

\bibitem[Croisille {\em et~al.\/}(1995)Croisille, Khanfir \&
  Chanteu]{Croisille95}
{\sc \au{Croisille, JP.}, \au{Khanfir, R.} \& \au{Chanteu, G.}} \yr{1995}
  \at{Numerical simulation of the {MHD} equations by a kinetic-type method}.
  \jt{J. Sci. Comput.}  \bvol{10},  \pg{81--92}.

\bibitem[{D'Amicis} {\em et~al.\/}(2021){D'Amicis}, {Bruno}, {Panasenco},
  {Telloni}, {Perrone}, {Marcucci}, {Woodham}, {Velli}, {De Marco},
  {Jagarlamudi}, {Coco}, {Owen}, {Louarn}, {Livi}, {Horbury}, {Andr{\'e}},
  {Angelini}, {Evans}, {Fedorov}, {Genot}, {Lavraud}, {Matteini}, {M{\"u}ller},
  {O'Brien}, {Pezzi}, {Rouillard}, {Sorriso-Valvo}, {Tenerani}, {Verscharen} \&
  {Zouganelis}]{2021A&A...656A..21D}
{\sc \au{{D'Amicis}, R.}, \au{{Bruno}, R.}, \au{{Panasenco}, O.},
  \au{{Telloni}, D.}, \au{{Perrone}, D.}, \au{{Marcucci}, M.~F.},
  \au{{Woodham}, L.}, \au{{Velli}, M.}, \au{{De Marco}, R.}, \au{{Jagarlamudi},
  V.}, \au{{Coco}, I.}, \au{{Owen}, C.}, \au{{Louarn}, P.}, \au{{Livi}, S.},
  \au{{Horbury}, T.}, \au{{Andr{\'e}}, N.}, \au{{Angelini}, V.}, \au{{Evans},
  V.}, \au{{Fedorov}, A.}, \au{{Genot}, V.}, \au{{Lavraud}, B.},
  \au{{Matteini}, L.}, \au{{M{\"u}ller}, D.}, \au{{O'Brien}, H.}, \au{{Pezzi},
  O.}, \au{{Rouillard}, A.~P.}, \au{{Sorriso-Valvo}, L.}, \au{{Tenerani}, A.},
  \au{{Verscharen}, D.} \& \au{{Zouganelis}, I.}} \yr{2021}  \at{{First {S}olar
  {O}rbiter observation of the {A}lfv{\'e}nic slow wind and identification of
  its solar source}}.  \jt{Astron. \& Astroph.}  \bvol{656},  \pg{A21}.

\bibitem[De~Rosis(2017)]{CM3D-derosis-2017}
{\sc \au{De~Rosis, Alessandro}} \yr{2017}  \at{Nonorthogonal
  central-moments-based lattice {B}oltzmann scheme in three dimensions}.
  \jt{Phys. Rev. E}  \bvol{95},  \pg{013310}.

\bibitem[De~Rosis \& Luo(2019)]{derosisHermite}
{\sc \au{De~Rosis, Alessandro} \& \au{Luo, Kai~H.}} \yr{2019}  \at{Role of
  higher-order {H}ermite polynomials in the central-moments-based lattice
  {B}oltzmann framework}.  \jt{Phys. Rev. E}  \bvol{99}~(1),  \pg{013301}.

\bibitem[{De Rosis} {\em et~al.\/}(2018){De Rosis}, Lévêque \&
  Chahine]{DeRosis18}
{\sc \au{{De Rosis}, Alessandro}, \au{Lévêque, Emmanuel} \& \au{Chahine,
  Robert}} \yr{2018}  \at{Advanced lattice {B}oltzmann scheme for
  high-{R}eynolds-number magneto-hydrodynamic flows}.  \jt{J. of Turb.}
  \bvol{19}~(6),  \pg{446--462}.

\bibitem[Dellar(2002)]{DELLAR2002}
{\sc \au{Dellar, Paul~J.}} \yr{2002}  \at{Lattice kinetic schemes for
  magnetohydrodynamics}.  \jt{J. of Comp. Phys.}  \bvol{179}~(1),
  \pg{95--126}.

\bibitem[Dellar(2009)]{Dellar2009}
{\sc \au{Dellar, P~J}} \yr{2009}  \at{Moment equations for
  magnetohydrodynamics}.  \jt{J. of Stat. Mech.: Th. and Exp.}
  \bvol{2009}~(06),  \pg{P06003}.

\bibitem[Dellar(2011)]{DELLAR2011}
{\sc \au{Dellar, Paul~J.}} \yr{2011}  \at{Lattice boltzmann formulation for
  braginskii magnetohydrodynamics}.  \jt{Comp. \& Fl.}  \bvol{46}~(1),
  \pg{201--205}, 10th ICFD Conference Series on Numerical Methods for Fluid
  Dynamics (ICFD 2010).

\bibitem[Dellar(2013)]{DELLAR2013115}
{\sc \au{Dellar, Paul~J.}} \yr{2013}  \at{Lattice {B}oltzmann
  magnetohydrodynamics with current-dependent resistivity}.  \jt{J. of Comp.
  Phys.}  \bvol{237},  \pg{115--131}.

\bibitem[Dudson {\em et~al.\/}(2015)Dudson, Allen, Breyiannis, Brugger,
  Buchanan, Easy, Farley, Joseph, Kim, McGann \& et~al.]{bout++}
{\sc \au{Dudson, B.~D.}, \au{Allen, A.}, \au{Breyiannis, G.}, \au{Brugger, E.},
  \au{Buchanan, J.}, \au{Easy, L.}, \au{Farley, S.}, \au{Joseph, I.}, \au{Kim,
  M.}, \au{McGann, A.~D.} \& \au{et~al.}} \yr{2015}  \at{Bout : Recent and
  current developments}.  \jt{Journal of Plasma Physics}  \bvol{81}~(1),
  \pg{365810104}.

\bibitem[d’Humieres(1994)]{Humiere1994}
{\sc \au{d’Humieres, D}} \yr{1994}  \at{Generalized lattice {B}oltzmann
  equations}.  \jt{Prog. Aeronaut. Astronaut.}  \bvol{159},  \pg{450--458}.

\bibitem[Feraco {\em et~al.\/}(2018)Feraco, Marino, Pumir, Primavera, Mininni,
  Pouquet \& Rosenberg]{Feraco2018}
{\sc \au{Feraco, Fabio}, \au{Marino, Raffaele}, \au{Pumir, Alain},
  \au{Primavera, Leonardo}, \au{Mininni, Pablo}, \au{Pouquet, Annick} \&
  \au{Rosenberg, Duane}} \yr{2018}  \at{Vertical drafts and mixing in
  stratified turbulence: {S}harp transition with {F}roude number}.
  \jt{Europhys. Lett.}  \bvol{123},  \pg{44002}.

\bibitem[Ferrand {\em et~al.\/}(2022)Ferrand, Sahraoui, Galtier, Andr{\'{e}}s,
  Mininni \& Dmitruk]{Ferrand2022}
{\sc \au{Ferrand, R.}, \au{Sahraoui, F.}, \au{Galtier, S.}, \au{Andr{\'{e}}s,
  N.}, \au{Mininni, P.} \& \au{Dmitruk, P.}} \yr{2022}  \at{An in-depth
  numerical study of exact laws for compressible {H}all magnetohydrodynamic
  turbulence}.  \jt{Astroph. J.}  \bvol{927}~(2),  \pg{205}.

\bibitem[Flint \& Vahala(2018)]{Vahala2018}
{\sc \au{Flint, Christopher} \& \au{Vahala, George}} \yr{2018}  \at{A partial
  entropic lattice {B}oltzmann {MHD} simulation of the {O}rszag–{T}ang
  vortex}.  \jt{Rad. Eff. and Def. in Sol.}  \bvol{173}~(1-2),  \pg{55--65}.

\bibitem[{Fox} {\em et~al.\/}(2016){Fox}, {Velli}, {Bale}, {Decker},
  {Driesman}, {Howard}, {Kasper}, {Kinnison}, {Kusterer}, {Lario}, {Lockwood},
  {McComas}, {Raouafi} \& {Szabo}]{2016SSRv..204....7F}
{\sc \au{{Fox}, N.~J.}, \au{{Velli}, M.~C.}, \au{{Bale}, S.~D.}, \au{{Decker},
  R.}, \au{{Driesman}, A.}, \au{{Howard}, R.~A.}, \au{{Kasper}, J.~C.},
  \au{{Kinnison}, J.}, \au{{Kusterer}, M.}, \au{{Lario}, D.}, \au{{Lockwood},
  M.~K.}, \au{{McComas}, D.~J.}, \au{{Raouafi}, N.~E.} \& \au{{Szabo}, A.}}
  \yr{2016}  \at{{The {S}olar {P}robe {P}lus {M}ission: {H}umanity's first
  visit to our star}}.  \jt{Space Sci. Rev.}  \bvol{204}~(1-4),  \pg{7--48}.

\bibitem[Galtier(2016)]{galtier2016}
{\sc \au{Galtier, S\'ebastien}} \yr{2016} {\em Introduction to modern
  magnetohydrodynamics\/}.  \publ{Cambridge University Press}.

\bibitem[Galtier \& Buchlin(2007)]{Galtier2007}
{\sc \au{Galtier, Sebastien} \& \au{Buchlin, Eric}} \yr{2007}  \at{Multiscale
  {H}all-magnetohydrodynamic turbulence in the solar wind}.  \jt{Astroph. J.}
  \bvol{656}~(1),  \pg{560--566}.

\bibitem[Geier {\em et~al.\/}(2007)Geier, Greiner \&
  Korvink]{geier2007properties}
{\sc \au{Geier, M.}, \au{Greiner, A.} \& \au{Korvink, J.G.}} \yr{2007}
  \at{Properties of the cascaded lattice {B}oltzmann automaton}.  \jt{Int. J.
  Mod. Phys. C}  \bvol{18}~(04),  \pg{455--462}.

\bibitem[Geier {\em et~al.\/}(2006)Geier, Greiner \& Korvink]{Geier2006}
{\sc \au{Geier, Martin}, \au{Greiner, Andreas} \& \au{Korvink, Jan~G.}}
  \yr{2006}  \at{Cascaded digital lattice {B}oltzmann automata for high
  {R}eynolds number flow}.  \jt{Phys. Rev. E}  \bvol{73},  \pg{066705}.

\bibitem[Geier {\em et~al.\/}(2015)Geier, Schönherr, Pasquali \&
  Krafczyk]{GEIERcumulant}
{\sc \au{Geier, Martin}, \au{Schönherr, Martin}, \au{Pasquali, Andrea} \&
  \au{Krafczyk, Manfred}} \yr{2015}  \at{The cumulant lattice {B}oltzmann
  equation in three dimensions: {T}heory and validation}.  \jt{Comp. \& Math.
  with App.}  \bvol{70}~(4),  \pg{507--547}.

\bibitem[G\'omez {\em et~al.\/}(2010)G\'omez, Mininni \& Dmitruk]{Gomez2010}
{\sc \au{G\'omez, Daniel~O.}, \au{Mininni, Pablo~D.} \& \au{Dmitruk, Pablo}}
  \yr{2010}  \at{{H}all-magnetohydrodynamic small-scale dynamos}.  \jt{Phys.
  Rev. E}  \bvol{82},  \pg{036406}.

\bibitem[Gonz{\'{a}}lez-Morales {\em et~al.\/}(2019)Gonz{\'{a}}lez-Morales,
  Khomenko \& Cally]{Gonzalez2019}
{\sc \au{Gonz{\'{a}}lez-Morales, P.~A.}, \au{Khomenko, E.} \& \au{Cally,
  P.~S.}} \yr{2019}  \at{Fast-to-{A}lfv{\'{e}}n mode conversion mediated by
  {H}all current. {II}. application to the solar atmosphere}.  \jt{Astroph. J.}
   \bvol{870}~(2),  \pg{94}.

\bibitem[He \& Luo(1997)]{He-Luo1997}
{\sc \au{He, Xiaoyi} \& \au{Luo, Li-Shi}} \yr{1997}  \at{Theory of the lattice
  {B}oltzmann method: {F}rom the {B}oltzmann equation to the lattice
  {B}oltzmann equation}.  \jt{Phys. Rev. E}  \bvol{56},  \pg{6811--6817}.

\bibitem[He {\em et~al.\/}(1998)He, Shan \& Doolen]{He1998}
{\sc \au{He, Xiaoyi}, \au{Shan, Xiaowen} \& \au{Doolen, Gary~D.}} \yr{1998}
  \at{Discrete {B}oltzmann equation model for nonideal gases}.  \jt{Phys. Rev.
  E}  \bvol{57},  \pg{R13--R16}.

\bibitem[Herbert {\em et~al.\/}(2016)Herbert, Marino, Rosenberg \&
  Pouquet]{herbert16}
{\sc \au{Herbert, C.}, \au{Marino, R.}, \au{Rosenberg, D.} \& \au{Pouquet, A.}}
  \yr{2016}  \at{Waves and vortices in the inverse cascade regime of stratified
  turbulence with or without rotation}.  \jt{J. of Fl. Mech.}  \bvol{806},
  \pg{165--204}.

\bibitem[Higuera {\em et~al.\/}(1989)Higuera, Succi \& Benzi]{Higuera_1989}
{\sc \au{Higuera, F.~J.}, \au{Succi, S.} \& \au{Benzi, R.}} \yr{1989}
  \at{Lattice gas dynamics with enhanced collisions}.  \jt{Europhysics Letters}
   \bvol{9}~(4),  \pg{345}.

\bibitem[Hoelzl {\em et~al.\/}(2021)Hoelzl, Huijsmans, Pamela, Bécoulet,
  Nardon, Artola, Nkonga, Atanasiu, Bandaru, Bhole, Bonfiglio, Cathey, Czarny,
  Dvornova, Fehér, Fil, Franck, Futatani, Gruca, Guillard, Haverkort, Holod,
  Hu, Kim, Korving, Kos, Krebs, Kripner, Latu, Liu, Merkel, Meshcheriakov,
  Mitterauer, Mochalskyy, Morales, Nies, Nikulsin, Orain, Pratt, Ramasamy,
  Ramet, Reux, Särkimäki, Schwarz, Verma, Smith, Sommariva, Strumberger, van
  Vugt, Verbeek, Westerhof, Wieschollek \& Zielinski]{JOREK}
{\sc \au{Hoelzl, M.}, \au{Huijsmans, G.T.A.}, \au{Pamela, S.J.P.},
  \au{Bécoulet, M.}, \au{Nardon, E.}, \au{Artola, F.J.}, \au{Nkonga, B.},
  \au{Atanasiu, C.V.}, \au{Bandaru, V.}, \au{Bhole, A.}, \au{Bonfiglio, D.},
  \au{Cathey, A.}, \au{Czarny, O.}, \au{Dvornova, A.}, \au{Fehér, T.},
  \au{Fil, A.}, \au{Franck, E.}, \au{Futatani, S.}, \au{Gruca, M.},
  \au{Guillard, H.}, \au{Haverkort, J.W.}, \au{Holod, I.}, \au{Hu, D.},
  \au{Kim, S.K.}, \au{Korving, S.Q.}, \au{Kos, L.}, \au{Krebs, I.},
  \au{Kripner, L.}, \au{Latu, G.}, \au{Liu, F.}, \au{Merkel, P.},
  \au{Meshcheriakov, D.}, \au{Mitterauer, V.}, \au{Mochalskyy, S.},
  \au{Morales, J.A.}, \au{Nies, R.}, \au{Nikulsin, N.}, \au{Orain, F.},
  \au{Pratt, J.}, \au{Ramasamy, R.}, \au{Ramet, P.}, \au{Reux, C.},
  \au{Särkimäki, K.}, \au{Schwarz, N.}, \au{Verma, P.~Singh}, \au{Smith,
  S.F.}, \au{Sommariva, C.}, \au{Strumberger, E.}, \au{van Vugt, D.C.},
  \au{Verbeek, M.}, \au{Westerhof, E.}, \au{Wieschollek, F.} \& \au{Zielinski,
  J.}} \yr{2021}  \at{The jorek non-linear extended mhd code and applications
  to large-scale instabilities and their control in magnetically confined
  fusion plasmas}.  \jt{Nuclear Fusion}  \bvol{61}~(6),  \pg{065001}.

\bibitem[{Horbury} {\em et~al.\/}(2020){Horbury}, {O'Brien}, {Carrasco
  Blazquez}, {Bendyk}, {Brown}, {Hudson}, {Evans}, {Oddy}, {Carr}, {Beek},
  {Cupido}, {Bhattacharya}, {Dominguez}, {Matthews}, {Myklebust}, {Whiteside},
  {Bale}, {Baumjohann}, {Burgess}, {Carbone}, {Cargill}, {Eastwood},
  {Erd{\"o}s}, {Fletcher}, {Forsyth}, {Giacalone}, {Glassmeier}, {Goldstein},
  {Hoeksema}, {Lockwood}, {Magnes}, {Maksimovic}, {Marsch}, {Matthaeus},
  {Murphy}, {Nakariakov}, {Owen}, {Owens}, {Rodriguez-Pacheco}, {Richter},
  {Riley}, {Russell}, {Schwartz}, {Vainio}, {Velli}, {Vennerstrom}, {Walsh},
  {Wimmer-Schweingruber}, {Zank}, {M{\"u}ller}, {Zouganelis} \&
  {Walsh}]{2020A&A...642A...9H}
{\sc \au{{Horbury}, T.~S.}, \au{{O'Brien}, H.}, \au{{Carrasco Blazquez}, I.},
  \au{{Bendyk}, M.}, \au{{Brown}, P.}, \au{{Hudson}, R.}, \au{{Evans}, V.},
  \au{{Oddy}, T.~M.}, \au{{Carr}, C.~M.}, \au{{Beek}, T.~J.}, \au{{Cupido},
  E.}, \au{{Bhattacharya}, S.}, \au{{Dominguez}, J.~A.}, \au{{Matthews}, L.},
  \au{{Myklebust}, V.~R.}, \au{{Whiteside}, B.}, \au{{Bale}, S.~D.},
  \au{{Baumjohann}, W.}, \au{{Burgess}, D.}, \au{{Carbone}, V.}, \au{{Cargill},
  P.}, \au{{Eastwood}, J.}, \au{{Erd{\"o}s}, G.}, \au{{Fletcher}, L.},
  \au{{Forsyth}, R.}, \au{{Giacalone}, J.}, \au{{Glassmeier}, K.~H.},
  \au{{Goldstein}, M.~L.}, \au{{Hoeksema}, T.}, \au{{Lockwood}, M.},
  \au{{Magnes}, W.}, \au{{Maksimovic}, M.}, \au{{Marsch}, E.}, \au{{Matthaeus},
  W.~H.}, \au{{Murphy}, N.}, \au{{Nakariakov}, V.~M.}, \au{{Owen}, C.~J.},
  \au{{Owens}, M.}, \au{{Rodriguez-Pacheco}, J.}, \au{{Richter}, I.},
  \au{{Riley}, P.}, \au{{Russell}, C.~T.}, \au{{Schwartz}, S.}, \au{{Vainio},
  R.}, \au{{Velli}, M.}, \au{{Vennerstrom}, S.}, \au{{Walsh}, R.},
  \au{{Wimmer-Schweingruber}, R.~F.}, \au{{Zank}, G.}, \au{{M{\"u}ller}, D.},
  \au{{Zouganelis}, I.} \& \au{{Walsh}, A.~P.}} \yr{2020}  \at{{The {S}olar
  {O}rbiter magnetometer}}.  \jt{Astron. \& Astroph.}  \bvol{642},  \pg{A9}.

\bibitem[Horstmann {\em et~al.\/}(2022)Horstmann, Touil, Vienne, Ricot \&
  Lévêque]{Horstmann2022}
{\sc \au{Horstmann, Jan}, \au{Touil, Hatem}, \au{Vienne, Lucien}, \au{Ricot,
  Denis} \& \au{Lévêque, Emmanuel}} \yr{2022}  \at{Consistent time-step
  optimization in the lattice {B}oltzmann method}.  \jt{J. of Comp. Phys.}
  \bvol{462},  \pg{111224}.

\bibitem[{Huba}(2003)]{Huba2003}
{\sc \au{{Huba}, J.~D.}} \yr{2003}  \at{{{H}all Magnetohydrodynamics - A
  tutorial}}.  \bt{In {\em Space plasma simulation\/}}, ,  \vol{vol. 615},
  \pg{pp. 166--192}.  \publ{{B{\"u}chner}, J. and {Dum}, C. and {Scholer}, M.}

\bibitem[Hénon(1987)]{Henon1987}
{\sc \au{Hénon, Michel}} \yr{1987}  \at{Viscosity of a lattice gas}.
  \jt{Compl. Sys.}  \bvol{462}.

\bibitem[{Iroshnikov}(1963)]{1963AZh....40..742I}
{\sc \au{{Iroshnikov}, P.~S.}} \yr{1963}  \at{{Turbulence of a conducting fluid
  in a strong magnetic field}}.  \jt{Astron. Zh.}  \bvol{40},  \pg{742}.

\bibitem[Kiyani {\em et~al.\/}(2015)Kiyani, Osman \& Chapman]{Kiyani2015}
{\sc \au{Kiyani, Khurom~H.}, \au{Osman, Kareem~T.} \& \au{Chapman, Sandra~C.}}
  \yr{2015}  \at{Dissipation and heating in solar wind turbulence: {F}rom the
  macro to the micro and back again}.  \jt{Philos. Trans. of the R. Soc. A:
  Math., Phys. and Eng. Sc.}  \bvol{373}~(2041),  \pg{20140155}.

\bibitem[{Kolmogorov}(1941)]{1941DoSSR..30..301K}
{\sc \au{{Kolmogorov}, A.}} \yr{1941}  \at{{The local structure of turbulence
  in incompressible viscous fluid for very large {R}eynolds' numbers}}.
  \jt{Dokl. Akad. Nauk SSSR}  \bvol{30},  \pg{301--305}.

\bibitem[K{\"o}rner {\em et~al.\/}(2006)K{\"o}rner, Pohl, R{\"u}de, Th{\"u}rey
  \& Zeiser]{Korner2006}
{\sc \au{K{\"o}rner, Carolin}, \au{Pohl, Thomas}, \au{R{\"u}de, Ulrich},
  \au{Th{\"u}rey, Nils} \& \au{Zeiser, Thomas}} \yr{2006} Parallel lattice
  {B}oltzmann methods for {CFD} applications.  \bt{In {\em Numerical solution
  of partial differential equations on parallel computers\/} (ed.
  \ed{Are~Magnus Bruaset \& Aslak Tveito})},  \pg{pp. 439--466}.  \publ{Berlin,
  Heidelberg: Springer Berlin Heidelberg}.

\bibitem[{Kraichnan}(1965)]{1965PhFl....8.1385K}
{\sc \au{{Kraichnan}, Robert~H.}} \yr{1965}  \at{{Inertial-range spectrum of
  hydromagnetic turbulence}}.  \jt{Phys. of Fl.}  \bvol{8}~(7),
  \pg{1385--1387}.

\bibitem[Krueger {\em et~al.\/}(2016)Krueger, Kusumaatmaja, Kuzmin, Shardt,
  Silva \& Viggen]{Kruger2016}
{\sc \au{Krueger, Timm}, \au{Kusumaatmaja, Halim}, \au{Kuzmin, Alexandr},
  \au{Shardt, Orest}, \au{Silva, Goncalo} \& \au{Viggen, Erlend~Magnus}}
  \yr{2016} {\em The lattice {B}oltzmann method: {P}rinciples and practice\/}.
  \publ{Springer}.

\bibitem[Kulsrud(2005)]{kulsrud}
{\sc \au{Kulsrud, Russell~M.}} \yr{2005} {\em Plasma Physics for
  Astrophysics\/}.  \publ{Princeton University Press}.

\bibitem[Lewy {\em et~al.\/}(1928)Lewy, Friedrichs \& Courant]{lewy1928}
{\sc \au{Lewy, H}, \au{Friedrichs, K} \& \au{Courant, R}} \yr{1928}
  \at{{\"U}ber die partiellen {D}ifferenzengleichungen der mathematischen
  {P}hysik}.  \jt{Mathematische annalen}  \bvol{100},  \pg{32--74}.

\bibitem[Ma {\em et~al.\/}(2018)Ma, Russell, Toth, Chen, Nagy, Harada,
  McFadden, Halekas, Lillis, Connerney, Espley, DiBraccio, Markidis, Peng, Fang
  \& Jakosky]{Ma2018}
{\sc \au{Ma, Yingjuan}, \au{Russell, Christopher~T.}, \au{Toth, Gabor},
  \au{Chen, Yuxi}, \au{Nagy, Andrew~F.}, \au{Harada, Yuki}, \au{McFadden,
  James}, \au{Halekas, Jasper~S.}, \au{Lillis, Rob}, \au{Connerney, John
  E.~P.}, \au{Espley, Jared}, \au{DiBraccio, Gina~A.}, \au{Markidis, Stefano},
  \au{Peng, Ivy~Bo}, \au{Fang, Xiaohua} \& \au{Jakosky, Bruce~M.}} \yr{2018}
  \at{Reconnection in the martian magnetotail: {H}all-{MHD} with embedded
  {P}article-in-{C}ell simulations}.  \jt{J. of Geoph. Res.: Sp. Phys.}
  \bvol{123}~(5),  \pg{3742--3763}.

\bibitem[Mahajan \& Krishan(2005)]{Mahajan05}
{\sc \au{Mahajan, S.~M.} \& \au{Krishan, V.}} \yr{2005}  \at{{Exact solution of
  the incompressible {H}all magnetohydrodynamics}}.  \jt{Mon. Not. of the R.
  Astron. Soc.: Lett.}  \bvol{359}~(1),  \pg{L27--L29}.

\bibitem[{Malara} \& {Velli}(1996)]{1996PhPl....3.4427M}
{\sc \au{{Malara}, F.} \& \au{{Velli}, M.}} \yr{1996}  \at{{Parametric
  instability of a large-amplitude nonmonochromatic {A}lfv{\'e}n wave}}.
  \jt{Phys. of Pl.}  \bvol{3}~(12),  \pg{4427--4433}.

\bibitem[Malaspinas(2015)]{malaspinas2015increasing}
{\sc \au{Malaspinas, Orestis}} \yr{2015}  \at{Increasing stability and accuracy
  of the lattice {B}oltzmann scheme: {R}ecursivity and regularization}.
  \jt{arXiv:1505.06900 preprint} .

\bibitem[{Marchand, P.} {\em et~al.\/}(2018){Marchand, P.}, {Commer\c{c}on, B.}
  \& {Chabrier, G.}]{CRAL2018}
{\sc \au{{Marchand, P.}}, \au{{Commer\c{c}on, B.}} \& \au{{Chabrier, G.}}}
  \yr{2018}  \at{Impact of the {H}all effect in star formation and the issue of
  angular momentum conservation}.  \jt{Astron. \& Astroph.}  \bvol{619},
  \pg{A37}.

\bibitem[{Marino} {\em et~al.\/}(2013){Marino}, {Mininni}, {Rosenberg} \&
  {Pouquet}]{Marino2013}
{\sc \au{{Marino}, R.}, \au{{Mininni}, P.~D.}, \au{{Rosenberg}, D.} \&
  \au{{Pouquet}, A.}} \yr{2013}  \at{{Inverse cascades in rotating stratified
  turbulence: {F}ast growth of large scales}}.  \jt{Europhys. Lett.}
  \bvol{102}~(4),  \pg{44006}.

\bibitem[Marino {\em et~al.\/}(2014)Marino, Mininni, Rosenberg \&
  Pouquet]{Marino2014}
{\sc \au{Marino, R.}, \au{Mininni, P.~D.}, \au{Rosenberg, D.~L.} \&
  \au{Pouquet, A.}} \yr{2014}  \at{Large-scale anisotropy in stably stratified
  rotating flows}.  \jt{Phys. Rev. E}  \bvol{90},  \pg{023018}.

\bibitem[Marino {\em et~al.\/}(2015{\natexlab{{\em a\/}}})Marino, Pouquet \&
  Rosenberg]{Marino2015}
{\sc \au{Marino, R.}, \au{Pouquet, A.} \& \au{Rosenberg, D.}}
  \yr{2015{\natexlab{{\em a\/}}}}  \at{Resolving the paradox of oceanic
  large-scale balance and small-scale mixing}.  \jt{Phys. Rev. Lett.}
  \bvol{114},  \pg{114504}.

\bibitem[Marino {\em et~al.\/}(2015{\natexlab{{\em b\/}}})Marino, Rosenberg,
  Herbert \& Pouquet]{Marino2015a}
{\sc \au{Marino, Raffaele}, \au{Rosenberg, Duane}, \au{Herbert, Corentin} \&
  \au{Pouquet, Annick}} \yr{2015{\natexlab{{\em b\/}}}}  \at{Interplay of waves
  and eddies in rotating stratified turbulence and the link with
  kinetic-potential energy partition}.  \jt{Europhys. Lett.}  \bvol{112}.

\bibitem[Marino \& Sorriso-Valvo(2023)]{Marino2023}
{\sc \au{Marino, Raffaele} \& \au{Sorriso-Valvo, Luca}} \yr{2023}  \at{Scaling
  laws for the energy transfer in space plasma turbulence}.  \jt{Phys. Rep.}
  \bvol{1006},  \pg{1--144}.

\bibitem[{Marino} {\em et~al.\/}(2008){Marino}, {Sorriso-Valvo}, {Carbone},
  {Noullez}, {Bruno} \& {Bavassano}]{Marino2008}
{\sc \au{{Marino}, R.}, \au{{Sorriso-Valvo}, L.}, \au{{Carbone}, V.},
  \au{{Noullez}, A.}, \au{{Bruno}, R.} \& \au{{Bavassano}, B.}} \yr{2008}
  \at{{Heating the solar wind by a magnetohydrodynamic turbulent energy
  cascade}}.  \jt{Astroph. J. Lett.}  \bvol{677}~(1),  \pg{L71}.

\bibitem[{Marino} {\em et~al.\/}(2011){Marino}, {Sorriso-Valvo}, {Carbone},
  {Veltri}, {Noullez} \& {Bruno}]{Marino2011}
{\sc \au{{Marino}, Raffaele}, \au{{Sorriso-Valvo}, Luca}, \au{{Carbone},
  Vincenzo}, \au{{Veltri}, Pierluigi}, \au{{Noullez}, Alain} \& \au{{Bruno},
  Roberto}} \yr{2011}  \at{{The magnetohydrodynamic turbulent cascade in the
  ecliptic solar wind: {S}tudy of {U}lysses data}}.  \jt{Plan. and Sp. Sc.}
  \bvol{59}~(7),  \pg{592--597}.

\bibitem[{Marino} {\em et~al.\/}(2012){Marino}, {Sorriso-Valvo}, {D'Amicis},
  {Carbone}, {Bruno} \& {Veltri}]{Marino2012}
{\sc \au{{Marino}, R.}, \au{{Sorriso-Valvo}, L.}, \au{{D'Amicis}, R.},
  \au{{Carbone}, V.}, \au{{Bruno}, R.} \& \au{{Veltri}, P.}} \yr{2012}  \at{{On
  the occurrence of the third-order scaling in high latitude solar wind}}.
  \jt{Astroph. J.}  \bvol{750}~(1),  \pg{41}.

\bibitem[Martínez {\em et~al.\/}(1994)Martínez, Chen \&
  Matthaeus]{Martinez1994}
{\sc \au{Martínez, Daniel~O.}, \au{Chen, Shiyi} \& \au{Matthaeus, William~H.}}
  \yr{1994}  \at{Lattice {B}oltzmann magnetohydrodynamics}.  \jt{Phys. of Pl.}
  \bvol{1}~(6),  \pg{1850--1867}.

\bibitem[{Matthaeus} {\em et~al.\/}(2008){Matthaeus}, {Weygand}, {Chuychai},
  {Dasso}, {Smith} \& {Kivelson}]{Matthaeus}
{\sc \au{{Matthaeus}, W.~H.}, \au{{Weygand}, J.~M.}, \au{{Chuychai}, P.},
  \au{{Dasso}, S.}, \au{{Smith}, C.~W.} \& \au{{Kivelson}, M.~G.}} \yr{2008}
  \at{{Interplanetary magnetic {T}aylor microscale and implications for plasma
  dissipation}}.  \jt{Astroph. J. Lett.}  \bvol{678}~(2),  \pg{L141}.

\bibitem[Mendoza \& Mu\~noz(2008)]{Mendoza2008}
{\sc \au{Mendoza, M.} \& \au{Mu\~noz, J.~D.}} \yr{2008}  \at{Three-dimensional
  lattice boltzmann model for magnetic reconnection}.  \jt{Phys. Rev. E}
  \bvol{77},  \pg{026713}.

\bibitem[Meyrand \& Galtier(2012)]{Meyrand2012}
{\sc \au{Meyrand, Romain} \& \au{Galtier, S\'ebastien}} \yr{2012}
  \at{Spontaneous chiral symmetry breaking of {H}all magnetohydrodynamic
  turbulence}.  \jt{Phys. Rev. Lett.}  \bvol{109},  \pg{194501}.

\bibitem[Mininni {\em et~al.\/}(2002)Mininni, G{\'{o}}mez \&
  Mahajan]{Mininni2002}
{\sc \au{Mininni, Pablo~D.}, \au{G{\'{o}}mez, Daniel~O.} \& \au{Mahajan,
  Swadesh~M.}} \yr{2002}  \at{Dynamo action in {H}all magnetohydrodynamics}.
  \jt{Astroph. J.}  \bvol{567}~(1),  \pg{L81--L83}.

\bibitem[Mininni {\em et~al.\/}(2003)Mininni, Gomez \& Mahajan]{Mininni2003}
{\sc \au{Mininni, Pablo~D.}, \au{Gomez, Daniel~O.} \& \au{Mahajan, Swadesh~M.}}
  \yr{2003}  \at{Dynamo action in magnetohydrodynamics and
  {H}all-magnetohydrodynamics}.  \jt{Astroph. J.}  \bvol{587}~(1),
  \pg{472--481}.

\bibitem[Mininni {\em et~al.\/}(2005)Mininni, Gomez \& Mahajan]{Mininni2005}
{\sc \au{Mininni, Pablo~D.}, \au{Gomez, Daniel~O.} \& \au{Mahajan, Swadesh~M.}}
  \yr{2005}  \at{Direct simulations of helical {H}all-{MHD} turbulence and
  dynamo action}.  \jt{Astroph. J.}  \bvol{619}~(2),  \pg{1019--1027}.

\bibitem[Mininni {\em et~al.\/}(2006)Mininni, Pouquet \&
  Montgomery]{Mininni2006}
{\sc \au{Mininni, P.~D.}, \au{Pouquet, A.~G.} \& \au{Montgomery, D.~C.}}
  \yr{2006}  \at{Small-scale structures in three-dimensional
  magnetohydrodynamic turbulence}.  \jt{Phys. Rev. Lett.}  \bvol{97},
  \pg{244503}.

\bibitem[Mininni {\em et~al.\/}(2011)Mininni, Rosenberg, Reddy \&
  Pouquet]{Mininni2011}
{\sc \au{Mininni, Pablo~D.}, \au{Rosenberg, Duane}, \au{Reddy, Raghu} \&
  \au{Pouquet, Annick}} \yr{2011}  \at{A hybrid {MPI}–{O}pen{MP} scheme for
  scalable parallel pseudospectral computations for fluid turbulence}.
  \jt{Parall. Comp.}  \bvol{37}~(6),  \pg{316--326}.

\bibitem[Miura \& Araki(2014)]{Miura2014}
{\sc \au{Miura, H.} \& \au{Araki, K.}} \yr{2014}  \at{Structure transitions
  induced by the {H}all term in homogeneous and isotropic magnetohydrodynamic
  turbulence}.  \jt{Phys. of Pl.}  \bvol{21}~(7),  \pg{072313}.

\bibitem[Montgomery \& Doolen(1987)]{Montgomery1987}
{\sc \au{Montgomery, David} \& \au{Doolen, Gary~D.}} \yr{1987}
  \at{Magnetohydrodynamic cellular automata}.  \jt{Phys. Lett. A}  \bvol{120},
  \pg{229--231}.

\bibitem[Morales {\em et~al.\/}(2005)Morales, Dasso \& Gómez]{Morales2005}
{\sc \au{Morales, Laura}, \au{Dasso, S.} \& \au{Gómez, Daniel}} \yr{2005}
  \at{{H}all effect in incompressible magnetic reconnection}.  \jt{J. of Geoph.
  Res.}  \bvol{110}.

\bibitem[{M{\"u}ller} {\em et~al.\/}(2020){M{\"u}ller}, {St. Cyr},
  {Zouganelis}, {Gilbert}, {Marsden}, {Nieves-Chinchilla}, {Antonucci},
  {Auch{\`e}re}, {Berghmans}, {Horbury}, {Howard}, {Krucker}, {Maksimovic},
  {Owen}, {Rochus}, {Rodriguez-Pacheco}, {Romoli}, {Solanki}, {Bruno},
  {Carlsson}, {Fludra}, {Harra}, {Hassler}, {Livi}, {Louarn}, {Peter},
  {Sch{\"u}hle}, {Teriaca}, {del Toro Iniesta}, {Wimmer-Schweingruber},
  {Marsch}, {Velli}, {De Groof}, {Walsh} \& {Williams}]{2020A&A...642A...1M}
{\sc \au{{M{\"u}ller}, D.}, \au{{St. Cyr}, O.~C.}, \au{{Zouganelis}, I.},
  \au{{Gilbert}, H.~R.}, \au{{Marsden}, R.}, \au{{Nieves-Chinchilla}, T.},
  \au{{Antonucci}, E.}, \au{{Auch{\`e}re}, F.}, \au{{Berghmans}, D.},
  \au{{Horbury}, T.~S.}, \au{{Howard}, R.~A.}, \au{{Krucker}, S.},
  \au{{Maksimovic}, M.}, \au{{Owen}, C.~J.}, \au{{Rochus}, P.},
  \au{{Rodriguez-Pacheco}, J.}, \au{{Romoli}, M.}, \au{{Solanki}, S.~K.},
  \au{{Bruno}, R.}, \au{{Carlsson}, M.}, \au{{Fludra}, A.}, \au{{Harra}, L.},
  \au{{Hassler}, D.~M.}, \au{{Livi}, S.}, \au{{Louarn}, P.}, \au{{Peter}, H.},
  \au{{Sch{\"u}hle}, U.}, \au{{Teriaca}, L.}, \au{{del Toro Iniesta}, J.~C.},
  \au{{Wimmer-Schweingruber}, R.~F.}, \au{{Marsch}, E.}, \au{{Velli}, M.},
  \au{{De Groof}, A.}, \au{{Walsh}, A.} \& \au{{Williams}, D.}} \yr{2020}
  \at{{The {S}olar {O}rbiter mission. {S}cience overview}}.  \jt{Astron.
  Astroph.}  \bvol{642},  \pg{A1}.

\bibitem[{Norman} \& {Heyvaerts}(1985)]{Norman1985}
{\sc \au{{Norman}, C.} \& \au{{Heyvaerts}, J.}} \yr{1985}  \at{{Anomalous
  magnetic field diffusion during star formation}}.  \jt{Astron. \& Astroph.}
  \bvol{147}~(2),  \pg{247--256}.

\bibitem[{Orszag} \& {Tang}(1979)]{Orszag1979}
{\sc \au{{Orszag}, S.~A.} \& \au{{Tang}, C.~M.}} \yr{1979}  \at{{Small-scale
  structure of two-dimensional magnetohydrodynamic turbulence}}.  \jt{J. of Fl.
  Mech.}  \bvol{90},  \pg{129--143}.

\bibitem[Pandey \& Wardle(2008)]{Pandey2008}
{\sc \au{Pandey, B.~P.} \& \au{Wardle, Mark}} \yr{2008}  \at{{H}all
  magnetohydrodynamics of partially ionized plasmas}.  \jt{Mon. Not. of the R.
  Astron. Soc.}  \bvol{385}~(4),  \pg{2269--2278}.

\bibitem[Papini {\em et~al.\/}(2019)Papini, Franci, Landi, Verdini, Matteini \&
  Hellinger]{Papini2019}
{\sc \au{Papini, Emanuele}, \au{Franci, Luca}, \au{Landi, Simone}, \au{Verdini,
  Andrea}, \au{Matteini, Lorenzo} \& \au{Hellinger, Petr}} \yr{2019}  \at{Can
  {H}all magnetohydrodynamics explain plasma turbulence at sub-ion scales?}
  \jt{Astroph. J.}  \bvol{870}~(1),  \pg{52}.

\bibitem[{Parashar} {\em et~al.\/}(2019){Parashar}, {Cuesta} \&
  {Matthaeus}]{Tulasi}
{\sc \au{{Parashar}, T.~N.}, \au{{Cuesta}, M.} \& \au{{Matthaeus}, W.~H.}}
  \yr{2019}  \at{{Reynolds number and intermittency in the expanding solar
  wind: {P}redictions based on Voyager observations}}.  \jt{Astroph. J. Lett.}
  \bvol{884}~(2),  \pg{L57}.

\bibitem[Patterson \& Orszag(1971)]{Patterson1971}
{\sc \au{Patterson, G.~S.} \& \au{Orszag, Steven~A.}} \yr{1971}  \at{Spectral
  calculations of isotropic turbulence: {E}fficient removal of aliasing
  interactions}.  \jt{Phys. of Fl.}  \bvol{14}~(11),  \pg{2538--2541}.

\bibitem[Pattison {\em et~al.\/}(2008)Pattison, Premnath, Morley \&
  Abdou]{PATTISON2008}
{\sc \au{Pattison, M.J.}, \au{Premnath, K.N.}, \au{Morley, N.B.} \& \au{Abdou,
  M.A.}} \yr{2008}  \at{Progress in lattice boltzmann methods for
  magnetohydrodynamic flows relevant to fusion applications}.  \jt{Fusion
  Engineering and Design}  \bvol{83}~(4),  \pg{557--572}.

\bibitem[Pouquet \& Marino(2013)]{Pouquet2013}
{\sc \au{Pouquet, A.} \& \au{Marino, R.}} \yr{2013}  \at{Geophysical turbulence
  and the duality of the energy flow across scales}.  \jt{Phys. Rev. Lett.}
  \bvol{111},  \pg{234501}.

\bibitem[Pouquet {\em et~al.\/}(2019)Pouquet, Rosenberg, Stawarz \&
  Marino]{pouquet_helicity}
{\sc \au{Pouquet, A.}, \au{Rosenberg, D.}, \au{Stawarz, J.} \& \au{Marino, R.}}
  \yr{2019}  \at{Helicity dynamics, inverse, and bidirectional cascades in
  fluid and magnetohydrodynamic turbulence: {A} brief review}.  \jt{Earth Space
  Sci.}  \bvol{6},  \pg{1--19}.

\bibitem[Riley {\em et~al.\/}(2008)Riley, Richard \& Girimaji]{Riley}
{\sc \au{Riley, B.}, \au{Richard, J.} \& \au{Girimaji, S.~S}} \yr{2008}
  \at{Progress in lattice boltzmann methods for magnetohydrodynamic schemes in
  turbulence and rectangular jets}.  \jt{Int. J. Mod. Phys. C}  \bvol{19},
  \pg{1211--1220}.

\bibitem[Rosenberg {\em et~al.\/}(2020)Rosenberg, Mininni, Reddy \&
  Pouquet]{rosenberg20}
{\sc \au{Rosenberg, D.}, \au{Mininni, P.D.}, \au{Reddy, R.} \& \au{Pouquet,
  A.}} \yr{2020}  \at{{G}{P}{U} parallelization of a hybrid pseudospectral
  geophysical turbulence framework using {C}{U}{D}{A}}.  \jt{Atm.}  \bvol{11},
  \pg{178}.

\bibitem[Sahraoui {\em et~al.\/}(2009)Sahraoui, Goldstein, Robert \&
  Khotyaintsev]{Sar2009}
{\sc \au{Sahraoui, F.}, \au{Goldstein, M.~L.}, \au{Robert, P.} \&
  \au{Khotyaintsev, Yu.~V.}} \yr{2009}  \at{Evidence of a cascade and
  dissipation of solar-wind turbulence at the electron gyroscale}.  \jt{Phys.
  Rev. Lett.}  \bvol{102},  \pg{231102}.

\bibitem[Shan \& He(1998)]{Shan1998}
{\sc \au{Shan, Xiaowen} \& \au{He, Xiaoyi}} \yr{1998}  \at{Discretization of
  the velocity space in the solution of the {B}oltzmann equation}.  \jt{Phys.
  Rev. Lett.}  \bvol{80},  \pg{65--68}.

\bibitem[Shen {\em et~al.\/}(2018)Shen, Li, Pullin, Samtaney \&
  Wheatley]{Shen2018}
{\sc \au{Shen, Naijian}, \au{Li, Yuan}, \au{Pullin, D.~I.}, \au{Samtaney, Ravi}
  \& \au{Wheatley, Vincent}} \yr{2018}  \at{On the magnetohydrodynamic limits
  of the ideal two-fluid plasma equations}.  \jt{Physics of Plasmas}
  \bvol{25}~(12),  \pg{122113}.

\bibitem[{Shi} {\em et~al.\/}(2021){Shi}, {Velli}, {Panasenco}, {Tenerani},
  {R{\'e}ville}, {Bale}, {Kasper}, {Korreck}, {Bonnell}, {Dudok de Wit},
  {Malaspina}, {Goetz}, {Harvey}, {MacDowall}, {Pulupa}, {Case}, {Larson},
  {Verniero}, {Livi}, {Stevens}, {Whittlesey}, {Maksimovic} \&
  {Moncuquet}]{2021A&A...650A..21S}
{\sc \au{{Shi}, C.}, \au{{Velli}, M.}, \au{{Panasenco}, O.}, \au{{Tenerani},
  A.}, \au{{R{\'e}ville}, V.}, \au{{Bale}, S.~D.}, \au{{Kasper}, J.},
  \au{{Korreck}, K.}, \au{{Bonnell}, J.~W.}, \au{{Dudok de Wit}, T.},
  \au{{Malaspina}, D.~M.}, \au{{Goetz}, K.}, \au{{Harvey}, P.~R.},
  \au{{MacDowall}, R.~J.}, \au{{Pulupa}, M.}, \au{{Case}, A.~W.}, \au{{Larson},
  D.}, \au{{Verniero}, J.~L.}, \au{{Livi}, R.}, \au{{Stevens}, M.},
  \au{{Whittlesey}, P.}, \au{{Maksimovic}, M.} \& \au{{Moncuquet}, M.}}
  \yr{2021}  \at{{Alfv{\'e}nic versus non-{A}lfv{\'e}nic turbulence in the
  inner heliosphere as observed by {P}arker {S}olar {P}robe}}.  \jt{Astron. \&
  Astroph.}  \bvol{650},  \pg{A21}.

\bibitem[Silva \& Semiao(2014)]{SILVA2014}
{\sc \au{Silva, Goncalo} \& \au{Semiao, Viriato}} \yr{2014}  \at{Truncation
  errors and the rotational invariance of three-dimensional lattice models in
  the lattice {B}oltzmann method}.  \jt{J. of Comp. Phys.}  \bvol{269},
  \pg{259--279}.

\bibitem[{Smith} {\em et~al.\/}(2006){Smith}, {Hamilton}, {Vasquez} \&
  {Leamon}]{2006ApJ...645L..85S}
{\sc \au{{Smith}, Charles~W.}, \au{{Hamilton}, Kathleen}, \au{{Vasquez},
  Bernard~J.} \& \au{{Leamon}, Robert~J.}} \yr{2006}  \at{{Dependence of the
  dissipation range spectrum of interplanetary magnetic fluctuations on the
  rate of energy cascade}}.  \jt{Astroph. J. Lett.}  \bvol{645}~(1),
  \pg{L85--L88}.

\bibitem[{Sorriso-Valvo, L.} {\em et~al.\/}(2023){Sorriso-Valvo, L.}, {Marino,
  R.}, {Foldes, R.}, {L\'ev\^eque, E.}, {D\'{}Amicis, R.}, {Bruno, R.},
  {Telloni, D.} \& {Yordanova, E.}]{Sorriso2023}
{\sc \au{{Sorriso-Valvo, L.}}, \au{{Marino, R.}}, \au{{Foldes, R.}},
  \au{{L\'ev\^eque, E.}}, \au{{D\'{}Amicis, R.}}, \au{{Bruno, R.}},
  \au{{Telloni, D.}} \& \au{{Yordanova, E.}}} \yr{2023}  \at{Helios 2
  observations of solar wind turbulence decay in the inner heliosphere}.
  \jt{Astron. \& Astroph.}  \bvol{672},  \pg{A13}.

\bibitem[Sovinec \& King(2010)]{NIMROD}
{\sc \au{Sovinec, C.R.} \& \au{King, J.R.}} \yr{2010}  \at{Analysis of a mixed
  semi-implicit/implicit algorithm for low-frequency two-fluid plasma
  modeling}.  \jt{Journal of Computational Physics}  \bvol{229}~(16),
  \pg{5803--5819}.

\bibitem[Succi {\em et~al.\/}(1991)Succi, Vergassola \&
  Benzi]{SucciVergassolaBenzi1991}
{\sc \au{Succi, S.}, \au{Vergassola, M.} \& \au{Benzi, R.}} \yr{1991}
  \at{Lattice {B}oltzmann scheme for two-dimensional magnetohydrodynamics}.
  \jt{Phys. Rev. A}  \bvol{43},  \pg{4521--4524}.

\bibitem[{Telloni} {\em et~al.\/}(2022{\natexlab{{\em a\/}}}){Telloni},
  {Adhikari}, {Zank}, {Hadid}, {S{\'a}nchez-Cano}, {Sorriso-Valvo}, {Zhao},
  {Panasenco}, {Shi}, {Velli}, {Susino}, {Verscharen}, {Milillo}, {Alberti},
  {Narita}, {Verdini}, {Grimani}, {Bruno}, {D'Amicis}, {Perrone}, {Marino},
  {Carbone}, {Califano}, {Malara}, {Stawarz}, {Laker}, {Liberatore}, {Bale},
  {Kasper}, {Heyner}, {de Wit}, {Goetz}, {Harvey}, {MacDowall}, {Malaspina},
  {Pulupa}, {Case}, {Korreck}, {Larson}, {Livi}, {Stevens}, {Whittlesey},
  {Auster} \& {Richter}]{2022ApJ...938L...8T}
{\sc \au{{Telloni}, Daniele}, \au{{Adhikari}, Laxman}, \au{{Zank}, Gary~P.},
  \au{{Hadid}, Lina~Z.}, \au{{S{\'a}nchez-Cano}, Beatriz}, \au{{Sorriso-Valvo},
  Luca}, \au{{Zhao}, Lingling}, \au{{Panasenco}, Olga}, \au{{Shi}, Chen},
  \au{{Velli}, Marco}, \au{{Susino}, Roberto}, \au{{Verscharen}, Daniel},
  \au{{Milillo}, Anna}, \au{{Alberti}, Tommaso}, \au{{Narita}, Yasuhito},
  \au{{Verdini}, Andrea}, \au{{Grimani}, Catia}, \au{{Bruno}, Roberto},
  \au{{D'Amicis}, Raffaella}, \au{{Perrone}, Denise}, \au{{Marino}, Raffaele},
  \au{{Carbone}, Francesco}, \au{{Califano}, Francesco}, \au{{Malara},
  Francesco}, \au{{Stawarz}, Julia~E.}, \au{{Laker}, Ronan}, \au{{Liberatore},
  Alessandro}, \au{{Bale}, Stuart~D.}, \au{{Kasper}, Justin~C.}, \au{{Heyner},
  Daniel}, \au{{de Wit}, Thierry~Dudok}, \au{{Goetz}, Keith}, \au{{Harvey},
  Peter~R.}, \au{{MacDowall}, Robert~J.}, \au{{Malaspina}, David~M.},
  \au{{Pulupa}, Marc}, \au{{Case}, Anthony~W.}, \au{{Korreck}, Kelly~E.},
  \au{{Larson}, Davin}, \au{{Livi}, Roberto}, \au{{Stevens}, Michael~L.},
  \au{{Whittlesey}, Phyllis}, \au{{Auster}, Hans-Ulrich} \& \au{{Richter},
  Ingo}} \yr{2022{\natexlab{{\em a\/}}}}  \at{{Observation and modeling of the
  solar wind turbulence evolution in the sub-{M}ercury inner heliosphere}}.
  \jt{Astroph. J. Lett.}  \bvol{938}~(2),  \pg{L8}.

\bibitem[{Telloni} {\em et~al.\/}(2019){Telloni}, {Carbone}, {Bruno}, {Zank},
  {Sorriso-Valvo} \& {Mancuso}]{2019ApJ...885L...5T}
{\sc \au{{Telloni}, Daniele}, \au{{Carbone}, Francesco}, \au{{Bruno}, Roberto},
  \au{{Zank}, Gary~P.}, \au{{Sorriso-Valvo}, Luca} \& \au{{Mancuso},
  Salvatore}} \yr{2019}  \at{{Ion cyclotron waves in field-aligned solar wind
  turbulence}}.  \jt{Astroph. J. Lett.}  \bvol{885}~(1),  \pg{L5}.

\bibitem[{Telloni} {\em et~al.\/}(2021){Telloni}, {Sorriso-Valvo}, {Woodham},
  {Panasenco}, {Velli}, {Carbone}, {Zank}, {Bruno}, {Perrone}, {Nakanotani},
  {Shi}, {D'Amicis}, {De Marco}, {Jagarlamudi}, {Steinvall}, {Marino},
  {Adhikari}, {Zhao}, {Liang}, {Tenerani}, {Laker}, {Horbury}, {Bale},
  {Pulupa}, {Malaspina}, {MacDowall}, {Goetz}, {de Wit}, {Harvey}, {Kasper},
  {Korreck}, {Larson}, {Case}, {Stevens}, {Whittlesey}, {Livi}, {Owen}, {Livi},
  {Louarn}, {Antonucci}, {Romoli}, {O'Brien}, {Evans} \&
  {Angelini}]{2021ApJ...912L..21T}
{\sc \au{{Telloni}, Daniele}, \au{{Sorriso-Valvo}, Luca}, \au{{Woodham},
  Lloyd~D.}, \au{{Panasenco}, Olga}, \au{{Velli}, Marco}, \au{{Carbone},
  Francesco}, \au{{Zank}, Gary~P.}, \au{{Bruno}, Roberto}, \au{{Perrone},
  Denise}, \au{{Nakanotani}, Masaru}, \au{{Shi}, Chen}, \au{{D'Amicis},
  Raffaella}, \au{{De Marco}, Rossana}, \au{{Jagarlamudi}, Vamsee~K.},
  \au{{Steinvall}, Konrad}, \au{{Marino}, Raffaele}, \au{{Adhikari}, Laxman},
  \au{{Zhao}, Lingling}, \au{{Liang}, Haoming}, \au{{Tenerani}, Anna},
  \au{{Laker}, Ronan}, \au{{Horbury}, Timothy~S.}, \au{{Bale}, Stuart~D.},
  \au{{Pulupa}, Marc}, \au{{Malaspina}, David~M.}, \au{{MacDowall}, Robert~J.},
  \au{{Goetz}, Keith}, \au{{de Wit}, Thierry~Dudok}, \au{{Harvey}, Peter~R.},
  \au{{Kasper}, Justin~C.}, \au{{Korreck}, Kelly~E.}, \au{{Larson}, Davin},
  \au{{Case}, Anthony~W.}, \au{{Stevens}, Michael~L.}, \au{{Whittlesey},
  Phyllis}, \au{{Livi}, Roberto}, \au{{Owen}, Christopher~J.}, \au{{Livi},
  Stefano}, \au{{Louarn}, Philippe}, \au{{Antonucci}, Ester}, \au{{Romoli},
  Marco}, \au{{O'Brien}, Helen}, \au{{Evans}, Vincent} \& \au{{Angelini},
  Virginia}} \yr{2021}  \at{{Evolution of solar wind turbulence from 0.1 to 1
  au during the first {P}arker {S}olar {P}robe-{S}olar {O}rbiter radial
  alignment}}.  \jt{Astroph. J. Lett.}  \bvol{912}~(2),  \pg{L21}.

\bibitem[{Telloni} {\em et~al.\/}(2022{\natexlab{{\em b\/}}}){Telloni}, {Zank},
  {Stangalini}, {Downs}, {Liang}, {Nakanotani}, {Andretta}, {Antonucci},
  {Sorriso-Valvo}, {Adhikari}, {Zhao}, {Marino}, {Susino}, {Grimani}, {Fabi},
  {D'Amicis}, {Perrone}, {Bruno}, {Carbone}, {Mancuso}, {Romoli}, {Deppo},
  {Fineschi}, {Heinzel}, {Moses}, {Naletto}, {Nicolini}, {Spadaro}, {Teriaca},
  {Frassati}, {Jerse}, {Landini}, {Pancrazzi}, {Russano}, {Sasso}, {Biondo},
  {Burtovoi}, {Capuano}, {Casini}, {Casti}, {Chioetto}, {Leo}, {Giarrusso},
  {Liberatore}, {Berghmans}, {Auch{\`e}re}, {Cuadrado}, {Chitta}, {Harra},
  {Kraaikamp}, {Long}, {Mandal}, {Parenti}, {Pelouze}, {Peter}, {Rodriguez},
  {Sch{\"u}hle}, {Schwanitz}, {Smith}, {Verbeeck} \&
  {Zhukov}]{2022ApJ...936L..25T}
{\sc \au{{Telloni}, Daniele}, \au{{Zank}, Gary~P.}, \au{{Stangalini}, Marco},
  \au{{Downs}, Cooper}, \au{{Liang}, Haoming}, \au{{Nakanotani}, Masaru},
  \au{{Andretta}, Vincenzo}, \au{{Antonucci}, Ester}, \au{{Sorriso-Valvo},
  Luca}, \au{{Adhikari}, Laxman}, \au{{Zhao}, Lingling}, \au{{Marino},
  Raffaele}, \au{{Susino}, Roberto}, \au{{Grimani}, Catia}, \au{{Fabi},
  Michele}, \au{{D'Amicis}, Raffaella}, \au{{Perrone}, Denise}, \au{{Bruno},
  Roberto}, \au{{Carbone}, Francesco}, \au{{Mancuso}, Salvatore}, \au{{Romoli},
  Marco}, \au{{Deppo}, Vania~Da}, \au{{Fineschi}, Silvano}, \au{{Heinzel},
  Petr}, \au{{Moses}, John~D.}, \au{{Naletto}, Giampiero}, \au{{Nicolini},
  Gianalfredo}, \au{{Spadaro}, Daniele}, \au{{Teriaca}, Luca}, \au{{Frassati},
  Federica}, \au{{Jerse}, Giovanna}, \au{{Landini}, Federico}, \au{{Pancrazzi},
  Maurizio}, \au{{Russano}, Giuliana}, \au{{Sasso}, Clementina}, \au{{Biondo},
  Ruggero}, \au{{Burtovoi}, Aleksandr}, \au{{Capuano}, Giuseppe~E.},
  \au{{Casini}, Chiara}, \au{{Casti}, Marta}, \au{{Chioetto}, Paolo},
  \au{{Leo}, Yara~De}, \au{{Giarrusso}, Marina}, \au{{Liberatore}, Alessandro},
  \au{{Berghmans}, David}, \au{{Auch{\`e}re}, Fr{\'e}d{\'e}ric},
  \au{{Cuadrado}, Regina~Aznar}, \au{{Chitta}, Lakshmi~P.}, \au{{Harra},
  Louise}, \au{{Kraaikamp}, Emil}, \au{{Long}, David~M.}, \au{{Mandal}, Sudip},
  \au{{Parenti}, Susanna}, \au{{Pelouze}, Gabriel}, \au{{Peter}, Hardi},
  \au{{Rodriguez}, Luciano}, \au{{Sch{\"u}hle}, Udo}, \au{{Schwanitz}, Conrad},
  \au{{Smith}, Phil~J.}, \au{{Verbeeck}, Cis} \& \au{{Zhukov}, Andrei~N.}}
  \yr{2022{\natexlab{{\em b\/}}}}  \at{{Observation of a magnetic switchback in
  the solar corona}}.  \jt{Astroph. J. Lett.}  \bvol{936}~(2),  \pg{L25}.

\bibitem[Tóth {\em et~al.\/}(2008)Tóth, Ma \& Gombosi]{TOTH2008}
{\sc \au{Tóth, Gábor}, \au{Ma, Yingjuan} \& \au{Gombosi, Tamas~I.}} \yr{2008}
   \at{{H}all magnetohydrodynamics on block-adaptive grids}.  \jt{J. of Comp.
  Phys.}  \bvol{227}~(14),  \pg{6967--6984}.

\bibitem[Wang {\em et~al.\/}(2001)Wang, Bhattacharjee \& Ma]{Wang2001}
{\sc \au{Wang, Xiaogang}, \au{Bhattacharjee, A.} \& \au{Ma, Z.~W.}} \yr{2001}
  \at{Scaling of collisionless forced reconnection}.  \jt{Phys. Rev. Lett.}
  \bvol{87},  \pg{265003}.

\bibitem[Xia \& Yang(2015)]{Xia15}
{\sc \au{Xia, Zhenwei} \& \au{Yang, Weihong}} \yr{2015}  \at{Exact solutions of
  the incompressible dissipative {H}all magnetohydrodynamics}.  \jt{Phys. of
  Plas.}  \bvol{22}~(3),  \pg{032306}.

\bibitem[{Yadav} {\em et~al.\/}(2022){Yadav}, {Miura} \& {Pandit}]{Yadav2022}
{\sc \au{{Yadav}, Sharad~K.}, \au{{Miura}, Hideaki} \& \au{{Pandit}, Rahul}}
  \yr{2022}  \at{{Statistical properties of three-dimensional {H}all
  magnetohydrodynamics turbulence}}.  \jt{Phys. of Fl.}  \bvol{34}~(9),
  \pg{095135}.

\end{thebibliography}

\bigskip

\begin{verbatim}
For the purpose of Open Access, a CC-BY public copyright licence has been 
applied by the authors to the present document and will be applied to all 
subsequent versions up to the Author Accepted Manuscript arising from this
submission
\end{verbatim}


\end{document}